\documentclass{article}

    \PassOptionsToPackage{numbers, compress}{natbib}

\usepackage[final]{neurips_2025}

\usepackage[utf8]{inputenc} 
\usepackage[T1]{fontenc}    
\usepackage{url}            
\usepackage{booktabs}       
\usepackage{amsfonts}       
\usepackage{nicefrac}       
\usepackage{microtype}      
\usepackage{xcolor}         
\usepackage{wrapfig}
\usepackage{times}
\usepackage{epsfig}
\usepackage{graphicx}
\usepackage{amsmath}
\usepackage{amssymb}

\usepackage{float}
\usepackage{pifont}
\usepackage{footnote}
\usepackage{enumitem}
\usepackage{bm}
\usepackage{arydshln}
\usepackage{booktabs}
\usepackage{multicol}
\usepackage{multirow}
\usepackage{color}
\usepackage{xcolor}     
\usepackage{colortbl}
\usepackage{soul}
\usepackage{bbding}
\usepackage{makecell}
\usepackage{mathtools}
\usepackage{imakeidx}
\usepackage{amssymb}
\usepackage{graphicx}
\usepackage{amsmath}
\usepackage{threeparttable}
\definecolor{citecolor}{HTML}{0071BC}
\definecolor{linkcolor}{HTML}{ED1C24}
\usepackage[colorlinks,
            anchorcolor=red,
            citecolor=citecolor, 
            linkcolor=linkcolor,
            ]{hyperref}

\usepackage{arydshln}
\usepackage{lipsum}
\usepackage[toc]{multitoc}
\usepackage[edges]{forest}
\usepackage[normalem]{ulem}

\usepackage{bbding}
\usepackage[most]{tcolorbox}

\usepackage{algorithm}
\usepackage{algorithmic}

\usepackage{minitoc}
\usepackage[toc,page,header]{appendix}

\definecolor{orchid}{rgb}{0.85, 0.44, 0.84}
\definecolor{rubinred}{rgb}{0.82, 0.0, 0.28}
\definecolor{flagship}{rgb}{0.93, 0.06, 0.41}
\definecolor{radiologist}{rgb}{0.50, 0.50, 1}
\definecolor{YT}{HTML}{002FA7}

\newcommand{\loss}{{\fontfamily{ppl}\selectfont CARE}}

\newcolumntype{P}[1]{>{\centering\arraybackslash}p{#1}}
\newlength\savewidth

\title{Are Pixel-Wise Metrics Reliable for Sparse-View Computed Tomography Reconstruction?}

\author{
Tianyu Lin\textsuperscript{1} \quad 
Xinran Li\textsuperscript{1,2} \quad 
Chuntung Zhuang\textsuperscript{1} \quad 
Qi Chen\textsuperscript{1} \quad 
Yuanhao Cai\textsuperscript{1} \quad \\
\textbf{Kai Ding}\textsuperscript{3} \quad 
\textbf{Alan L. Yuille}\textsuperscript{1} \quad 
\textbf{Zongwei Zhou}\textsuperscript{1,}\thanks{Correspondence to Zongwei Zhou (\href{mailto:zzhou82@jh.edu}{\textsc{zzhou82@jh.edu}})} \\[2.5mm]
\textsuperscript{1}Johns Hopkins University \quad
\textsuperscript{2}Yale University \quad
\textsuperscript{3}Johns Hopkins Medicine \\[1.5mm]%
{\small Code, dataset, and models:~\href{https://github.com/MrGiovanni/CARE}{https://github.com/MrGiovanni/CARE}}
}
\begin{document}
\maketitle

\doparttoc 
\faketableofcontents 
 
\begin{figure}[ht]
    \includegraphics[width=\linewidth]{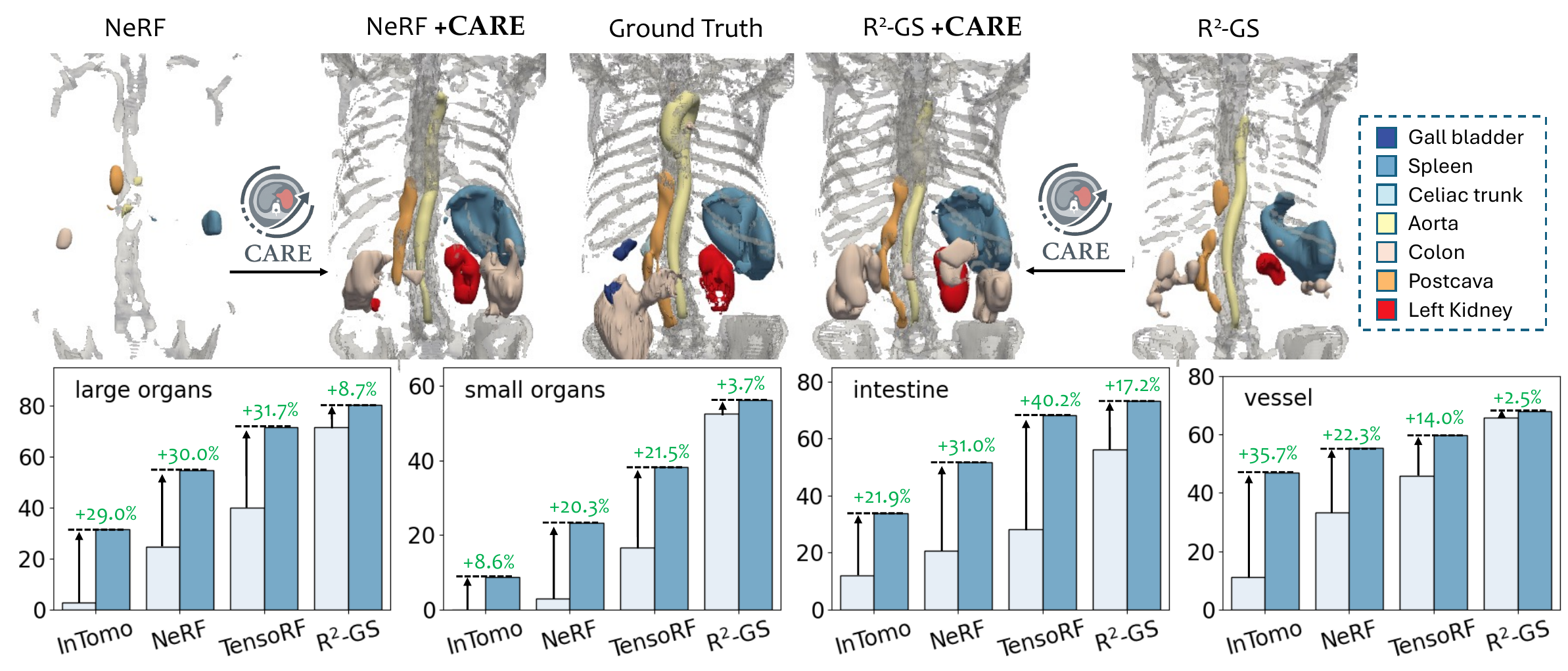}
    \caption{
        \textbf{\loss\ Improves Structural Completeness in Sparse-View CT Reconstruction.}
        \emph{Top}: Qualitative comparison of reconstructed CT scans from NeRF and R$^2$-GS, with and without our proposed \loss\ on clinically important structures. \emph{Bottom}: Quantitative gains in structural completeness across four reconstruction methods. \loss\ consistently improves results, confirming its model-agnostic and plug-and-play nature.
        }
    \label{fig:visual}
\end{figure} 

\begin{abstract} 
Widely adopted evaluation metrics for sparse-view CT reconstruction---such as Structural Similarity Index Measure and Peak Signal-to-Noise Ratio---prioritize pixel-wise fidelity but often fail to capture the completeness of critical anatomical structures, particularly small or thin regions that are easily missed. To address this limitation, we propose a suite of novel anatomy-aware evaluation metrics designed to assess structural completeness across anatomical structures, including large organs, small organs, intestines, and vessels. Building on these metrics, we introduce \loss, a Completeness-Aware Reconstruction Enhancement framework that incorporates structural penalties during training to encourage anatomical preservation of significant structures. \loss\ is model-agnostic and can be seamlessly integrated into analytical, implicit, and generative methods. 
When applied to these methods, \loss\ substantially improves structural completeness in CT reconstructions, achieving up to \textbf{+32\%} improvement for large organs, \textbf{+22\%} for small organs, \textbf{+40\%} for intestines, and \textbf{+36\%} for vessels.

\end{abstract}

\section{Introduction}\label{sec:introduction}
Reducing radiation dose is a major concern in computed tomography (CT), where a standard image acquisition typically requires 800 to 1500 X-ray projections~\cite{mostafapour2024ultra,davidolawade2025aidriven}. These high doses pose long-term health risks, especially for vulnerable patients such as children, pregnant individuals, and those undergoing frequent imaging~\cite{zhang2018regularization,cao2023large,koch2024radiation}. One promising research direction to lower radiation is \emph{sparse-view CT}, which aims to reconstruct high-quality images from a significantly reduced number of projection views---often fewer than 50 views~\cite{zhang2025xlrm,tan2024msdiff,yang2024ct}. While this approach can dramatically lower radiation dose, it introduces significant challenges: reconstructions from limited data often lose fine anatomical details~\cite{lantz2024detectability}. These details are critical for clinical applications such as surgical planning, radiotherapy, and longitudinal disease monitoring~\cite{yang2024biplanar,xu2024plpd}. This raises a central research question: \emph{how can we ensure the completeness of clinically important anatomical structures in sparse-view CT reconstruction?}

Over the past decade, researchers have reported steady improvements in sparse-view CT reconstruction by optimizing pixel-wise metrics---e.g., Peak Signal-to-Noise Ratio (PSNR)~\cite{hore2010image} and Structural Similarity Index Measure (SSIM)~\cite{wang2004image}. \emph{However, higher PSNR or SSIM scores do not guarantee better clinical utility (analyzed in \S\ref{sec:method-motivation}).} These metrics compute global averages of per-voxel intensity differences, which means they are largely insensitive to the presence or absence of small but clinically important anatomical structures. For example, missing the gall-bladder, adrenal glands, celiac artery, or a segmental vein affects only a tiny fraction of the total scan volume---often less than 0.0001\%---and may change PSNR/SSIM by just the third or fourth decimal place. As a result, the reconstructed CT scans that appear high-quality under PSNR/SSIM may still miss important anatomical structures and be unsuitable for clinical applications \cite{chow2016review}. Our internal reader study at Johns Hopkins University, involving 21 board-certified radiologists, assessed CT reconstructions produced by state-of-the-art CT reconstruction methods. The results raised significant concerns, including missing anatomical structures, implausible hallucinations, and severe artifacts, suggesting a growing disconnect between algorithmic novelty and clinical realism.
This highlights a pressing need for new evaluation metrics that can directly assess whether anatomical structures are preserved in sparse-view CT reconstruction.

To assess the anatomical structures in the reconstructed CT scans, conventional practice requires radiologists to manually inspect the scans---a process that is subjective, time-consuming, and impractical to scale. Fortunately, recent advances in \textit{medical image segmentation}, powered by strong model architectures~\cite{bassi2024touchstone,isensee2024nnu,liu2023clip,liu2024universal,bassi2025scaling} and large-scale annotated datasets~\cite{wasserthal2022totalsegmentator,li2024abdomenatlas,qu2023annotating,li2025scalemai,chen2025scaling}, now enable AI models to automatically segment multiple anatomical structures with high accuracy and consistency. These advances greatly reduce the need for manual inspection and motivate a key hypothesis: automated segmentation can serve as a scalable metric to assess whether anatomical structures are preserved in reconstructed scans.

We introduce \loss, a \emph{Completeness-Aware Reconstruction Enhancement} framework that defines new evaluation metrics—derived from medical image segmentation—and uses them to directly supervise diffusion models, enhancing the structural completeness of preexisting sparse-view CT reconstruction methods. Our contributions are:
\begin{enumerate}
    \item \textbf{Anatomy-Aware CT Reconstruction Metric}: We leverage nnU-Net (under Apache-2.0 license)~\cite{isensee2021nnu} as an anatomy segmentator, trained on more than 3,000 voxel-wise annotated CT scans \cite{li2024well}, to build a suite of anatomy-aware metrics that measure the anatomical completeness of four clinically important categories of anatomical structures (detailed in \figureautorefname~\ref{fig:metric-pitfall-and-organs}) in reconstructed CT scans. 
    \item \textbf{Anatomy-Aware CT Reconstruction Framework}: We go beyond evaluation by using the anatomy-aware metrics as direct supervision. During training, \loss\ compares the segmentation output of the reconstructed CT scans with the segmentator's prediction and back-propagates a penalty whenever any anatomical structure is missing or incomplete. This structural penalty is fully differentiable and agnostic to the underlying forward model, allowing \loss\ to be applied effectively to a range of preexisting reconstruction methods, including Neural Radiance Fields (NeRF), Gaussian Splatting (GS), and their variants. Instead of training a separate reconstruction model for each patient, using diffusion models to enhance the CT scans reconstructed by existing methods enables the model to learn anatomical priors from a large population and generalize across patients, thus making \loss\ patient-agnostic.
\end{enumerate}

Extensive experiments show that applying \loss\ to preexisting sparse-view CT reconstruction methods produces striking gains (summarized in \figureautorefname~\ref{fig:visual}; detailed in \S\ref{sec:experiment_result}). Compared with nine preexisting methods, our \loss\ achieves up to \textbf{32\%} for large organs, \textbf{+22\%} for small organs, \textbf{+40\%} for intestinal structures, and \textbf{+36\%} for vascular structures. These gains are both statistically significant and clinically important, but worryingly, such improvements are entirely hidden when evaluated using widely adopted pixel-wise metrics like PSNR and SSIM (see Table~\ref{tab:CARE-results}). This highlights a critical pitfall in current evaluation practices and raises concerns about their continued usage in clinical reconstruction studies.

\section{Related Works}\label{sec:related_work}

\textbf{Anatomy-Aware Evaluation.}
Conventional image quality metrics, such as PSNR \cite{hore2010image} and SSIM \cite{wang2004image}, focus primarily on pixel-wise fidelity but fail to reliably reflect clinical significance, particularly regarding the completeness of fine anatomical structures. This limitation is especially pronounced under challenging imaging conditions, including sparse-view or low-dose acquisitions. Despite recent efforts, research explicitly targeting anatomy-aware evaluation metrics remains scarce~\cite{sun2025eagle}.

Beyond pixel-wise fidelity, several perceptual image quality metrics have been proposed to better align with human visual judgment. The Fréchet Inception Distance (FID)~\cite{heusel2017gans} and Kernel Inception Distance (KID)~\cite{binkowski2018demystifying} evaluate distributional similarity between sets of real and generated images using features from a pre-trained Inception network. Similarly, LPIPS~\cite{zhang2018unreasonable} leverages deep features to measure perceptual dissimilarity between individual image pairs. GMSD~\cite{xue2013gradient} assesses quality based on gradient magnitude similarity, capturing structural distortions more effectively than PSNR or SSIM. More recently, GLIPS~\cite{aziz2025global} combines global and local perceptual cues to evaluate photorealism in AI-generated images. While these metrics improve upon traditional pixel-wise measures in natural image domains, they remain agnostic to anatomical semantics and do not explicitly assess the presence, completeness, or geometric fidelity of clinically relevant structures in medical imaging.

\textbf{CT Reconstruction Methods.}
Traditional reconstruction methods, such as Filtered Back Projection (FBP) \cite{ramachandran1971three} and iterative algorithms like Simultaneous Algebraic Reconstruction Technique (SART) \cite{andersen1984simultaneous} and Adaptive Steepest Descent-Projection onto Convex Sets (ASD-POCS) \cite{sidky2008image}, are widely employed due to their simplicity and effectiveness. However, these methods exhibit significant quality degradation under sparse-view or low-dose scenarios \cite{clement2025ai}. Early work by Zhang~\cite{zhang2018regularization} and Chen~\cite{chen2008prior} mitigated this issue by incorporating a prior CT scan as strong regularisation, enabling acceptable image quality with substantially fewer projections. 
To overcome the remaining limitations, recent advances have leveraged neural rendering methods, including NeRF \cite{mildenhall2021nerf}, InTomo \cite{zang2021intratomo}, TensoRF \cite{TensoRF}, Neural Attenuation Fields (NAF) \cite{zha2022naf}, and SAX-NeRF \cite{cai2024structure}, to synthesize views and significantly improve 3D reconstruction fidelity. In parallel, Gaussian splatting methods such as R$^2$-GS \cite{zha2024r} have also emerged, demonstrating promising results in achieving high-quality reconstructions with enhanced anatomical details. 
However, their capability to accurately reconstruct anatomical structures remains unverified, primarily due to the absence of suitable evaluation metrics.

\textbf{Generative AI Models.}
Denoising Diffusion Probabilistic Models~\cite{ho2020denoising,chen2024towards,li2024text} have become strong alternatives to GANs~\cite{goodfellow2014generative} and VAEs~\cite{kingma2014auto} for high-fidelity image synthesis.  
In medicine, Stable-Diffusion–style implementations~\cite{diffusers, lin2024stable} are already used to suppress CT noise and streak artefacts.  
Very recent work shows that diffusion priors can also tackle \emph{sparse-view} CT, reporting state-of-the-art PSNR/SSIM with as few projections~\cite{tan2024msdiff,yang2024ct,sun2024difr3ct}.  
Yet these methods still optimise only pixel-wise losses and may hallucinate or omit small organs and vessels.  
Our \loss\ closes this gap by adding an anatomy-aware segmentation loss to the diffusion objective, explicitly safeguarding clinically critical structures while preserving the low-dose advantage of sparse-view acquisition.

\section{\loss: Completeness-Aware Reconstruction Enhancement}
\label{sec:method}

\begin{figure}[ht]
\centering
    \includegraphics[width=\linewidth]{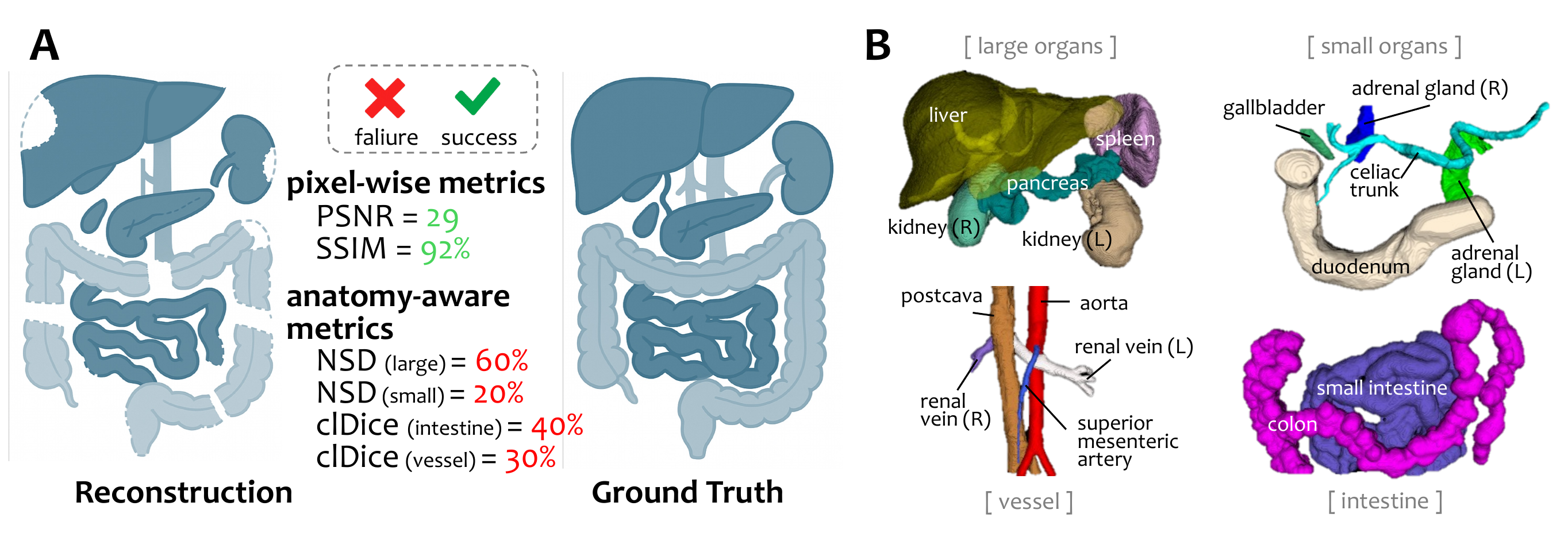}
    \caption{\textbf{Pixel-wise Metrics Overlook Structural Errors in the Focused Anatomical Structures.} \textbf{A.} Example pixel-wise metrics' pitfall. Pixel-wise metrics are insensitive to anatomical preservation failure. \textbf{B.} The four types of organs evaluated by \loss.}
    \label{fig:metric-pitfall-and-organs}
\end{figure}

\subsection{Motivation: Pitfalls of Conventional Pixel–Wise Metrics}
\label{sec:method-motivation}
PSNR and SSIM dominate sparse-view CT reconstruction benchmarks because they are easy to compute and correlate well with global visual fidelity on natural images.  In the clinical setting, however, these per‑voxel averages are blind to errors that matter most.  
\figureautorefname~\ref{fig:metric-pitfall-and-organs} shows a common situation that even though having reasonable pixel-wise metrics, reconstruction could still fail in anatomical preservation.
The reason is straightforward: small organs, intestines and vessels occupy far less than 0.01~\% of the volume, so their absence barely changes the third decimal place of SSIM and PSNR.  Clinical evaluation, by contrast, hinges on whether these structures are present at all.  
Our own experiments (detailed in \S\ref{sec:results_anatomy}) confirm the gap: classical analytical methods such as FDK and SART rank mid‑pack on pixel metrics but top on anatomy‑aware scores, whereas several neural renderers achieve the opposite pattern.  These observations motivate a shift from intensity similarity towards metrics that directly measure anatomical completeness.

\subsection{Anatomy‑Aware CT Reconstruction Metrics}
\label{sec:method-metrics}

\begin{wrapfigure}{r}{0.5\linewidth}
    \includegraphics[width=\linewidth]{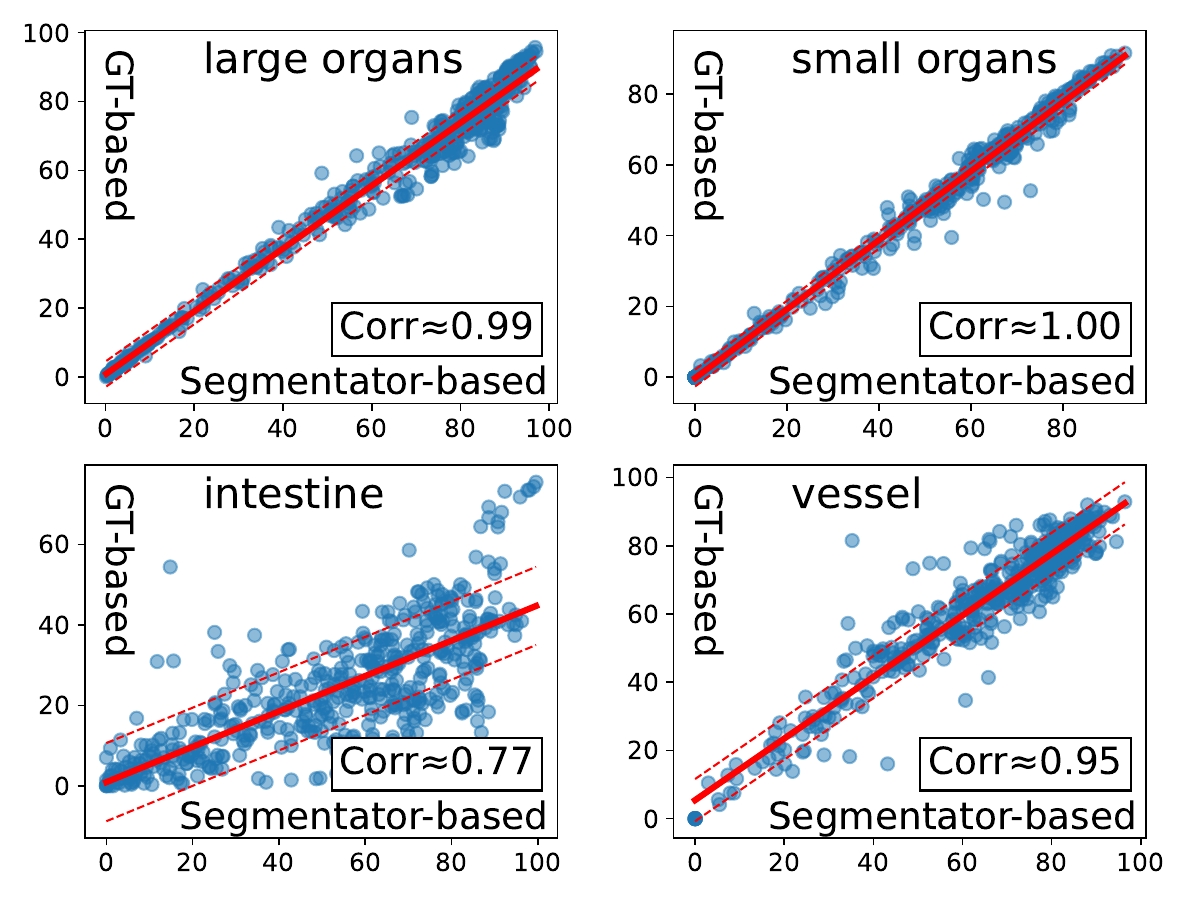}
    \caption{\textbf{Correlation Between Ground Truth (GT) Based and Segmentator Based Anatomy-Aware Metrics.} High correlation scores indicate that the anatomy segmentator is a strong substitute for human expert.}
    \label{fig:fig_seg_corr}
\end{wrapfigure}

We propose segmentation-based anatomy-aware CT reconstruction metrics to evaluate CT reconstruction quality. 
We utilize a frozen multi-organ segmentation model to compute metrics for four different kinds of anatomical structures.
For large and small organs, we use the NSD metric due to its sensitivity to boundary accuracy and tolerance for minor shape deviations.
For vessels and intestines, we apply the clDice~\cite{shit2021cldice} metric, given its effectiveness in capturing tubular structures and preserving topological connectivity.
Finally, we report these metrics separately for each anatomical category\footnote{These anatomical categories are divided based on clinical guidelines (see details in \S\ref{sec:supp-experiments-datasets}).}.

\textbf{Anatomy Segmentator.} 
A great CT reconstruction result should preserve the anatomical structures of the original CT scans, thus having similar segmentation results from a fixed segmentator between the original and reconstructed CT scans.
Instead of relying on manually annotated ground-truth labels to derive our anatomy-aware metrics, we employ a frozen nnU-Net~\cite{isensee2021nnu}---termed the \textit{anatomy segmentator}---to capture clinically relevant fidelity. We choose nnU-Net framework for its self-configuring architecture and consistently state-of-the-art segmentation performance across diverse medical-imaging tasks. It was trained internally on the JHH dataset\footnote{A private CT segmentation dataset from Johns Hopkins Hospital (see details in \S\ref{sec:supp-experiments-datasets-private}).} with 3,151 CT scans annotated over three years by a team of 21 radiologists, with all annotations independently reviewed by one of three senior radiologists not involved in the initial annotation process to ensure quality \cite{li2025scalemai,xia2022felix}. The anatomy segmentator will be publicly released.
Leveraging a segmentator to construct our anatomy-aware metrics enables effortless deployment across large-scale CT reconstruction evaluations, eliminating the dependency on manual annotations. \figureautorefname~\ref{fig:fig_seg_corr} shows that the anatomy-aware metrics building on human annotated ground truth and segmentator's prediction have high correlation, further indicating the robustness of segmentator-based metrics. This further allows our anatomy-aware metrics to be applied to CT scans in the absence of segmentation labels.

\textbf{Normalized Surface Dice (NSD).}  
Both large organs and small organs can be viewed as compact, surface‑dominated objects.  NSD evaluates their geometric agreement by asking what fraction of the predicted surface \(S_P\) and reference surface \(S_G\) lie within a tolerance band of width \(\tau\):
\begin{equation}
\mathrm{NSD}_\tau =
\frac{\bigl|\{x\in S_P: d(x,S_G)\le\tau\}\bigr| + \bigl|\{x\in S_G: d(x,S_P)\le\tau\}\bigr|}
     {|S_P| + |S_G|}
\end{equation}

NSD is \emph{scale‑invariant} since every surface element located farther than $\tau$ incurs the same penalty.  For example, a uniform 3 mm displacement with $\tau=2$ mm therefore drives $\mathrm{NSD}_{2\text{mm}}$ to $0$ regardless how big the organ is.

On the contrary, the widely used Dice similarity coefficient (DSC) is scale-variant. DSC measures volumetric overlap via
$\mathrm{DSC}(P,G)=\frac{2|P\cap G|}{|P|+|G|}$.
As a result, the sensitivity of DSC to the same boundary shift is diluted by organ size.  The expected loss satisfies
$\Delta \mathrm{DSC} \approx\delta/(2R_{\mathrm{eff}})$ with \(R_{\mathrm{eff}}=V_G/S_G\). For instance, with a 50 mm‑radius liver, a 3 mm outer‑shell error lowers DSC by merely 0.06—far too small to flag what clinicians deem a major discrepancy.

By focusing on boundary proximity rather than voxel volume, NSD retains millimetre‑level sensitivity and offers a more faithful assessment of non‑tubular organ segmentations than DSC.

\textbf{Center‑line Dice (clDice).}  
Vessels and intestine are elongated, branching and hollow.  Their clinical relevance stems from \emph{connectivity}: a single broken segment in the superior‑mesenteric artery may compromise whole‑organ perfusion.  clDice operates on skeletonised centre‑lines and is sensitive to both continuity and topology, penalising even thin discontinuities that DSC or NSD might overlook.  
However, the ground-truth annotations for intestinal structures in most public datasets are often noisy or incomplete~\cite{Gibson2018AbdomenFCN}, which likely depresses the intestine–metric correlation by about 0.77 shown in \figureautorefname~\ref{fig:fig_seg_corr}. Curating higher-quality bowel annotations—and re-evaluating clDice under those labels—will therefore be an important direction for future work.

Our anatomy‑aware formulation thus matches the geometric character of each organ class: surface‑sensitive NSD for compact organs, topology‑sensitive clDice for slender, branching tubes.  Together, they provide an automatic, quantitative, and clinically meaningful view through which to evaluate reconstruction algorithms.

\subsection{Anatomy-Aware CT Reconstruction Framework}\label{sec:method-CARE}
\begin{figure}[ht]
    \centering
    \includegraphics[width=\linewidth]{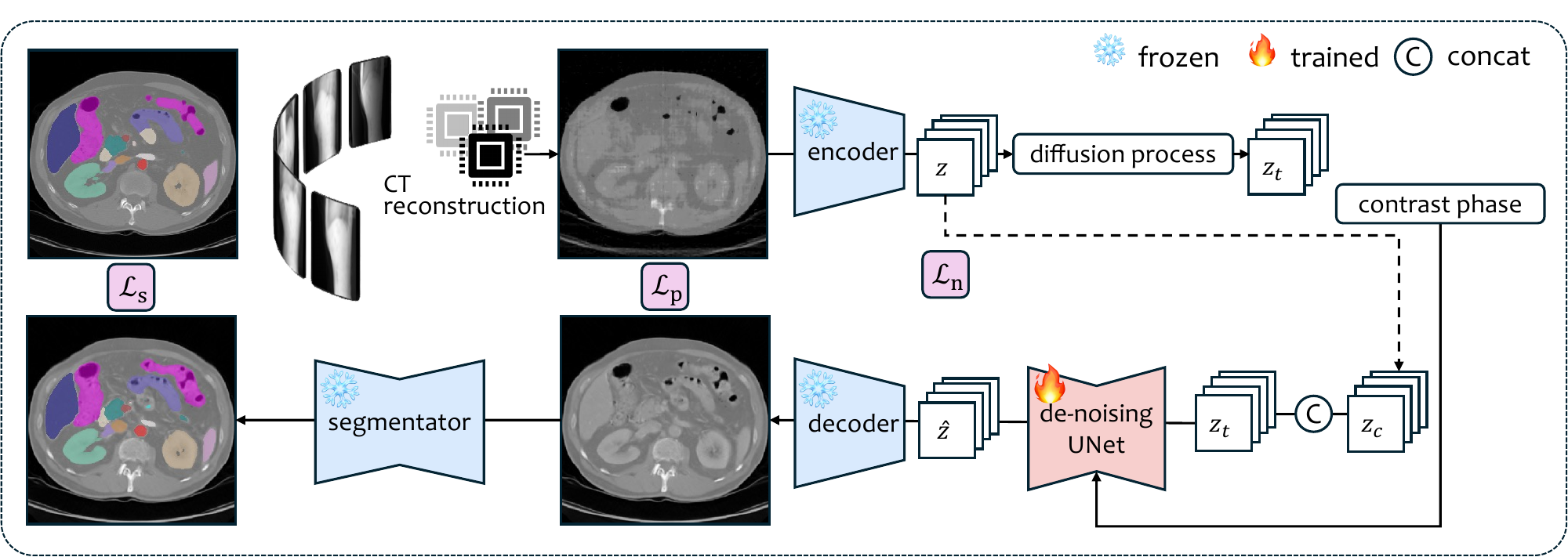}
    \caption{\textbf{\loss\ Framework.} 
    Given the frozen anatomy segmentator, autoencoder, and the pretrained latent diffusion model, we then adapt the latent diffusion model to real reconstructed CT scans. \loss\ can be integrated into any reconstruction method to perform its enhancement capability. The overall training is supervised by three loss terms: the noise-prediction loss $\mathcal{L}_n$, pixel-space reconstruction loss $\mathcal{L}_p$, and anatomy-guidance loss $\mathcal{L}_s$.}   
    \label{fig:framework}
\end{figure}

We propose a simple latent diffusion framework named \loss\ for CT enhancement, as shown in \figureautorefname~\ref{fig:framework}. \loss\ starts by training an autoencoder and a latent diffusion model using a high-quality CT dataset, then integrating anatomy-guided supervision with additional pixel space constraints.

\textbf{Preliminary.}
We begin by adapting a KL‑regularized variational autoencoder to CT domain with the JHH dataset. Let $X$ denote an input CT image in pixel space. Note that to consider inter-frame consistency and retain the original architecture of the autoencoder network, we stack three adjacent CT slices to be the three channels of the input image, giving $X\in\mathbb R^{H\times W\times 3}$. 
The encoder $\mathcal E_{\theta_E}$ maps $X$ to a latent tensor $z=\mathcal E_{\theta_E}(X)\in\mathbb R^{h\times w\times c}$.
A paired decoder $\mathcal D_{\theta_D}$ reconstructs the input as $\hat X=\mathcal D_{\theta_D}(z)$.
The resulting autoencoder model reconstructs scans faithfully yet produces well‑behaved latents that remain fixed for all later stages.

Using this frozen autoencoder, every CT image is mapped to the latent space for the diffusion process. 
During training, a de-noising UNet sees the noisy latent of the CT image concatenated with a deterministic, anatomy‑preserving degradation of the original latent. The text embedding encodes contrast phase information of the CT scan. Its sole objective is to predict the injected noise, conditioning on the degraded latent steers the learning process toward anatomically consistent de‑noising. Therefore, the model acquires an internal ability to transform low‑quality latents to high‑quality, structure‑preserving counterparts.
The autoencoder and latent diffusion model build the bedrock of our \loss\ framework.

\textbf{\loss.} With the pretrained latent diffusion model, we then adapt the de-noising UNet to real CT reconstructions by introducing anatomy-guided supervision.

Each CT image $X\in\mathbb{R}^{H\times W\times 3}$ is first compressed by the frozen autoencoder into a latent representation through $z=\mathcal{E}_{\theta_\text{E}}(X)\in\mathbb{R}^{h\times w\times c}$.  In the forward diffusion process, a noisy sequence $\{z_t\}_{t=0}^{T}$ is obtained by
\begin{equation}\label{eq:add-noise}
    z_t=\sqrt{\bar\alpha_t}z_0+\sqrt{1-\bar\alpha_t}\epsilon,
    \qquad
    \epsilon\sim\mathcal{N}(0,I),t\in\{1,\dots,T\},
\end{equation}
where $T=1,000$, $z_0\triangleq z$ and $\{\bar\alpha_t\}_{t=1}^{T}$ is a fixed variance‑preserving schedule. 

For each training data pair, we take a ground‑truth CT image $X_{\mathrm{gt}}$ and its sparse‑view reconstruction counterpart $X_{\mathrm{rec}}$ (e.g.\ FDK‑50 views) and encode them as $z_{\mathrm{gt}}=\mathcal{E}_{\theta_\text{E}}(X_{\mathrm{gt}})$ and $z_{\mathrm{rec}}=\mathcal{E}_{\theta_\text{E}}(X_{\mathrm{rec}})$.  The forward diffusion step is applied to $z_{\mathrm{gt}}$ to produce $z_t$, after which the input of the de-noising UNet is formed by concatenation with the actual reconstruction latent,
\begin{equation}
    \tilde z_t=\operatorname{Concat}[z_t,z_{\mathrm{rec}}]\in\mathbb{R}^{h\times w\times 2c}.
\end{equation}

The de-noising UNet $\epsilon_\theta$ receives $\tilde z_t$, the diffusion timestep $t$, and a fixed text embedding $c_{\mathrm{phase}}$ of the contrast phase information, and predicts the noise residual $\hat{\epsilon}=\epsilon_\theta(\hat{z}_t, t, c_{\mathrm{phase}})$. Then the first supervision term of the de-noising UNet is the standard \emph{Noise-prediction loss} $\mathcal{L}_{\text n}$
\begin{equation}\label{eq:noise-prediction-loss}
    \mathcal L_{\text n}
    =\mathbb E_{t,\epsilon,X_{\mathrm{gt}}}
      \left\|
        \epsilon-\epsilon_\theta\bigl(
          \operatorname{Concat}[z_t,z_{\mathrm{rec}}],t,c_{\mathrm{phase}}
        \bigr)
      \right\|_2^{2}.
\end{equation}
$\mathcal{L}_{\text n}$ focuses on accurate latent de-noising; $z_{\mathrm{rec}}$ supplies anatomy information and provides pro-structural cues during the de-noising process, thereby allowing the diffusion model to anchor the global CT shape and spatial layout while generating fine structural details.

To introduce anatomy-guided supervision, we augment the latent-space objective by incorporating two additional pixel space constraints. Specifically, after getting the predicted noise $\hat \theta$, we then reverse the diffusion process in \equationautorefname~\eqref{eq:add-noise} to estimate the input latent via
\begin{equation}
    \hat z_0=\frac{1}{\sqrt{\bar\alpha_t}}\left(z_t-\sqrt{1-\bar\alpha_t}\hat\epsilon\right).
\end{equation}
Then, $\hat z_0$ is decoded to the pixel space via the autoencoder. The overall fidelity in pixel space is supervised by an L1 regularization term named \emph{Pixel-space reconstruction loss} $\mathcal L_{\text p}$
\begin{equation}\label{eq:pixel-space-reconstruction-loss}
    \mathcal L_{\text p}=\|\mathcal D_{\theta_\text{D}}(\hat z_0)-X_{\mathrm{gt}}\|_1.
\end{equation}
Most importantly, an anatomical completeness supervision is promoted with a segmentation guidance term from the anatomy segmentator $S(\cdot)$ mentioned in \S\ref{sec:method-metrics}, named \emph{Anatomy-guidance loss} $\mathcal{L}_{\text s}$
\begin{equation}\label{eq:anatomy-guidance-loss}
    \mathcal L_{\text s}=\ell_{\text{seg}}\bigl(S(\mathcal D_{\theta_\text{D}}(\hat z_0)),S(X_{\mathrm{gt}})\bigr),
\end{equation}
where $\ell_{\text seg}$ denotes cross entropy loss function.

Overall, the training objective of \loss\ is the weighted sum of \emph{Noise-prediction loss} $\mathcal{L}_{\text n}$ (\equationautorefname~\ref{eq:noise-prediction-loss}), \emph{Pixel-space reconstruction loss} $\mathcal L_{\text p}$ (\equationautorefname~\ref{eq:pixel-space-reconstruction-loss}), and \emph{Anatomy-guidance loss} $\mathcal{L}_{\text s}$ (\equationautorefname~\ref{eq:anatomy-guidance-loss}),
\begin{equation}
  \mathcal L_{\text{\loss}}
  =\mathcal L_{\text n}
  +\lambda_{\text p}\mathcal L_{\text p}
  +\lambda_{\text s}\mathcal L_{\text s},
\end{equation}
where $\lambda_{\text p}=1$ and $\lambda_{\text s}=0.001$ are the weights of $\mathcal{L}_{\text p}$ and $\mathcal{L}_{\text s}$ respectively. 
The de-noising UNet receives noisy latents of the ground‑truth CT image and the concatenated latents of the reconstructed CT image. 
Meanwhile, adding image and anatomy supervision extends the single-term noise-prediction loss used in the latent diffusion model into a composite objective that jointly optimizes pixel-space reconstruction fidelity and anatomy-aware consistency.

During inference \loss\ starts from pure Gaussian noise $z_T$ concatenated with a latent from an unseen reconstruction CT image, iteratively denoises to $z_0$, and decodes it through $\mathcal D_{\theta_{\mathrm{D}}}$ to deliver an anatomically complete CT image.

\section{Experiments and Results}\label{sec:experiment_result}
\subsection{Pixel-Wise Metric Pitfalls}
\begin{figure}[ht]
    \centering
    \includegraphics[width=\linewidth]{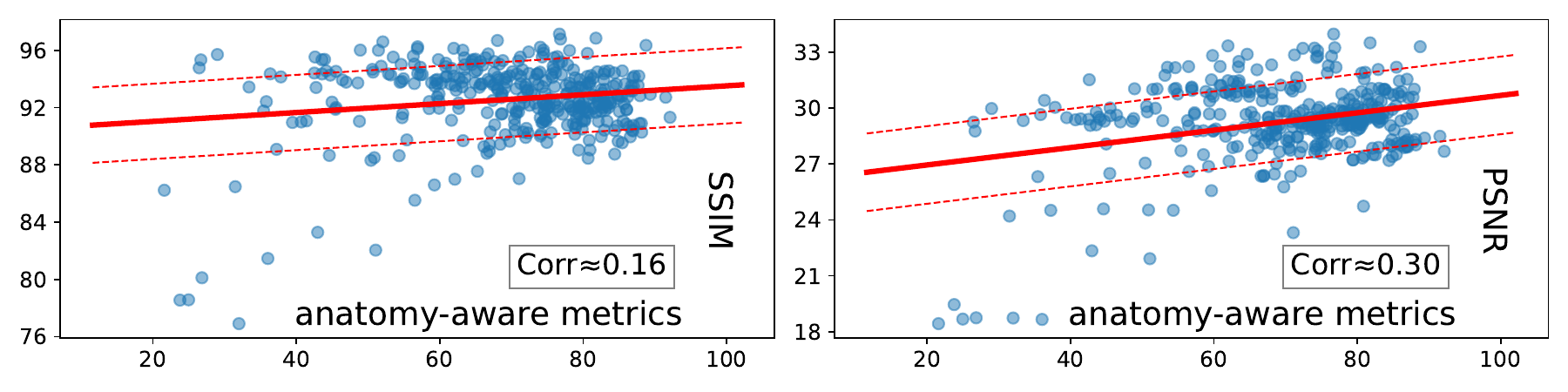}
    \caption{\textbf{Correlation Between Anatomy-Aware Metrics and Pixel-Wise Metrics.}  The solid red line is a linear fit; dashed lines denote \(\pm1\sigma\) of the residuals. Pearson correlation coefficients (\(Corr_{\mathrm{SSIM}}\approx0.16\), \(Corr_{\mathrm{PSNR}}\approx0.30\)) are low, indicating that better anatomical fidelity does not necessarily yield higher SSIM or PSNR.}
    \label{fig:pixel-seg-corr}
\end{figure}

\figureautorefname~\ref{fig:pixel-seg-corr} demonstrates that our anatomy-aware metrics capture reconstruction quality beyond what pixel-wise criteria can reveal.  The horizontal axis shows the average anatomy-aware metrics, computed as the mean of the four groups of anatomy-aware metrics (large organs, small organs, vessel, and intestine). The vertical axes show SSIM (left) and PSNR (right). 
Reconstructions judged by the proposed anatomy-aware metrics to possess more faithful organ and vessel geometry show only weak association with SSIM and PSNR, confirming that superior anatomical fidelity does \emph{not} guarantee higher pixel-similarity scores.

A further trend is the flattening of the regression lines: once a moderate anatomy-aware metric score is achieved, additional structural refinement yields progressively smaller gains in SSIM, and even less in PSNR.  This saturation effect indicates that pixel-wise metrics quickly reach a ceiling, whereas the anatomy-aware metrics continue to differentiate models, underscoring the need for structure-centric assessment in clinical utility.

\subsection{Benchmarking on Anatomy‐Aware Metrics}
\label{sec:results_anatomy}

We evaluated nine state‐of‐the‐art CT reconstruction methods—including traditional methods (FDK, SART, ASD‐POCS), neural rendering (InTomo, NeRF, TensoRF, SAX‐NeRF), and Gaussian‐splatting (R2‐GS)—on our high‐quality test set of 61 CT, which is by far the \emph{largest CT reconstruction benchmark dataset}. Evaluation is conducted on both conventional pixel‐wise metrics (SSIM, PSNR) and the proposed anatomy‐aware metrics. Results are summarized in \tableautorefname~\ref{tab:Benchmarking-results}.

\begin{table}[ht]
    \centering
    \begin{threeparttable}{}
    \scriptsize
    \caption{\textbf{Benchmarking CT Reconstruction Methods.} We evaluate preexisting CT reconstruction methods on high-quality CT scans using both pixel-wise metrics and our anatomy-aware metrics. We report the median and interquartile range (IQR) of these metrics. Cells are marked in \textcolor{blue}{blue}, where a deeper color denotes a greater value. Anatomy-aware metrics reveal that poorly performing methods fail in anatomical preservation, especially in small organs, intestines, and vessels.} 
    \label{tab:Benchmarking-results}
    \begin{tabular}{p{0.13\linewidth}P{0.11\linewidth}P{0.11\linewidth}P{0.11\linewidth}P{0.11\linewidth}P{0.11\linewidth}P{0.11\linewidth}}
    \toprule
      & \multicolumn{2}{c}{pixel-wise metric} & \multicolumn{4}{c}{\textbf{anatomy-aware metric (ours)}}\\
    \cmidrule(lr){2-3}\cmidrule(lr){4-7}
    method  & SSIM  & PSNR  & \makecell{ NSD$^\dag_\text{large}$} & \makecell{NSD$^\ddag_\text{small}$} & \makecell{clDice$^\bigtriangleup_\text{intestine}$}  & \makecell{ clDice$^*_\text{vessel}$}\\
    \midrule
    InTomo~\cite{zang2021intratomo}         &\cellcolor{blue!41.0}82.7\tiny{~(80.6,84.7)}  &\cellcolor{blue!36.45322648617175}24.3\tiny{~(23.1,25.2)}  &\cellcolor{blue!1.0}2.5\tiny{~(1.6,4.0)}  &\cellcolor{blue!0.0}0.0\tiny{~(0.0,0.0)}  &\cellcolor{blue!3.0}7.3\tiny{~(3.0,14.3)}  &\cellcolor{blue!8.0}18.0\tiny{~(0.0,30.9)} \\
    NeRF~\cite{mildenhall2021nerf}          &\cellcolor{blue!43.0}86.8\tiny{~(84.4,89.0)}  &\cellcolor{blue!38.581192872975826}25.7\tiny{~(24.7,27.1)}  &\cellcolor{blue!11.0}22.0\tiny{~(9.3,37.3)}  &\cellcolor{blue!0.0}0.2\tiny{~(0.0,4.1)}  &\cellcolor{blue!8.0}17.0\tiny{~(9.6,26.3)}  &\cellcolor{blue!17.0}34.5\tiny{~(20.1,59.0)} \\
    TensoRF~\cite{chen2022tensorf}                  &\cellcolor{blue!44.0}88.4\tiny{~(87.0,90.7)}  &\cellcolor{blue!41.64348063564633}27.8\tiny{~(26.9,28.9)}  &\cellcolor{blue!20.0}41.3\tiny{~(21.2,55.8)}  &\cellcolor{blue!5.0}11.4\tiny{~(0.0,28.7)}  &\cellcolor{blue!14.0}29.2\tiny{~(20.4,37.7)}  &\cellcolor{blue!23.0}47.2\tiny{~(34.7,58.4)} \\
    R$^2$-GS~\cite{zha2024r}             &\cellcolor{blue!46.0}93.2\tiny{~(91.9,93.9)}  &\cellcolor{blue!42.84789951357505}28.6\tiny{~(27.0,29.5)}  &\cellcolor{blue!38.0}76.7\tiny{~(55.8,86.8)}  &\cellcolor{blue!26.0}53.0\tiny{~(41.8,68.8)}  &\cellcolor{blue!27.0}54.5\tiny{~(44.9,68.0)}  &\cellcolor{blue!34.0}69.7\tiny{~(57.6,78.5)} \\
    NAF~\cite{zha2022naf}                   &\cellcolor{blue!46.0}92.3\tiny{~(91.1,94.1)}  &\cellcolor{blue!44.1268279757467}29.4\tiny{~(28.8,30.6)}  &\cellcolor{blue!42.0}84.3\tiny{~(76.4,87.7)}  &\cellcolor{blue!30.0}60.9\tiny{~(51.3,72.0)}  &\cellcolor{blue!35.0}71.3\tiny{~(61.3,79.0)}  &\cellcolor{blue!38.0}76.1\tiny{~(67.2,82.8)} \\
    FDK~\cite{feldkamp1984practical}                          &\cellcolor{blue!46.0}92.4\tiny{~(91.1,94.2)}  &\cellcolor{blue!44.23771676019473}29.5\tiny{~(28.8,30.8)}  &\cellcolor{blue!41.0}83.9\tiny{~(78.6,87.6)}  &\cellcolor{blue!31.0}63.3\tiny{~(52.7,70.1)}  &\cellcolor{blue!35.0}70.1\tiny{~(60.1,80.8)}  &\cellcolor{blue!38.0}77.4\tiny{~(69.4,82.3)} \\
    SART~\cite{andersen1984simultaneous}    &\cellcolor{blue!46.0}92.5\tiny{~(91.1,94.1)}  &\cellcolor{blue!44.08483814672893}29.4\tiny{~(28.8,30.6)}  &\cellcolor{blue!42.0}84.7\tiny{~(76.8,88.6)}  &\cellcolor{blue!31.0}62.3\tiny{~(53.2,75.3)}  &\cellcolor{blue!35.0}71.1\tiny{~(59.5,79.9)}  &\cellcolor{blue!39.0}78.8\tiny{~(70.4,84.4)} \\
    ASD-POCS~\cite{sidky2008image}          &\cellcolor{blue!46.0}92.5\tiny{~(91.3,94.1)}  &\cellcolor{blue!44.22636306514291}29.5\tiny{~(28.7,30.7)}  &\cellcolor{blue!42.0}84.9\tiny{~(77.3,88.5)}  &\cellcolor{blue!30.0}61.9\tiny{~(51.0,74.8)}  &\cellcolor{blue!36.0}72.0\tiny{~(60.4,80.0)}  &\cellcolor{blue!37.0}75.8\tiny{~(67.4,82.8)} \\
    SAX-NeRF~\cite{cai2024structure}                &\cellcolor{blue!46.0}93.7\tiny{~(92.8,95.0)}  &\cellcolor{blue!44.49139274215548}29.7\tiny{~(29.0,30.8)}  &\cellcolor{blue!42.0}84.8\tiny{~(73.9,89.5)}  &\cellcolor{blue!32.0}65.8\tiny{~(53.0,78.2)}  &\cellcolor{blue!35.0}71.5\tiny{~(56.8,82.5)}  &\cellcolor{blue!39.0}78.8\tiny{~(63.0,85.1)} \\
    \bottomrule
    \end{tabular}
    \begin{tablenotes}
    \item $^\dag$large organs include liver, kidney, pancreas, and spleen. The results are given in NSD ($\uparrow$).
    \item $^\ddag$small organs include gallbladder, adrenal gland, celiac trunk, and duodenum. The results are given in NSD ($\uparrow$).
    \item $^\bigtriangleup$intestine refers to colon and small intestine. The results are given in clDice ($\uparrow$).
    \item $^*$vessel include aorta, postcava, superior mesenteric artery, veins, and renal vein.  The results are given in clDice ($\uparrow$).
    \end{tablenotes}
    \end{threeparttable}
\end{table}

Table~\ref{tab:Benchmarking-results} shows that conventional pixel-wise scores can be overly optimistic: neural rendering methods such as InTomo, NeRF, and TensoRF report SSIM values above 0.83 and PSNR exceeding 24 dB, yet their anatomy-aware scores collapse—small-organ NSD approaches zero and vessel clDice rarely surpasses 35, revealing poor recovery of clinically critical structures.  
When evaluation pivots to anatomy fidelity, the methods ranking is effectively reshuffled: the traditional reconstruction baselines (FDK, SART, ASD-POCS) achieve the best or near-best performance across all four anatomical categories despite mediocre SSIM/PSNR, whereas high-SSIM models like R$^{2}$-GS or SAX-NeRF lag on organ and vessel integrity. 
The results further highlight a persistent difficulty in reconstructing fine anatomy: while large-organ NSD reaches the mid-80s for the strongest methods, small-organ surfaces remain challenging—even the top performer attains a median NSD below 70, underscoring the need for structure-aware objectives if CT reconstructions are to be clinically reliable.  These findings demonstrate that pixel-wise metrics alone are insufficient and anatomy-aware evaluation provides a more discriminative and clinically meaningful assessment of reconstruction quality.

\textbf{Reader Study.} To validate the clinical relevance of our anatomy-aware metrics, we conducted a reader study with 21 board-certified radiologists who independently evaluated a subset of reconstructed scans from the nine benchmarked methods. Radiologists were presented with anonymized image pairs (reconstruction vs. ground truth) and asked to rate anatomical completeness across the four structure categories defined. All 21 radiologists agreed that higher pixel-wise metrics does not guarantee better anatomy preservation. The rankings derived from their assessments showed strong agreement with our anatomy-aware metrics. This confirms that our proposed metrics align closely with expert clinical judgment, whereas conventional pixel-wise scores do not reflect perceptual or diagnostic fidelity. 

\textbf{Discussion.} Conventional pixel–wise criteria markedly over‑estimate the usefulness of sparse‑view reconstructions.  For instance, InTomo records an SSIM of 0.83 yet recovers virtually none of the small‑organ anatomy (median NSD $=0.0$) in \tableautorefname~\ref{tab:Benchmarking-results}.  Such disparity underscores the need for task‑aligned evaluation: the proposed anatomy‑aware metrics reveal structural failures that remain invisible to SSIM and PSNR.  

\subsection{Diffusion‐Based CT Enhancement}
\label{sec:results_diffusion}

\begin{figure}[ht]
    \centering
    \includegraphics[width=\linewidth]{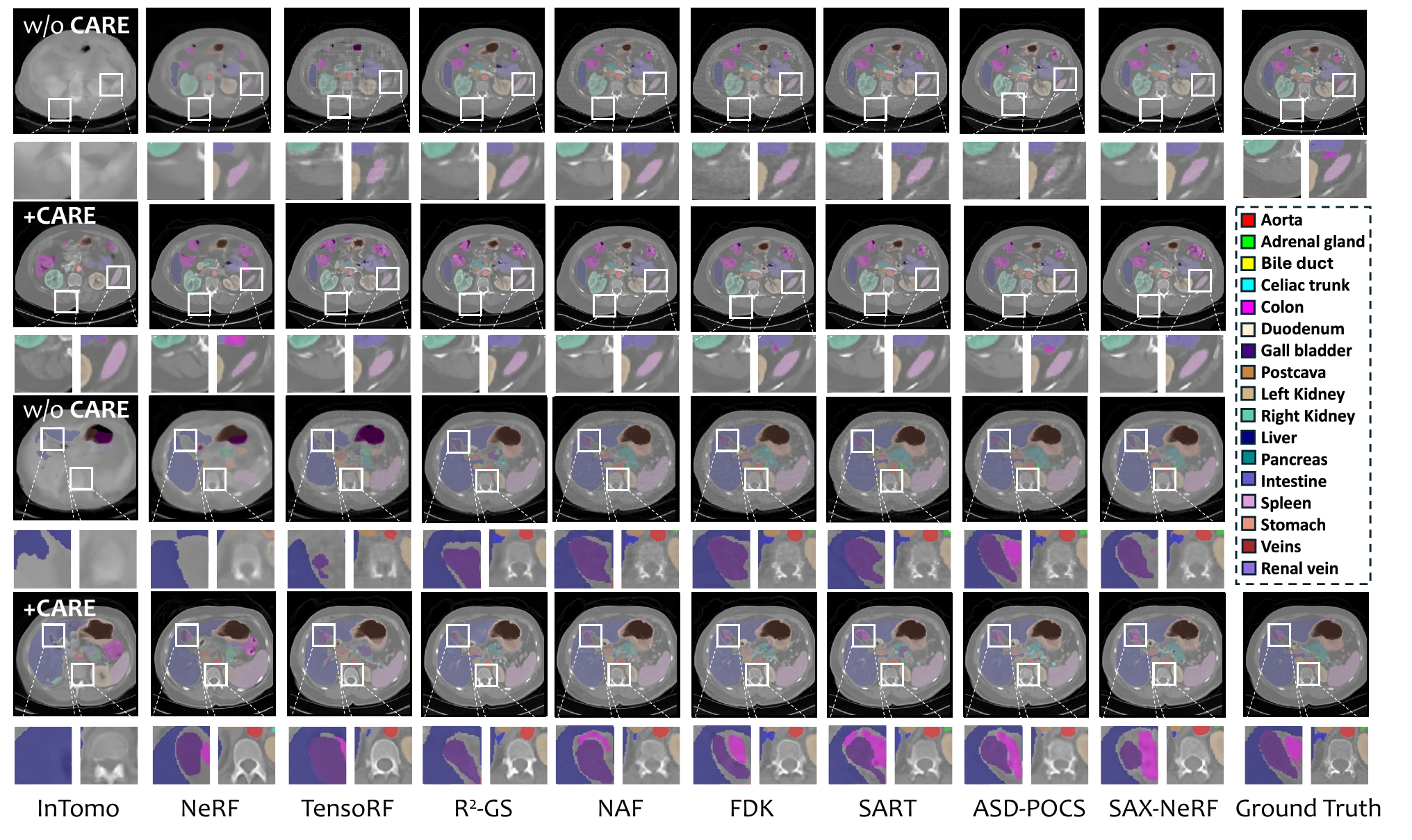}
    \caption{\textbf{Visual Comparison.} Visualizations of CT reconstruction with or without \loss\ on two example CT images. Please zoom in for a better view. These maps reveal that the most prominent changes occur at organ boundaries and within bone structures, where \loss more effectively reduces artifacts and enhances contrast. See more detailed comparison in \S\ref{sec:supp-qualitative}.}
    \label{fig:visual_compare}
\end{figure}

To demonstrate the enhancement performance of \loss, we integrated it into each reconstruction method under a sparse‐view reconstruction setting, i.e., CT reconstruction with 50 views of X-ray images. \tableautorefname~\ref{tab:CARE-results} reports the quantitative gains in both pixel‐wise and anatomy‐aware metrics. \figureautorefname~\ref{fig:visual_compare} shows the qualitative results of reconstruction with and without \loss. \loss\ significantly improves reconstruction quality both in pixel-level fidelity and anatomical structures.

\begin{table*}[ht]
    \centering
    \begin{threeparttable}{}
    \scriptsize
    \caption{\textbf{\loss-enhanced Reconstruction Evaluation.} Evaluate preexisting CT reconstruction methods on high-quality CT scans with \loss\ using both pixel-wise metrics and our anatomy-aware metrics. Note that the results are based on 36 CT scans that \loss\ has never been trained on. We report the median and interquartile range (IQR) of these metrics and perform the Mann-Whitney U test for statistical analysis. Cells are marked in color only if \loss\ shows there is a significant difference ($p$ < 0.05) with the original reconstruction, while \textcolor{green}{green} if the \loss\ enhancement results have improvement, and \textcolor{red}{red} otherwise. Deeper color represents greater difference.
    } 
    \label{tab:CARE-results}
    \begin{tabular}{p{0.08\linewidth}p{0.08\linewidth}P{0.1\linewidth}P{0.1\linewidth}P{0.1\linewidth}P{0.1\linewidth}P{0.1\linewidth}P{0.1\linewidth}}
    \toprule
    &  & \multicolumn{2}{c}{pixel-wise metric} & \multicolumn{4}{c}{\textbf{anatomy-aware metric (ours)}}\\
    \cmidrule(lr){3-4}\cmidrule(lr){5-8}
    method &   & SSIM&PSNR& \makecell{ NSD$_\text{large}$} & \makecell{NSD$_\text{small}$} & clDice$_\text{intestine}$  & \makecell{ clDice$_\text{vessel}$}\\
    
    \midrule
    \multirow{2}{*}{InTomo} 
    & 50 views  &82.7\tiny{~(80.9,84.7)}  &24.5\tiny{~(23.3,25.1)}  &2.4\tiny{~(1.6,3.6)}  &0.0\tiny{~(0.0,0.0)}  &6.3\tiny{~(1.2,17.4)}  &4.7\tiny{~(0.0,21.6)} \\
& \multicolumn{1}{r}{\textbf{+\loss}}  &\cellcolor{red!10}76.7\tiny{~(74.0,79.7)}  &\cellcolor{red!10}22.0\tiny{~(20.6,22.8)}  &\cellcolor{green!30}30.9\tiny{~(26.4,40.6)}  &\cellcolor{green!10}7.1\tiny{~(3.3,13.3)}  &\cellcolor{green!30}33.5\tiny{~(24.4,43.6)}  &\cellcolor{green!50}47.9\tiny{~(35.4,62.2)}\\ 
    
    \midrule
    \multirow{2}{*}{NeRF}   
    & 50 views  &87.7\tiny{~(84.5,89.4)}  &26.6\tiny{~(24.7,27.4)}  &22.1\tiny{~(8.2,35.8)}  &0.2\tiny{~(0.0,4.3)}  &17.2\tiny{~(9.0,28.0)}  &29.0\tiny{~(19.8,42.1)} \\
& \multicolumn{1}{r}{\textbf{+\loss}}  &\cellcolor{red!10}80.5\tiny{~(78.5,83.0)}  &\cellcolor{red!10}23.4\tiny{~(21.9,24.4)}  &\cellcolor{green!40}58.0\tiny{~(49.1,63.0)}  &\cellcolor{green!30}25.3\tiny{~(14.2,30.1)}  &\cellcolor{green!40}53.5\tiny{~(47.4,59.6)}  &\cellcolor{green!30}55.0\tiny{~(45.3,65.6)}\\ 
    
    \midrule
    \multirow{2}{*}{TensoRF}  
    & 50 views  &89.3\tiny{~(87.3,91.1)}  &28.0\tiny{~(27.3,29.7)}  &44.6\tiny{~(21.8,53.5)}  &12.1\tiny{~(1.1,28.4)}  &29.5\tiny{~(19.2,36.4)}  &46.1\tiny{~(36.8,58.5)} \\
& \multicolumn{1}{r}{\textbf{+\loss}}  &\cellcolor{red!0}88.8\tiny{~(87.2,90.1)}  &\cellcolor{red!0}27.5\tiny{~(27.2,28.2)}  &\cellcolor{green!40}75.3\tiny{~(65.7,79.3)}  &\cellcolor{green!30}38.2\tiny{~(30.2,46.4)}  &\cellcolor{green!40}67.9\tiny{~(62.8,75.8)}  &\cellcolor{green!20}58.8\tiny{~(51.0,68.4)}\\ 
    
    \midrule
    \multirow{2}{*}{{R$^2$-GS}}   
    & 50 views  &93.3\tiny{~(92.1,94.4)}  &28.8\tiny{~(27.4,29.7)}  &76.3\tiny{~(65.9,84.6)}  &59.1\tiny{~(41.1,68.3)}  &53.9\tiny{~(45.6,68.1)}  &64.5\tiny{~(59.4,78.6)} \\
& \multicolumn{1}{r}{\textbf{+\loss}}  &\cellcolor{red!10}89.6\tiny{~(87.9,90.7)}  &\cellcolor{red!10}27.3\tiny{~(26.2,28.9)}  &\cellcolor{green!0}82.5\tiny{~(74.0,88.3)}  &\cellcolor{red!0}55.7\tiny{~(42.5,67.3)}  &\cellcolor{green!30}75.4\tiny{~(68.0,82.8)}  &\cellcolor{green!0}67.6\tiny{~(61.8,76.9)}\\ 
    
    \midrule
    \multirow{2}{*}{NAF}   
    & 50 views  &92.5\tiny{~(91.1,94.0)}  &29.4\tiny{~(28.9,30.8)}  &83.7\tiny{~(76.2,86.8)}  &61.0\tiny{~(51.1,70.2)}  &71.6\tiny{~(62.2,77.8)}  &75.4\tiny{~(70.1,81.8)} \\
& \multicolumn{1}{r}{\textbf{+\loss}}  &\cellcolor{green!0}92.8\tiny{~(91.5,93.5)}  &\cellcolor{green!0}29.6\tiny{~(28.6,30.3)}  &\cellcolor{green!10}87.9\tiny{~(83.8,91.3)}  &\cellcolor{green!0}68.1\tiny{~(52.4,76.7)}  &\cellcolor{green!20}81.8\tiny{~(74.1,85.9)}  &\cellcolor{red!0}71.8\tiny{~(63.4,80.8)}\\
    \midrule
    \multirow{2}{*}{FDK}
    & 50 views  &92.5\tiny{~(91.1,94.2)}  &29.5\tiny{~(28.9,30.9)}  &83.8\tiny{~(77.3,86.8)}  &62.4\tiny{~(52.4,70.6)}  &68.2\tiny{~(59.8,76.9)}  &76.5\tiny{~(66.2,80.3)} \\
& \multicolumn{1}{r}{\textbf{+\loss}}  &\cellcolor{red!0}92.4\tiny{~(91.5,93.4)}  &\cellcolor{green!0}29.7\tiny{~(29.0,30.6)}  &\cellcolor{green!10}87.1\tiny{~(82.3,90.9)}  &\cellcolor{green!0}67.0\tiny{~(55.6,77.1)}  &\cellcolor{green!20}81.5\tiny{~(75.5,85.7)}  &\cellcolor{red!0}71.7\tiny{~(66.4,83.2)}\\ 
    
    \midrule
    \multirow{2}{*}{SART}   
    & 50 views  &92.5\tiny{~(91.1,94.2)}  &29.5\tiny{~(28.9,31.0)}  &84.6\tiny{~(76.9,88.1)}  &61.3\tiny{~(53.7,75.4)}  &70.0\tiny{~(61.0,77.6)}  &78.7\tiny{~(65.6,83.6)} \\
& \multicolumn{1}{r}{\textbf{+\loss}}  &\cellcolor{green!0}93.3\tiny{~(92.7,94.3)}  &\cellcolor{green!0}30.3\tiny{~(29.5,30.9)}  &\cellcolor{green!10}88.2\tiny{~(84.2,90.8)}  &\cellcolor{green!0}68.5\tiny{~(56.0,75.5)}  &\cellcolor{green!20}83.5\tiny{~(77.8,87.5)}  &\cellcolor{red!0}75.2\tiny{~(62.2,83.3)}\\
    
    \midrule
    \multirow{2}{*}{ASD-POCS}   
    & 50 views  &92.3\tiny{~(91.6,94.2)}  &29.5\tiny{~(28.9,30.7)}  &83.8\tiny{~(76.5,87.2)}  &61.0\tiny{~(51.4,73.6)}  &71.3\tiny{~(61.8,77.4)}  &74.7\tiny{~(67.4,81.1)} \\
& \multicolumn{1}{r}{\textbf{+\loss}}  &\cellcolor{green!0}92.4\tiny{~(91.8,93.5)}  &\cellcolor{red!0}29.4\tiny{~(28.4,30.1)}  &\cellcolor{green!10}88.1\tiny{~(82.6,90.9)}  &\cellcolor{green!0}67.8\tiny{~(56.2,75.4)}  &\cellcolor{green!10}80.7\tiny{~(74.9,84.9)}  &\cellcolor{green!0}76.4\tiny{~(66.6,79.9)}\\

    \midrule
    \multirow{2}{*}{SAX-NeRF}   
    & 50 views  &93.7\tiny{~(92.8,95.0)}  &29.6\tiny{~(29.0,31.0)}  &82.4\tiny{~(73.9,88.9)}  &66.5\tiny{~(55.0,78.0)}  &68.0\tiny{~(56.3,77.9)}  &79.3\tiny{~(62.4,85.1)} \\
& \multicolumn{1}{r}{\textbf{+\loss}}  &\cellcolor{red!10}92.5\tiny{~(91.2,93.3)}  &\cellcolor{red!0}29.4\tiny{~(28.8,30.1)}  &\cellcolor{green!10}88.7\tiny{~(81.9,90.9)}  &\cellcolor{green!0}67.1\tiny{~(54.5,75.1)}  &\cellcolor{green!20}81.5\tiny{~(76.0,86.8)}  &\cellcolor{red!0}72.0\tiny{~(62.9,84.1)}\\ 
    
    \bottomrule
    \end{tabular}
    \end{threeparttable}
\end{table*}

Under extreme sparsity in CT reconstruction, \loss‐enhanced reconstructions exhibit remarkable anatomical completeness: for example, InTomo+\loss\ increases large‐organ NSD from 2.4\% to 30.9\%, vessel clDice from 4.7\% to 47.9\%, and intestine clDice from 6.3\% to 33.5\%. NeRF+\loss\ and TensoRF+\loss\ show similar uplifts, confirming that our anatomy‐guided loss robustly transfers structural priors even in low‐data regimes. 

These results highlight \loss’s transformative impact: by embedding anatomical supervision directly into the diffusion framework, our method consistently elevates structural fidelity across diverse reconstruction backbones and sampling densities, without sacrificing general applicability or requiring additional scanning hardware.

\textbf{Reader Study.} We further evaluated the perceptual and clinical impact of \loss through a blinded reader study involving the same cohort of 21 radiologists. Participants assessed side-by-side reconstructions—with and without \loss—across diverse anatomical regions and reconstruction backbones. In all cases, radiologists preferred the \loss-enhanced version, citing improved visibility of small organs (e.g., gallbladder, adrenal glands), vascular continuity (e.g., celiac trunk, renal veins), and intestinal integrity. Notably, even when baseline reconstructions achieved high PSNR/SSIM, radiologists consistently identified missing or distorted anatomy that \loss successfully restored. These findings underscore \loss’s ability to translate quantitative gains in anatomy-aware metrics into tangible clinical improvements.

\textbf{Discussion.}
The proposed \loss\ framework offers a model- and patient-agnostic avenue to inject anatomical priors into existing pipelines. Paired with \loss, diverse backbones—from classical FDK to NeRF and Gaussian-splatting—achieve substantial gains across all anatomy-aware metrics.

\section{Discussion and Conclusion}\label{sec:disscussion-and-conclusion}
\textbf{Discussion.}
Our experiments in \S\ref{sec:results_anatomy} and \S\ref{sec:results_diffusion} show that \loss\  consistently improves the anatomical fidelity of sparse-view CT reconstructions across a wide range of baselines, including analytical, implicit, and generative methods. Further analysis in \S\ref{sec:supp-AV} demonstrates that \loss\ performs similar enhancement ability across arterial and portal venous phase CT scans, revealing \loss's potential in expanding to other CT phases. These improvements are particularly notable in small and complex structures like small organs and vessels, where segmentation-derived metrics such as NSD and clDice reveal gains of up to 40\%, even when pixel-wise metrics remain nearly unchanged. Notably, \loss\ is trained once and applied directly to unseen cases without fine-tuning, demonstrating strong patient-agnostic generalization. This robustness arises from its design: segmentation-guided supervision encodes population-level shape priors, and the diffusion model integrates them without requiring additional annotations or per-case adaptation. Its straightforward design enables seamless integration into existing reconstruction pipelines as a plug-in enhancement, without requiring modifications to or retraining of the underlying models. Moreover, \loss features a modular architecture that facilitates the incorporation of future advances in segmentation models, thereby enhancing the reliability of anatomical guidance and enabling the evaluation of an expanding set of anatomical structures.
Overall, \loss\  offers a flexible solution to enforce anatomical plausibility in CT reconstruction, particularly in settings where high-contrast structures are under-represented in pixel-domain loss functions.

\textbf{Conclusion.}
This work reframes CT reconstruction quality through an anatomy‑centric viewpoint and introduces \loss, a diffusion‑based enhancement module that can be integrated into any reconstruction algorithm.  A new suite of segmentation‑driven metrics exposes structural deficiencies overlooked by SSIM and PSNR, and extensive experiments on nine baselines under sparse-view settings demonstrate that \loss\ consistently elevates anatomical fidelity, delivering up to 35\% improvements in vessel {clDice} under 50‑view settings—without additional acquisition burden.  
A limitation of our method is that its improvement becomes marginal when the baseline reconstruction already preserves most anatomical structures. In such high-quality cases, i.e., the reconstructions from SAX-NeRF, \loss\  offers limited added benefit relative to its computational cost.
Our method brings algorithm goals in line with real clinical needs, making it easier to produce anatomy-aware CT reconstructions under sparse-view settings and sets a new standard for future reconstruction research.

\begin{ack}
This work was supported by the Lustgarten Foundation for Pancreatic Cancer Research, the Patrick J. McGovern Foundation Award, and the National Institutes of Health (NIH) under Award Number R01EB037669. We would like to thank the Johns Hopkins Research IT team in \href{https://researchit.jhu.edu/}{IT@JH} for their support and infrastructure resources where some of these analyses were conducted; especially \href{https://researchit.jhu.edu/research-hpc/}{DISCOVERY HPC}. We thank Jaimie Patterson for writing a news article about this project; thank Hamed Hooshangnejad, Heng Li, Wenxuan Li, and Guofeng Zhang for their helpful suggestions throughout the project. Paper content is covered by patents pending.
\end{ack}

{\small
\bibliographystyle{ieee_fullname}
\bibliography{refs,zzhou}
}

\clearpage
\appendix
\setcounter{page}{1}
\onecolumn
\renewcommand \thepart{}
\renewcommand \partname{}
\part{Appendix} 
\setcounter{secnumdepth}{4}
\setcounter{tocdepth}{4}
\parttoc 

\clearpage\section{Datasets, Models and Implementation Details}\label{sec:supp-datasets-details} 
\subsection{Datasets}\label{sec:supp-experiments-datasets}
We evaluate four different kinds of anatomical structures: large organs, small organs, intestine and vessel. Large organs and small organs are divided according to clinical guidelines, showing in \tableautorefname~\ref{tab:supp_anatomy_division}.
\begin{table}[htbp]
\centering
\scriptsize
\caption{\textbf{Grouping strategy for large and small organs.} We partition organs by typical in vivo volumes reported in radiologic volumetry studies. The 100 mL and 10 mm cutoff aligns with standard thresholds for distinguishing major organs from accessory structures in abdominal atlases.}
\label{tab:supp_anatomy_division}
\begin{tabular}{p{0.3\textwidth}p{0.5\textwidth}p{0.1\textwidth}}
\toprule
\textbf{Group} & \textbf{Representative mean volume (mL)} & \textbf{References} \\
\midrule
\textbf{Large organs} (> $\sim$100 mL) & Liver $\approx$1200–1700; Kidney $\approx$150–200; Spleen $\approx$215; Pancreas $\approx$71–83 & ~\cite{geraghty2004normal} \\
\textbf{Small organs / structures} (< $\sim$50 mL or < 10 mm diameter) & Gallbladder $\approx$30–50 mL; Each adrenal gland $\approx$4–6 mL; Celiac trunk diameter $\approx$6–12 mm; Duodenum lumen 2–3 cm & ~\cite{schick2022automatic} \\
\bottomrule
\end{tabular}
\end{table}

\subsubsection{{Large-scale Reconstruction Benchmark Dataset}}\label{sec:supp-experiments-datasets-recon} 
\begin{table}[htbp]
\centering
\scriptsize
\caption{\textbf{Evaluation dataset size of the CT reconstruction methods.} Among all the CT reconstruction methods we evaluate, NeRF~\cite{mildenhall2021nerf} and TensoRF~\cite{TensoRF} uses 20 CT scans but no human body CT included, SAX-NeRF~\cite{cai2024structure} use 15 CT scans with only 9 human body CT scans. Our reconstruction benchmark dataset is by far the \emph{largest} in the field of CT reconstruction.}
\label{tab:supp_largest_ct_dataset}
\begin{tabular}{lcc}
\toprule
\textbf{Method} & \textbf{Total CT Scan} & \textbf{Human Body CT Scan} \\
\midrule
InTomo~\cite{zang2021intratomo} & 7 & 4 \\
NeRF~\cite{mildenhall2021nerf} & 20 & 0 \\
TensoRF~\cite{TensoRF} & 20 & 0 \\
R2-GS~\cite{zha2024r} & 15 & 5 \\
NAF~\cite{zha2022naf} & 5 & 4 \\
FDK~\cite{feldkamp1984practical} & 1 & 0 \\
SART~\cite{andersen1984simultaneous} & 1 & 1 \\
ASD-POCS~\cite{sidky2008image} & 2 & 0 \\
SAX-NeRF~\cite{cai2024structure} & 15 & 9 \\
\midrule
\textbf{CARE} & \textbf{61} & \textbf{61} \\
\bottomrule
\end{tabular}
\end{table}

As shown in \tableautorefname~\ref{tab:supp_largest_ct_dataset}, our reconstruction benchmark dataset contains 61 high-quality CT scans, with 36 in arterial phase and 25 in portal venous phase, which is by far the largest CT reconstruction dataset. The real projections of these CT scans are not available. Thus, we use TIGRE~\cite{biguri2016tigre, biguri2019arbitrarily} package to generate synthetic projections following previous works~\cite{cai2024structure, zha2024r}.
The detailed meta information about our reconstruction evaluation dataset is listed in \tableautorefname~\ref{tab:supp_recon_meta}.

\begin{table}[htbp]
\centering
\scriptsize
\caption{\textbf{Statistics of the CT reconstruction benchmark dataset.} }
\label{tab:supp_recon_meta}
\begin{tabular}{p{0.1\linewidth}P{0.1\linewidth}P{0.1\linewidth}P{0.1\linewidth}P{0.12\linewidth}P{0.14\linewidth}P{0.13\linewidth}}
\toprule
\textbf{Dataset} & \textbf{CT scan$^*$} & \textbf{Age$^\dag$} & \textbf{Female$^*$} & \textbf{Male$^*$} & \textbf{In-plane spacing$^\ddag$} & \textbf{White$^*$} \\
\midrule
Training & 25 & 64.2 $\pm$ 8.9 & 10 (40.0) & 15 (60.0) & 0.70 (0.68, 0.79) & 6 (24.0) \\
Test     & 36 & 64.8 $\pm$ 9.2 & 16 (44.4) & 20 (55.6) & 0.77 (0.70, 0.98) & 4 (11.1) \\
\midrule
\textbf{Dataset} & \textbf{Black$^*$} & \textbf{Asian$^*$} & \textbf{Other$^*$} & \textbf{Resolution$^\bigtriangleup$} & \textbf{Height$^\bigtriangleup$} & \textbf{Voxel spacing$^\ddag$} \\
\midrule
Training & 4 (16.0) & 5 (20.0) & 6 (24.0) & 512 $\times$ 512 & 350 $\pm$ 302 & 0.74 $\pm$ 0.12 \\
Test     & 7 (19.4) & 10 (27.8) & 5 (13.9) & 512 $\times$ 512 & 427 $\pm$ 294 & 1.20 $\pm$ 1.04 \\
\bottomrule
\end{tabular}
\begin{tablenotes}
\item $^\dag$ Age in years (mean $\pm$ SD).
\item $^\ddag$ Spacing in millimeters. Applies to \textit{In-plane spacing} and \textit{Voxel spacing}. Values reported as median (IQR) or mean $\pm$ SD as shown.
\item $^\bigtriangleup$ Image resolution and height in pixels. Resolution reported as width $\times$ height; height reported as mean $\pm$ SD.
\item $^*$ Counts; some columns also include percentages in parentheses, e.g., \textit{Female}, \textit{Male}, and race categories. Applies to \textit{CT scan}, \textit{Female}, \textit{Male}, \textit{White}, \textit{Black}, \textit{Asian}, \textit{Other}.
\end{tablenotes}
\end{table}

For the anatomy-aware CT reconstruction metrics (\S\ref{sec:method-metrics}), we report the results of all 61 CT scans. For the anatomy-aware CT reconstruction framework \loss\ (\S\ref{sec:method-CARE}), we report the results of a subset of 36 CT scans (23 in arterial phase and 13 in portal venous phase), and the remaining 25 CT scans (13 in arterial phase and 12 in portal venous phase) were used for the training set. The detailed meta information about our reconstruction evaluation dataset is listed in \tableautorefname~\ref{tab:supp_seg_meta}.

\subsubsection{{JHH Dataset for {\loss} 
Pretraining}}\label{sec:supp-experiments-datasets-private} 
\begin{table}[htbp]
\centering
\scriptsize
\caption{\textbf{Statistics of the JHH dataset.} All the CT scans in the dataset have a resolution of 512 $\times$ 512. This dataset is collected using four major vendors: GE (39\%), Siemens (38\%), Phillips (12\%), and Toshiba (11\%).}
\label{tab:supp_seg_meta}
\begin{tabular}{p{0.08\linewidth}P{0.08\linewidth}P{0.08\linewidth}P{0.1\linewidth}P{0.11\linewidth}P{0.14\linewidth}P{0.08\linewidth}P{0.08\linewidth}}
\toprule
\textbf{Dataset} & \textbf{CT scan$^*$} & \textbf{Age$^\dag$} & \textbf{Female$^*$} & \textbf{Male$^*$} & \textbf{In-plane spacing$^\ddag$} & \textbf{White$^*$} & \textbf{Black$^*$} \\
\midrule
Training & 3151 & 64.9 $\pm$ 8.9 & 1524 (48.4) & 1627 (51.6) & 0.71 (0.66, 0.77) & 623 (19.8) & 619 (19.6) \\
Test     & 1958 & 64.9 $\pm$ 8.9 & 1008 (51.5) & 950 (48.5)  & 0.73 (0.68, 0.79) & 409 (20.9) & 394 (20.1) \\
\midrule
\textbf{Dataset} & \textbf{Asian$^*$} & \textbf{Other$^*$} & \textbf{Height$^\bigtriangleup$} & \textbf{Voxel spacing$^\ddag$} & \textbf{PDAC$^*$} & \textbf{PNET$^*$} & \textbf{Cyst$^*$} \\
\midrule
Training & 632 (20.1) & 661 (21.0) & 678 $\pm$ 201 & 0.72 $\pm$ 0.09 & 1119 & 591 & 429 \\
Test     & 389 (19.9) & 397 (20.3) & 690 $\pm$ 196 & 0.74 $\pm$ 0.09 & 554  & 327 & 512 \\
\bottomrule
\end{tabular}
\begin{tablenotes}
\item $^\dag$ Age in years (mean $\pm$ SD).
\item $^\ddag$ Spacing in millimeters. Applies to \textit{In-plane spacing} and \textit{Voxel spacing}. Values reported as median (IQR) or mean $\pm$ SD as shown.
\item $^\bigtriangleup$ Height in pixels (mean $\pm$ SD).
\item $^*$ Counts; some columns also include percentages in parentheses, e.g., \textit{Female}, \textit{Male}, and race categories. Applies to \textit{CT scan}, \textit{Female}, \textit{Male}, \textit{White}, \textit{Black}, \textit{Asian}, \textit{Other}, and pancreatic tumor types: \textit{PDAC}, \textit{PNET}, \textit{Cyst}.
\end{tablenotes}
\end{table}
The JHH dataset contains 6,212 high-quality artifact-free CT scans of 2,870 patients, where 3,107 are in the arterial phase and 3,105 are in the venous phase. The CT scans in this dataset have a spacing of 0.5 mm, providing 4 million CT slices in total. This dataset can be split into a training set of 3,151 CT scans and a testing set of 1,958 CT scans. For all 6,212 CT scans of the dataset, a total of 25 anatomical structures are annotated by a group of 21 radiologists for over three years, including the 17 structures we use as anatomy-aware CT reconstruction metrics. The 25 annotated anatomical structures are: aorta, left adrenal gland, right adrenal gland, common bile duct, celiac artery, colon, duodenum, gall bladder, postcava, left kidney, right kidney, liver, pancreas, pancreatic duct, superior mesenteric artery, intestine, spleen, stomach, veins, left renal vein, right renal vein, common bile duct, pancreatic PDAC, pancreatic cyst, and pancreatic PNET.

This dataset is used both in training the anatomy segmentator, autoencoder, and the latent diffusion model to build the anatomy-aware CT reconstruction metrics and is prepared for the \loss\ framework. Note that the anatomy segmentator is trained on the training set, whereas the autoencoder and latent diffusion model are trained on the entire 6,212 CT scans.

\subsection{Models}\label{sec:experiments-models}
We provide full access to our model checkpoints to contribute to the open-source community. We release the model checkpoints of our anatomy segmentator, autoencoder, and the \loss\ models in our codebase.

\subsection{Implementation Details}\label{sec:experiments-details}
\subsubsection{Anatomy Segmentator}
The proposed anatomy segmentator used in both the anatomy-aware CT reconstruction metrics and the \loss\ framework is implemented as a nnU-Net~\cite{isensee2021nnu}. We extend the training plan of nnU-Net to be trained on a 48GB NVIDIA RTX 6000 GPU, with all data preprocessing, training, and inference settings set as default. This model is trained on the training set of our JHH dataset, with 3,151 expertly annotated CT scans.

\subsubsection{Reconstruction Baselines}
The experiments are conducted on three traditional reconstruction methods (FDK~\cite{feldkamp1984practical}, SART~\cite{andersen1984simultaneous}, ASD-POCS~\cite{sidky2008image}), five NeRF-based reconstruction methods (InTomo~\cite{zang2021intratomo}, NeRF~\cite{mildenhall2021nerf}, TensoRF~\cite{TensoRF}, NAF~\cite{zha2022naf}, SAX-NeRF~\cite{cai2024structure}), and a Gaussian-Spaltting-based method R$^2$-GS~\cite{zha2024r}. We implement these methods using the CT reconstruction toolbox provided by SAX-NeRF\footnote{\href{https://github.com/caiyuanhao1998/SAX-NeRF}{https://github.com/caiyuanhao1998/SAX-NeRF}}~\cite{cai2024structure} and R$^2$-GS\footnote{\href{https://github.com/Ruyi-Zha/r2_gaussian}{https://github.com/Ruyi-Zha/r2$\_$gaussian}}~\cite{zha2024r}.  For each algorithm, the training setting is set as default. All of these reconstruction experiments are run on an eight NVIDIA RTX 6000 GPU server, each with 48 GB of memory.

\subsubsection{Autoencoder Training}\label{sec:supp-stage1}
The autoencoder model is initialized with the checkpoints provided by Stable Diffusion v1.5~\cite{rombach2021highresolution}. We set the weight of reconstruction loss and perceptual loss (detailed in \S\ref{sec:supp-technical-details-autoencoder}) to be $\lambda_{\text rec}=\lambda_{\text per}=1$. The weight of the KL regularization term is set to $\beta=1\times 10^{-6}$. The autoencoder model is trained on the JHH CT dataset (\S\ref{sec:supp-experiments-datasets-private}) for 150,000 iterations. We use AdamW optimizer during training. 

\subsubsection{Latent Diffusion Training}\label{sec:supp-stage2}
When training the latent diffusion model, we adopted the super resolution training scheme for the model to gain intrinsic enhancement ability. To create low-quality CT images, we followed SR3~\cite{saharia2022image} to use bicubic interpolation with anti-aliasing enabled to downsample the CT image and upsample it back to the original resolution. The downsampling factor is set to 4.

During training, the text prompt regarding different phases is fixed: the prompt of an arterial phase CT image is ``An Arterial CT slice.'', and the prompt of a portal phase CT image is ``A Portal-venous CT slice.''.  The de-noising UNet is trained on the same JHH dataset as used in the autoencoder training, with AdamW optimizer and 50,000 training iterations.

\subsubsection{\loss\ Training}\label{sec:supp-stage3}
The text prompt of \loss\ is identical to the latent diffusion training stage. 
The model is finetuned on 25 CT scans (as mentioned in \S\ref{sec:supp-experiments-datasets-recon}) for 50,000 iterations with the AdamW optimizer for each given CT reconstruction method. We set the weights of the losses to $\lambda_\text{p}=1$ and $\lambda_\text{s}=0.001$.

Our diffusion model is implemented by the diffusers~\cite{diffusers} package with a backbone of Stable Diffusion v1.5~\cite{rombach2021highresolution}. All three training stages of \loss\ (\S\ref{sec:supp-stage1}, \S\ref{sec:supp-stage2} and \S\ref{sec:supp-stage3}) are done on an eight RTX 8000 GPUs server, each with 48 GB of memory.

\clearpage\section{Anatomy Segmentator}\label{sec:supp-segmentation}
As mentioned in \S\ref{sec:supp-stage1}, the anatomy segmentator is trained on the training set of our JHH dataset with 3,151 CT scans. In \S\ref{sec:method-metrics} of the paper, we showed the high correlation (in \figureautorefname~\ref{fig:fig_seg_corr}) of the anatomy-aware CT reconstruction metrics with the segmentator's pseudolabel and the ground truth label, to support our assertion that employing the anatomy segmentator for these metrics is justifiable. 

Here, we also provide the segmentation results of our anatomy segmentator on the testing set of our JHH dataset with 1,958 CT scans. \tableautorefname~\ref{fig:supp-seg-jhh-test} shows that our anatomy segmentator achieves excellent performance over the large-scale testing set comparing to three state-of-the-art segmentation methods in the Touchstone 1.0~\cite{bassi2024touchstone} benchmark: MedNeXt~\cite{roy2023mednext}, MedFormer{~\cite{wang2024medformer}} and STU-Net-B{~\cite{huang2023stu}}.

\begin{table}[htbp]
\centering
\scriptsize
\caption{\textbf{Anatomy Segmentator's Performance on the Testing Set of JHH Dataset.} Segmentation performance is reported as the median and interquartile range (IQR) of Dice and IoU (\%) for selected anatomical structures across the testing set with 1,958 CT scans. For each structure, the best result is \textbf{bolded} and the sencond best \underline{underlined}.}\label{fig:supp-seg-jhh-test}
\begin{tabular}{p{0.25\linewidth}P{0.15\linewidth}P{0.15\linewidth}P{0.15\linewidth}P{0.15\linewidth}}
\toprule
\textbf{Anatomical Structures} & \textbf{Segmentator (Ours)}& \textbf{MedNeXt~\cite{roy2023mednext}} & \textbf{MedFormer~\cite{wang2024medformer}} & \textbf{STU-Net-B~\cite{huang2023stu}} \\
\midrule
\multicolumn{5}{l}{\textbf{Large Organs}} \\
\quad Liver & \underline{96.9}\tiny{~(96.4,97.3)}& 96.5\tiny{~(95.8,96.9)} & 96.7\tiny{~(96.2,97.1)} & \textbf{97.2}\tiny{~(96.8,97.5)} \\
\quad Kidney Left & \textbf{97.7}\tiny{~(97.4,98.0)} & \underline{97.2}\tiny{~(96.8,97.6)}& 97.1\tiny{~(96.7,97.5)} & 96.9\tiny{~(96.4,97.3)} \\
\quad Kidney Right & \textbf{97.7}\tiny{~(97.5,98.0)} & \underline{97.3}\tiny{~(97.0,97.7)}& 97.0\tiny{~(96.5,97.4)} & 96.8\tiny{~(96.3,97.1)} \\
\quad Spleen & 96.4\tiny{~(95.5,97.1)} & \underline{96.5}\tiny{~(95.7,97.2)}& \textbf{96.8}\tiny{~(96.0,97.3)} & 96.0\tiny{~(95.4,96.9)} \\
\quad Pancreas & 86.5\tiny{~(81.5,89.3)} & \textbf{88.1}\tiny{~(85.3,90.5)} & \underline{87.2}\tiny{~(84.0,89.6)}& 85.9\tiny{~(82.1,88.3)} \\
\midrule
\multicolumn{5}{l}{\textbf{Small Organs}} \\
\quad Gall Bladder & 90.9\tiny{~(86.1,93.4)} & 90.5\tiny{~(85.7,92.8)} & \underline{91.3}\tiny{~(87.4,93.5)}& \textbf{91.7}\tiny{~(88.2,93.8)} \\
\quad Adrenal Gland Left & 84.6\tiny{~(75.2,88.7)} & \textbf{86.1}\tiny{~(77.3,89.4)} & \underline{85.4}\tiny{~(76.9,88.2)}& 83.2\tiny{~(75.0,87.0)} \\
\quad Adrenal Gland Right & 80.1\tiny{~(71.7,83.5)} & \textbf{81.4}\tiny{~(73.9,86.5)} & \underline{81.2}\tiny{~(72.8,85.4)}& 79.5\tiny{~(70.7,83.2)} \\
\quad Celiac Artery & 65.6\tiny{~(56.5,73.4)} & \underline{66.8}\tiny{~(57.2,74.3)}& \textbf{67.1}\tiny{~(59.5,75.2)} & 65.9\tiny{~(57.0,73.1)} \\
\quad Duodenum & \textbf{86.4}\tiny{~(82.4,89.5)} & \underline{85.3}\tiny{~(80.5,88.1)}& 84.8\tiny{~(80.2,87.4)} & 85.1\tiny{~(81.2,88.3)} \\
\midrule
\multicolumn{5}{l}{\textbf{Intestinal Structures}} \\
\quad Colon & 84.4\tiny{~(82.6,91.1)} & \underline{85.1}\tiny{~(83.0,91.3)}& \textbf{85.2}\tiny{~(83.7,92.1)} & 84.2\tiny{~(82.1,90.5)} \\
\quad Intestine & 75.6\tiny{~(69.1,81.4)} & \underline{76.8}\tiny{~(70.4,82.3)}& \textbf{76.9}\tiny{~(71.3,83.0)} & 75.0\tiny{~(68.9,80.7)} \\
\midrule
\multicolumn{5}{l}{\textbf{Vascular Structures$^\dag$}} \\
\quad Aorta & \textbf{92.1}\tiny{~(87.4,95.0)} & 91.3\tiny{~(86.8,94.2)} & 90.9\tiny{~(86.1,93.9)} & \underline{91.5}\tiny{~(87.0,94.5)}\\
\quad Postcava & \underline{85.9}\tiny{~(79.3,87.4)}& \textbf{86.0}\tiny{~(80.4,88.2)} & 85.2\tiny{~(78.9,86.9)} & 84.7\tiny{~(78.1,86.3)} \\
\quad Superior Mesenteric Artery & 66.5\tiny{~(58.2,74.5)} & \underline{67.1}\tiny{~(59.1,75.2)}& \textbf{67.3}\tiny{~(60.3,76.5)} & 66.8\tiny{~(58.4,74.8)} \\
\bottomrule
\end{tabular}

\begin{tablenotes}
\item $^\dag$Since there is no ground truth label for renal veins (left or right) in the test set, we don't report the corresponding metrics here.
\end{tablenotes}
\end{table}

\clearpage
\section{Details and Disscussions of Anatomy-Aware Metrics}\label{sec:supp-technical-metrics}
\subsection{NSD vs. DSC: an Example of Sensitivity to Boundary Shifts}
Consider an organ with volume $V$ and surface area $S$. When the entire boundary shifts uniformly by a distance $\delta$ (e.g., organ boundary displaced by 3 mm), the volume discrepancy can be approximated as:
\begin{equation}
\Delta V \approx S \cdot \delta
\end{equation}
The corresponding drop in Dice Similarity Coefficient (DSC) can be approximated by:
\begin{equation}
\vert \Delta\text{DSC}\vert \approx \frac{\Delta V}{2V} = \frac{S \cdot \delta}{2V} = \frac{\delta}{2R_{\text{eff}}}, \quad \text{where} \quad R_{\text{eff}} \equiv \frac{V}{S}
\end{equation}

For a spherical organ with radius $R = 50 \text{mm}$ and $\delta = 3 \text{mm}$, we compute the effective radius $R_{\text{eff}}$ as:
\begin{equation}
R_{\text{eff}} = \frac{\frac{4}{3}\pi R^3}{4\pi R^2} = \frac{R}{3} \approx 16.67 \text{mm}
\end{equation}
This results in:
\begin{equation}
\vert \Delta\text{DSC}\vert \approx \frac{\delta}{2R_{\text{eff}}}=\frac{3}{2 \times 16.67} \approx 0.09
\end{equation}
This demonstrates that even with clinically significant displacement, DSC may only decrease a little, not reflecting the severe anatomical misalignment.
In contrast, Normalized Surface Dice (NSD) is independent of organ volume and instead penalizes boundary mismatches directly. For a mismatch tolerance $\tau = 2 \text{mm}$, a uniform 3 mm shift would lead NSD to fall to zero, reflecting the anatomical error more faithfully.

Overall, DSC is volume-biased and may underestimate boundary errors in large structures, while NSD maintains consistent sensitivity across anatomical scales.

\clearpage\subsection{High Correlation of GT-based and Segmentator-based Anatomy-Aware Metrics}
\figureautorefname~\ref{fig:fig_seg_corr} 
summarizes the high correlation between GT-based and segmentator-based anatomy-aware metrics over all nine CT reconstruction methods, which indicates that the proposed anatomy segmentator is a reasonable substitute for ground truth labels to build the anatomy-aware metrics.

\begin{figure}[ht]
    \centering
    \includegraphics[width=\linewidth]{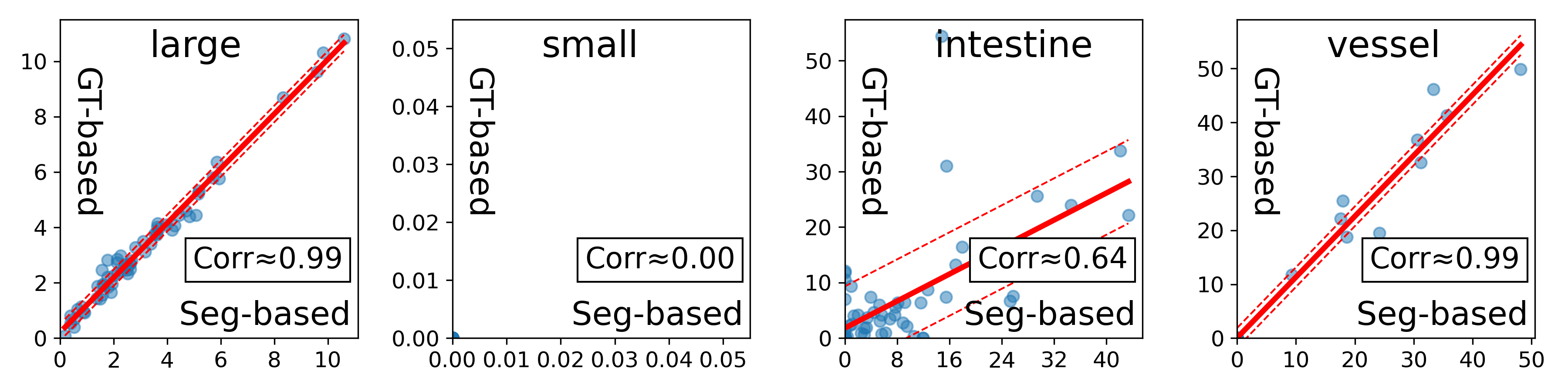}
    \caption{\textbf{Correlation Between GT-based and Segmentator-based Anatomy-Aware Metrics on InTomo~\cite{zang2021intratomo}.} The solid red line is a linear fit and the dashed lines denote $\pm 1\sigma$ of the residuals. High correlation shows that the proposed anatomy segmentator is a strong substitute for ground truth labels when building anatomy-aware metrics. }
    \label{fig:supp_pitfall_results-intraeg_corrtomo}
\end{figure}

Here, we also provide the detailed correlation scatter plots of all nine CT reconstruction methods on GT-based and segmentator-based anatomy-aware metrics. These correlation plots consistently give the same conclusion.
Note that the correlation coefficient of small organs in the InTomo method is 0.0 (in \figureautorefname~\ref{fig:supp_pitfall_results-intraeg_corrtomo}) because all the segmentation metrics are zeros. Moreover, the inadequate correlation coefficient of the intestine highlights an existing issue with the annotation of intestinal structures: the clinical experts concentrate solely on the volumetric precision of the annotation, overlooking the tubular form and connectivity.

\begin{figure}[ht]
    \centering
    \includegraphics[width=\linewidth]{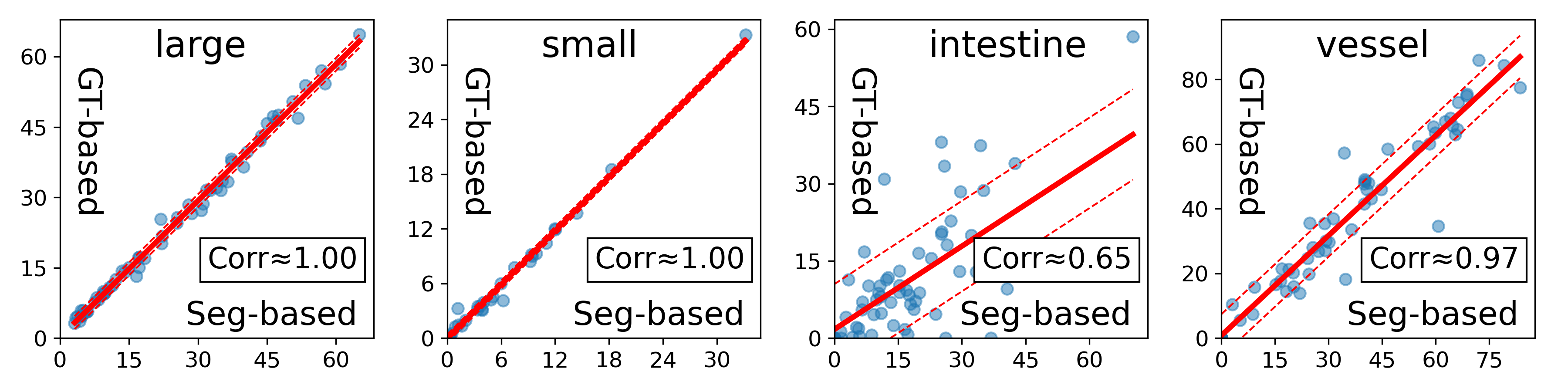}
    \caption{\textbf{Correlation Between GT-based and Segmentator-based Anatomy-Aware Metrics on NeRF~\cite{mildenhall2021nerf}.} The solid red line is a linear fit and the dashed lines denote $\pm 1\sigma$ of the residuals. High correlation shows that the proposed anatomy segmentator is a strong substitute for ground truth labels when building anatomy-aware metrics. }
    \label{fig:supp_seg_corr-nerf}
\end{figure}
\begin{figure}[ht]
    \centering
    \includegraphics[width=\linewidth]{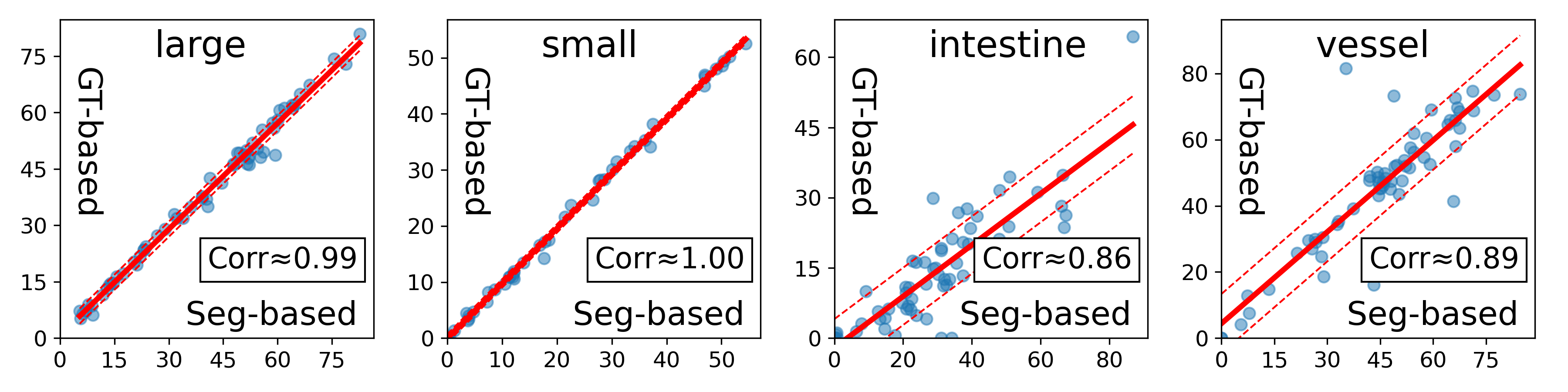}
    \caption{\textbf{Correlation Between GT-based and Segmentator-based Anatomy-Aware Metrics on TensoRF~\cite{chen2022tensorf}.} The solid red line is a linear fit and the dashed lines denote $\pm 1\sigma$ of the residuals. High correlation shows that the proposed anatomy segmentator is a strong substitute for ground truth labels when building anatomy-aware metrics. }
    \label{fig:supp_seg_corr-tensorf}
\end{figure}
\begin{figure}[ht]
    \centering
    \includegraphics[width=\linewidth]{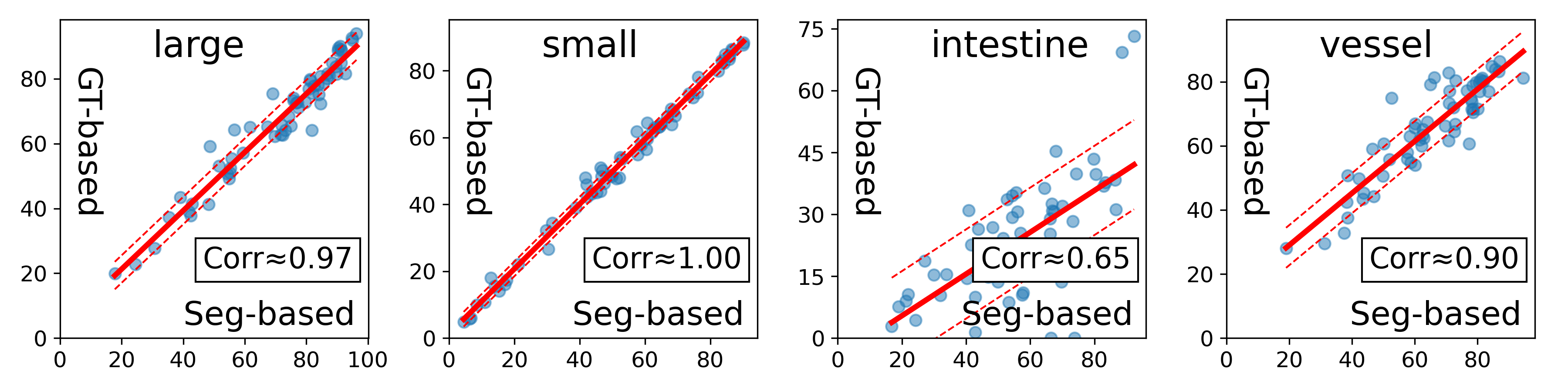}
    \caption{\textbf{Correlation Between GT-based and Segmentator-based Anatomy-Aware Metrics on R$^2$-GS~\cite{zha2024r}.} The solid red line is a linear fit and the dashed lines denote $\pm 1\sigma$ of the residuals. High correlation shows that the proposed anatomy segmentator is a strong substitute for ground truth labels when building anatomy-aware metrics. }
    \label{fig:supp_seg_corr-r2GS}
\end{figure}
\begin{figure}[ht]
    \centering
    \includegraphics[width=\linewidth]{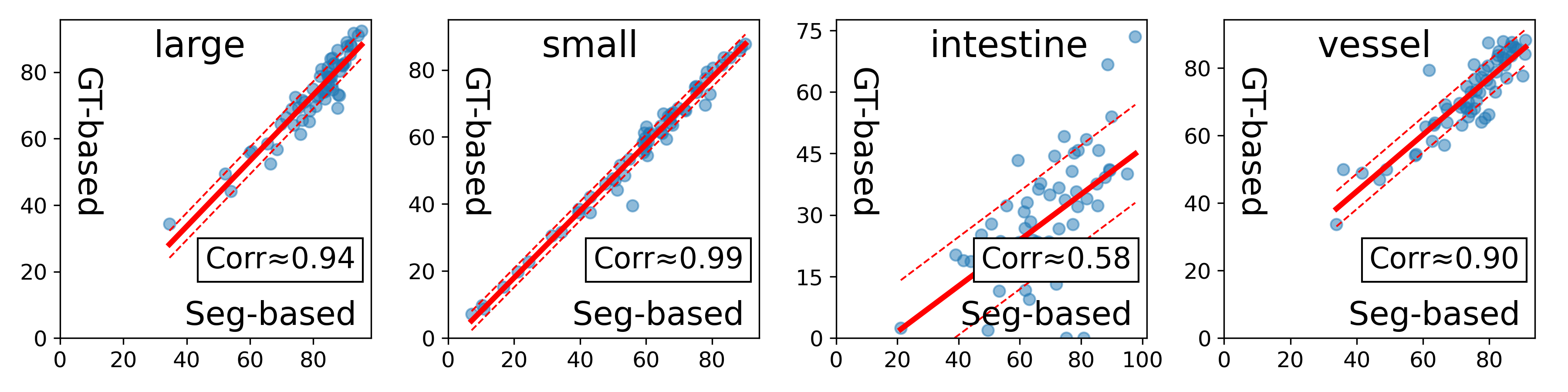}
    \caption{\textbf{Correlation Between GT-based and Segmentator-based Anatomy-Aware Metrics on NAF~\cite{zha2022naf}.} The solid red line is a linear fit and the dashed lines denote $\pm 1\sigma$ of the residuals. High correlation shows that the proposed anatomy segmentator is a strong substitute for ground truth labels when building anatomy-aware metrics. }
    \label{fig:supp_seg_corr-naf}
\end{figure}
\begin{figure}[ht]
    \centering
    \includegraphics[width=\linewidth]{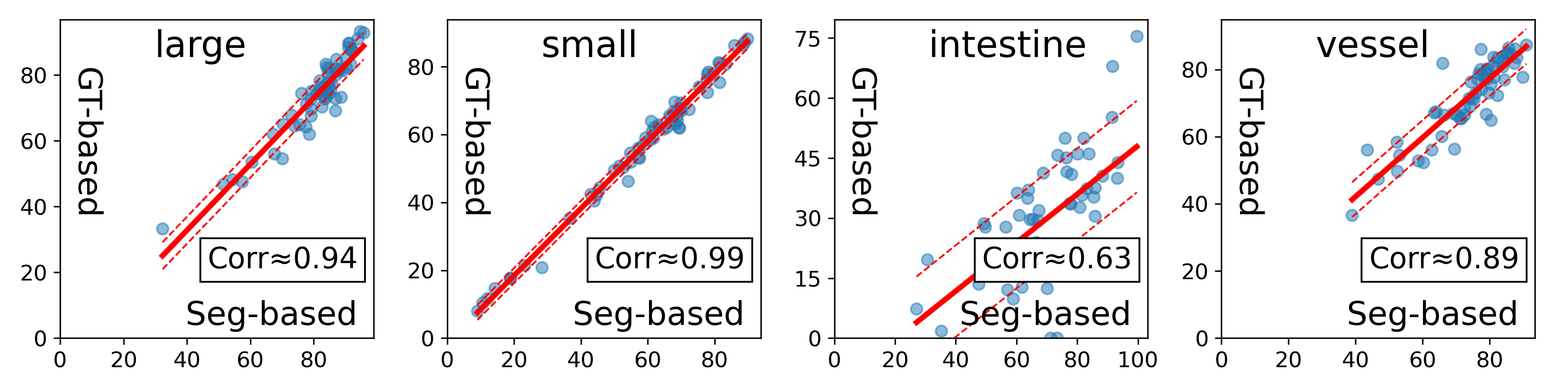}
    \caption{\textbf{Correlation Between GT-based and Segmentator-based Anatomy-Aware Metrics on FDK~\cite{feldkamp1984practical}.} The solid red line is a linear fit and the dashed lines denote $\pm 1\sigma$ of the residuals. High correlation shows that the proposed anatomy segmentator is a strong substitute for ground truth labels when building anatomy-aware metrics. }
    \label{fig:supp_seg_corr-FDK}
\end{figure}
\begin{figure}[ht]
    \centering
    \includegraphics[width=\linewidth]{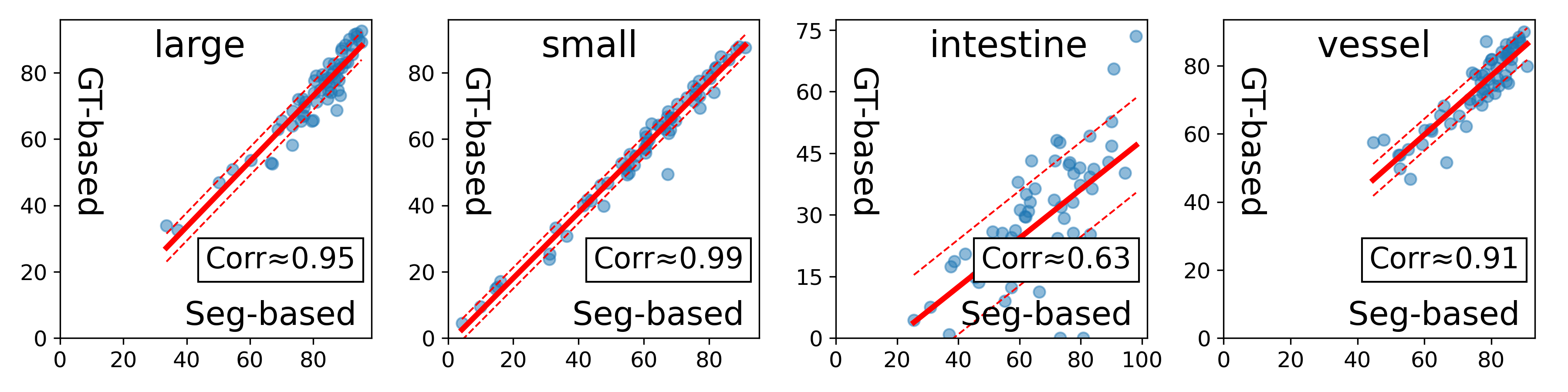}
    \caption{\textbf{Correlation Between GT-based and Segmentator-based Anatomy-Aware Metrics on SART~\cite{andersen1984simultaneous}.} The solid red line is a linear fit and the dashed lines denote $\pm 1\sigma$ of the residuals. High correlation shows that the proposed anatomy segmentator is a strong substitute for ground truth labels when building anatomy-aware metrics. }
    \label{fig:supp_seg_corr-SART}
\end{figure}
\begin{figure}[ht]
    \centering
    \includegraphics[width=\linewidth]{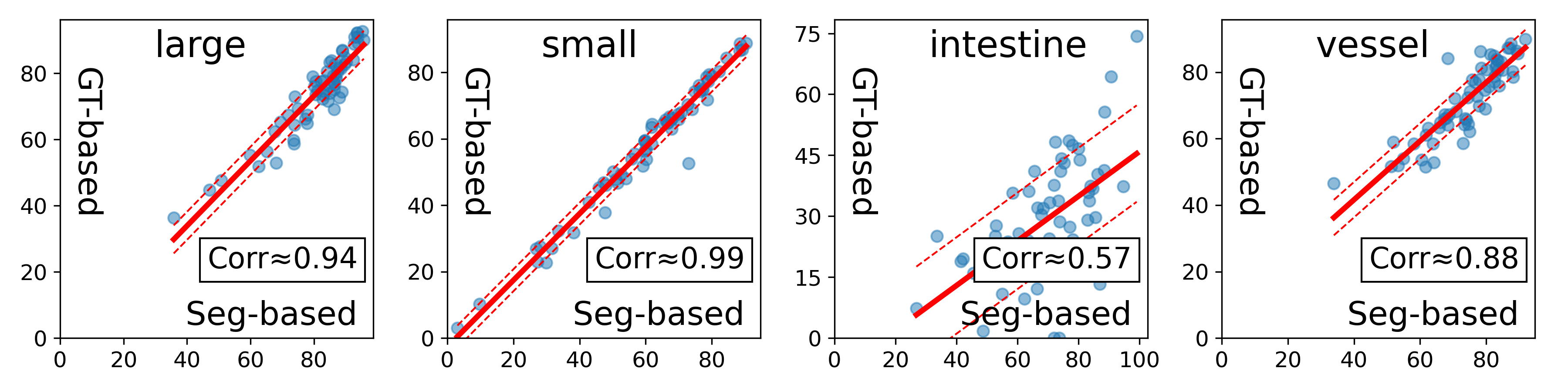}
    \caption{\textbf{Correlation Between GT-based and Segmentator-based Anatomy-Aware Metrics on ASD-POCS~\cite{sidky2008image}.} The solid red line is a linear fit and the dashed lines denote $\pm 1\sigma$ of the residuals. High correlation shows that the proposed anatomy segmentator is a strong substitute for ground truth labels when building anatomy-aware metrics. }
    \label{fig:supp_seg_corr-ASD}
\end{figure}
\begin{figure}[ht]
    \centering
    \includegraphics[width=\linewidth]{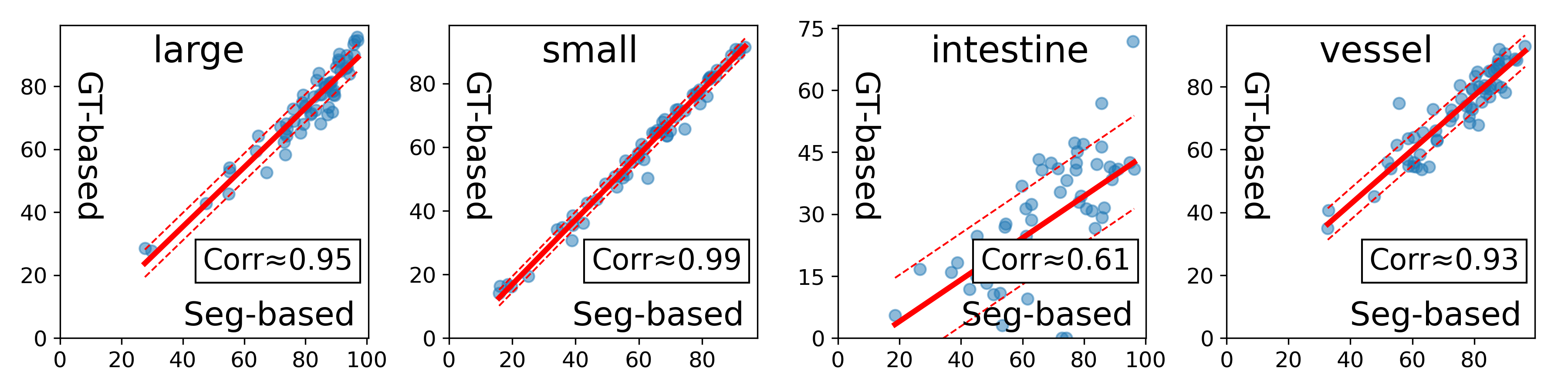}
    \caption{\textbf{Correlation Between GT-based and Segmentator-based Anatomy-Aware Metrics on SAX-NeRF~\cite{cai2024structure}.} The solid red line is a linear fit and the dashed lines denote $\pm 1\sigma$ of the residuals. High correlation shows that the proposed anatomy segmentator is a strong substitute for ground truth labels when building anatomy-aware metrics. }
    \label{fig:supp_seg_corr-sax}
\end{figure}

\clearpage\subsection{Poor Correlation between Pixel-Wise Metrics and Anatomical Preservation Performance}
\figureautorefname~\ref{fig:pixel-seg-corr} summarizes the poor correlation between pixel-wise metrics and anatomy-aware metrics over six CT reconstruction methods: FDK~\cite{feldkamp1984practical}, SART~\cite{andersen1984simultaneous}, ASD-POCS~\cite{sidky2008image}, NAF~\cite{zha2022naf}, SAX-NeRF~\cite{cai2024structure}, R$^2$-GS~\cite{zha2024r}, where NeRF~\cite{mildenhall2021nerf}, TensoRF~\cite{TensoRF}, and InTomo~\cite{zang2021intratomo} are eliminated due to large number of zero metric samples. This indicates that better anatomical preservation does not ensure higher pixel-wise metrics.

Here, we also provide the detailed correlation scatter plots of all nine CT reconstruction methods on all pixel-wise and anatomy-aware metrics. These correlation plots consistently give the same conclusion.

\begin{figure}[ht]
    \centering
    \includegraphics[width=\linewidth]{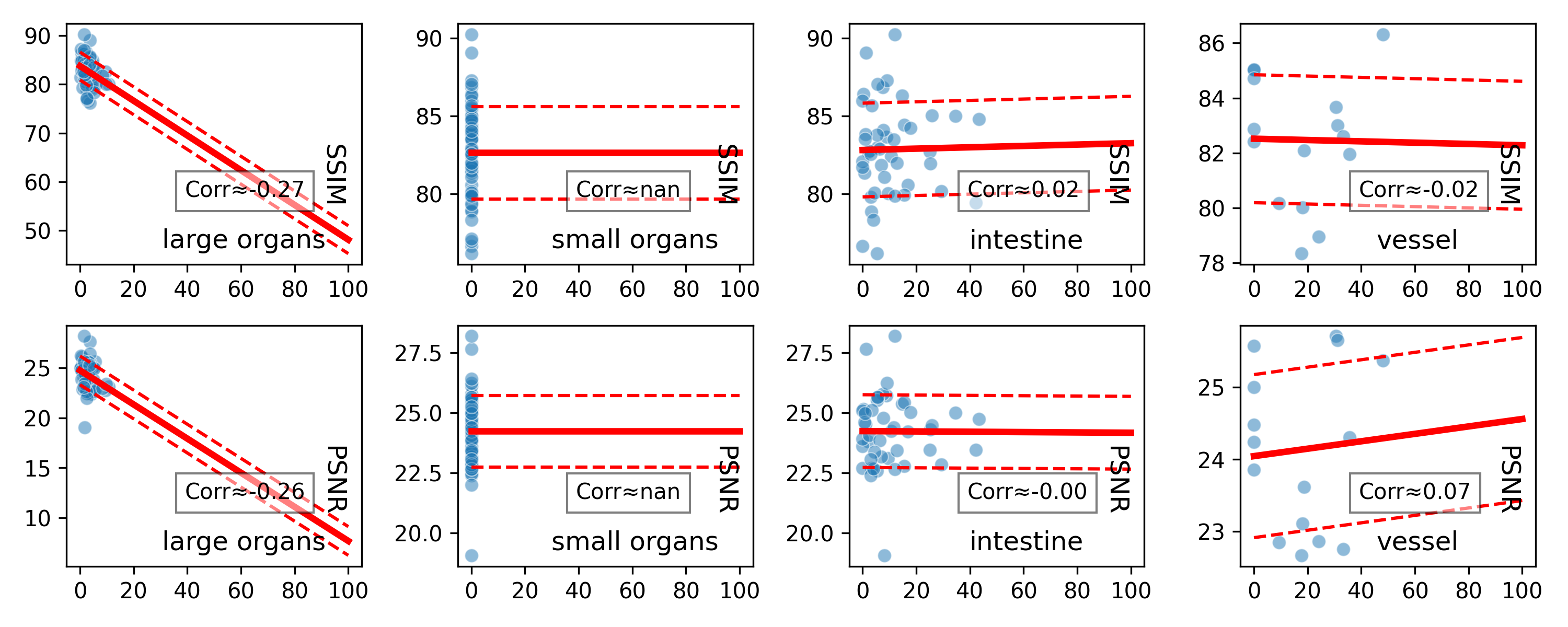}
    \caption{\textbf{Correlation Between Anatomy-Aware Metrics and Pixel-Wise Metrics of InTomo~\cite{zang2021intratomo}.} The solid red line is a linear fit and the dashed lines denote $\pm 1\sigma$ of the residuals. Low correlation indicates that better anatomical preservation does not guarantee higher pixel-wise metrics.}
    \label{fig:supp_pitfall_results-intratomo}
\end{figure}
\begin{figure}[ht]
    \centering
    \includegraphics[width=\linewidth]{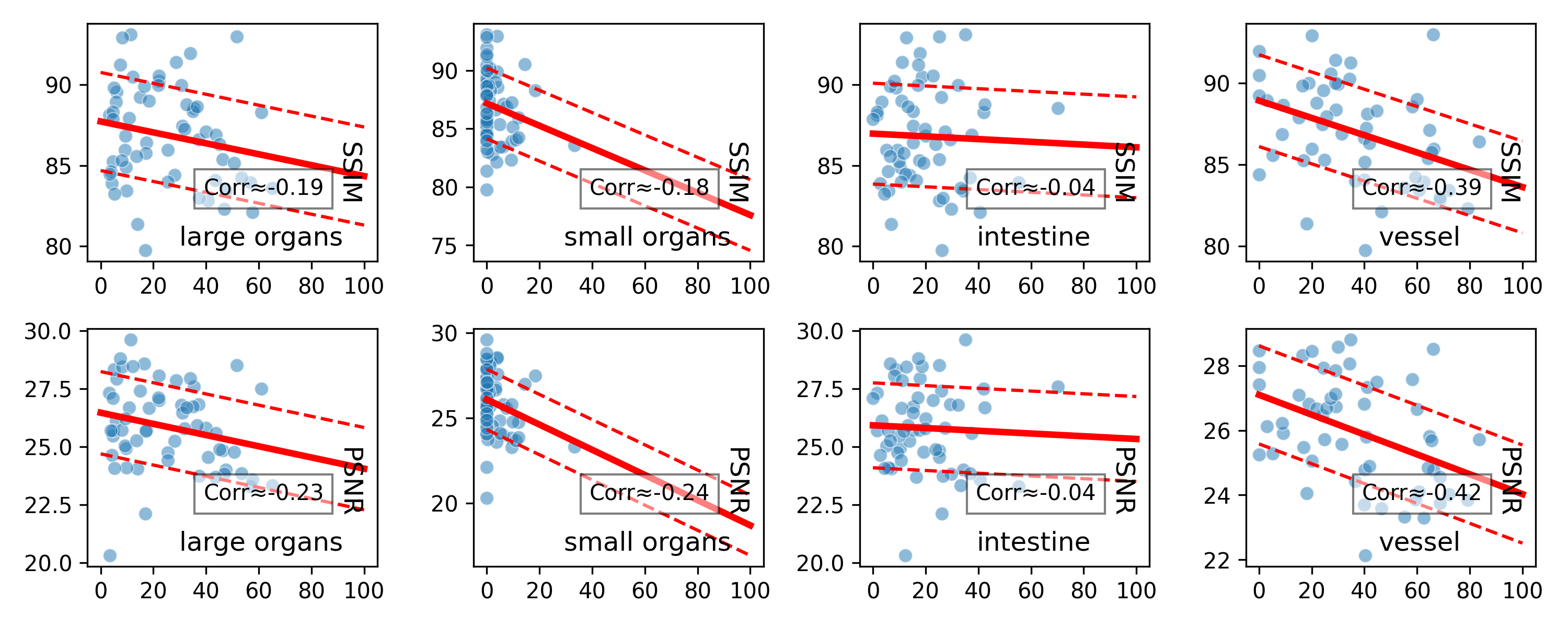}
    \caption{\textbf{Correlation Between Anatomy-Aware Metrics and Pixel-Wise Metrics of NeRF~\cite{mildenhall2021nerf}.} The solid red line is a linear fit and the dashed lines denote $\pm 1\sigma$ of the residuals. Low correlation indicates that better anatomical preservation does not guarantee higher pixel-wise metrics.}
    \label{fig:supp_pitfall_results-nerf}
\end{figure}
\begin{figure}[ht]
    \centering
    \includegraphics[width=\linewidth]{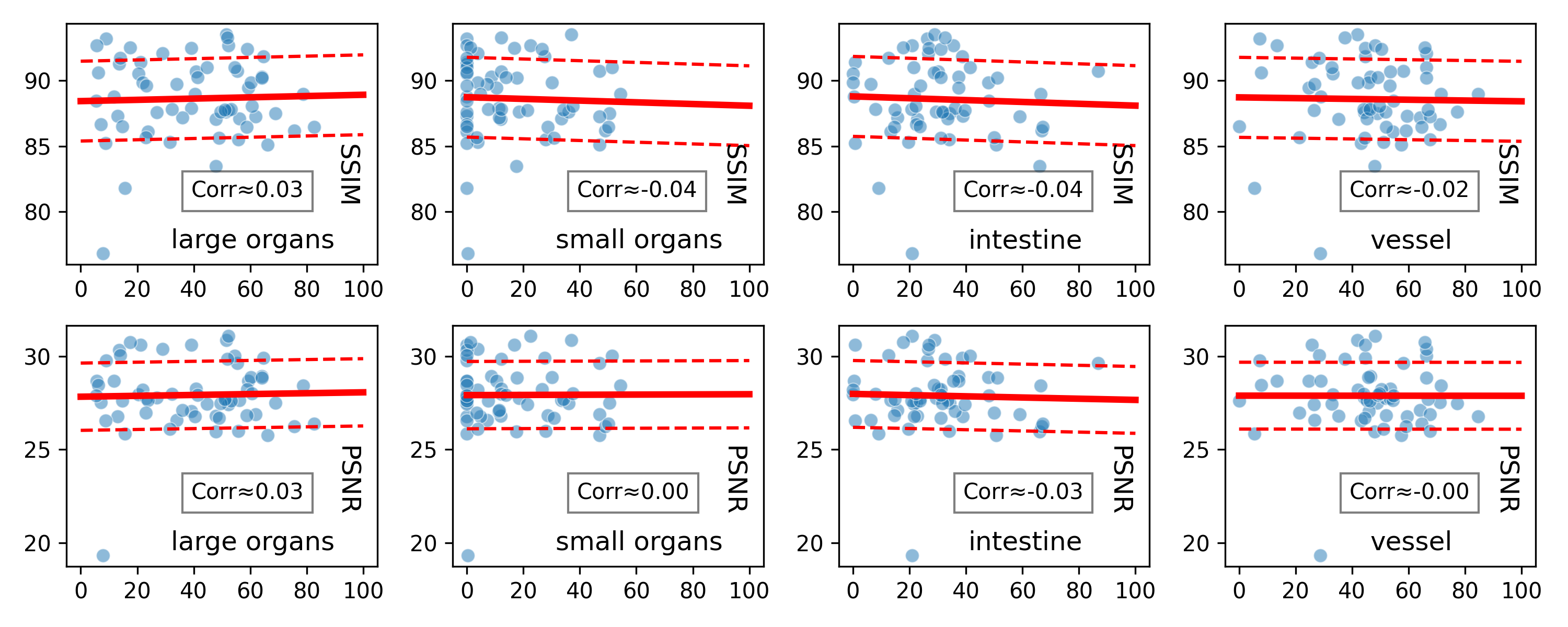}
    \caption{\textbf{Correlation Between Anatomy-Aware Metrics and Pixel-Wise Metrics of TensoRF~\cite{chen2022tensorf}.} The solid red line is a linear fit and the dashed lines denote $\pm 1\sigma$ of the residuals. Low correlation indicates that better anatomical preservation does not guarantee higher pixel-wise metrics.}
    \label{fig:supp_pitfall_results-tensorf}
\end{figure}
\begin{figure}[ht]
    \centering
    \includegraphics[width=\linewidth]{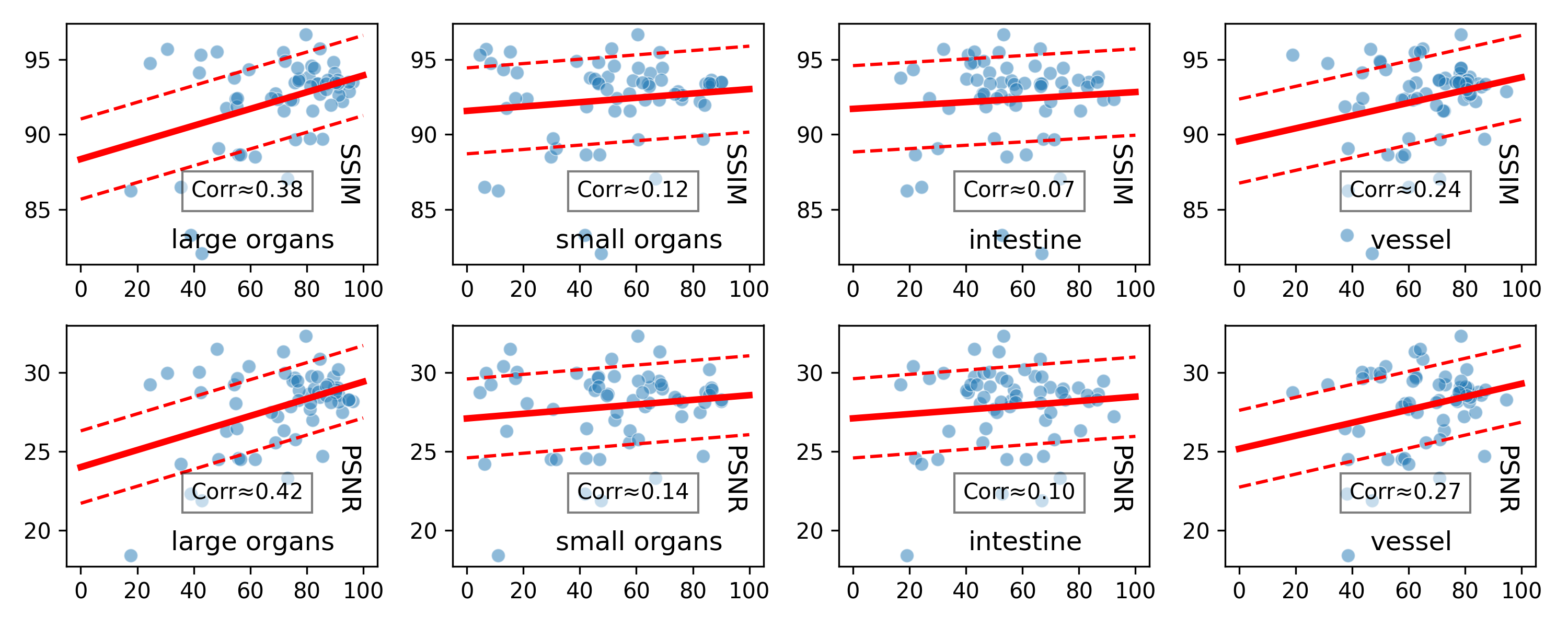}
    \caption{\textbf{Correlation Between Anatomy-Aware Metrics and Pixel-Wise Metrics of R$^2$-GS~\cite{zha2024r}.} The solid red line is a linear fit and the dashed lines denote $\pm 1\sigma$ of the residuals. Low correlation indicates that better anatomical preservation does not guarantee higher pixel-wise metrics.}
    \label{fig:supp_pitfall_results-r2GS}
\end{figure}
\begin{figure}[ht]
    \centering
    \includegraphics[width=\linewidth]{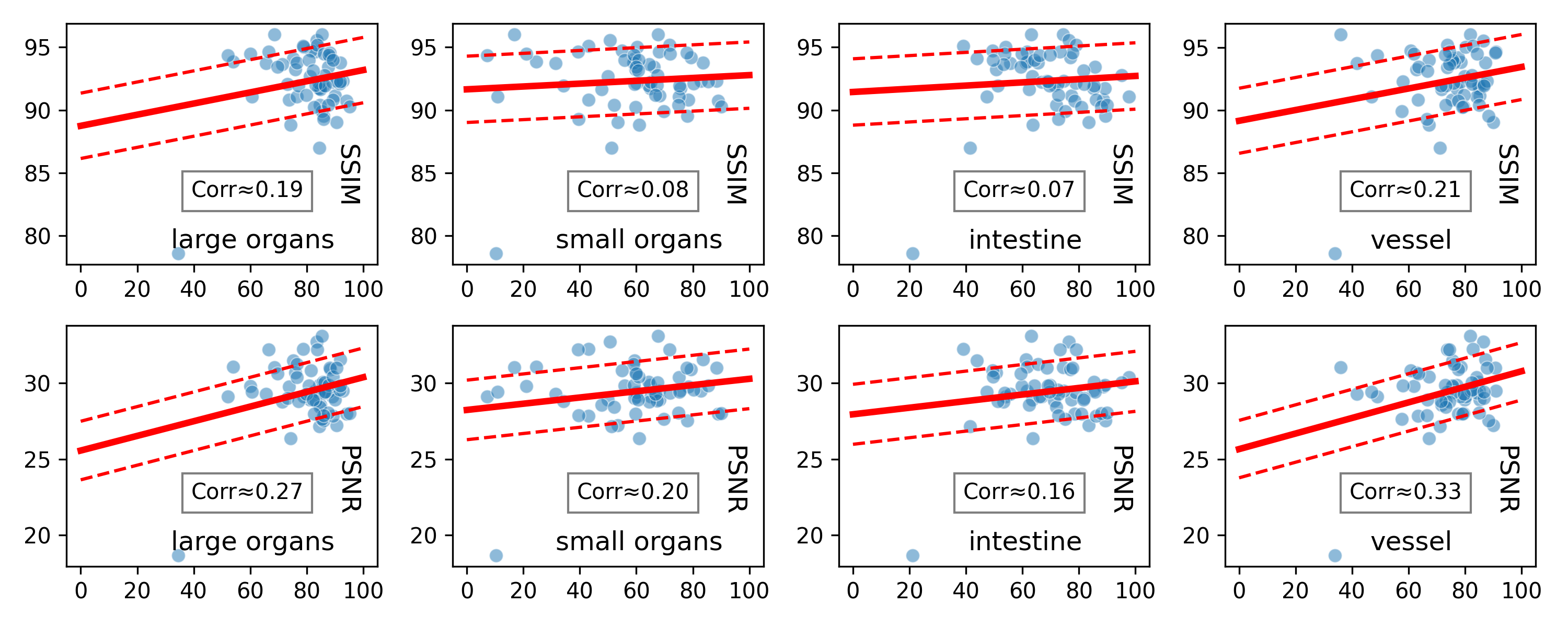}
    \caption{\textbf{Correlation Between Anatomy-Aware Metrics and Pixel-Wise Metrics of NAF~\cite{zha2022naf}.} The solid red line is a linear fit and the dashed lines denote $\pm 1\sigma$ of the residuals. Low correlation indicates that better anatomical preservation does not guarantee higher pixel-wise metrics.}
    \label{fig:supp_pitfall_results-naf}
\end{figure}
\begin{figure}[ht]
    \centering
    \includegraphics[width=\linewidth]{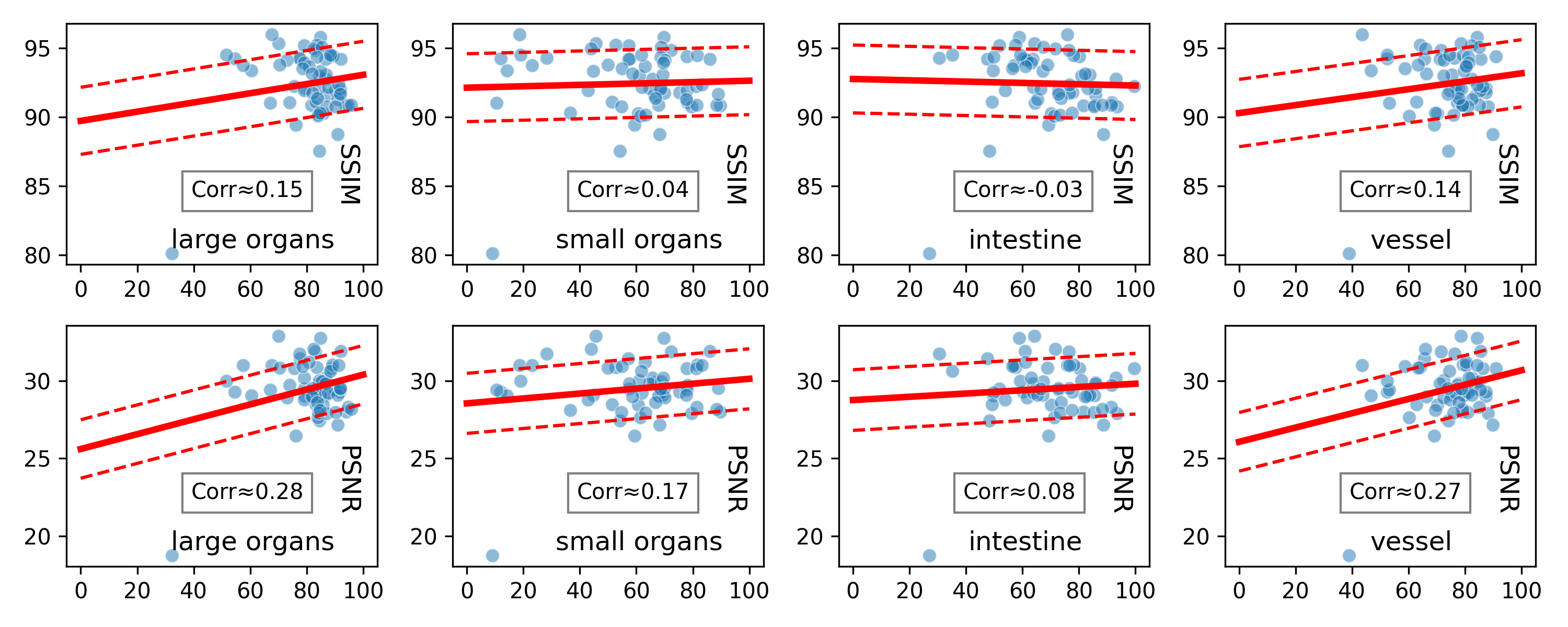}
    \caption{\textbf{Correlation Between Anatomy-Aware Metrics and Pixel-Wise Metrics of FDK~\cite{feldkamp1984practical}.} The solid red line is a linear fit and the dashed lines denote $\pm 1\sigma$ of the residuals. Low correlation indicates that better anatomical preservation does not guarantee higher pixel-wise metrics.}
    \label{fig:supp_pitfall_results-FDK}
\end{figure}
\begin{figure}[ht]
    \centering
    \includegraphics[width=\linewidth]{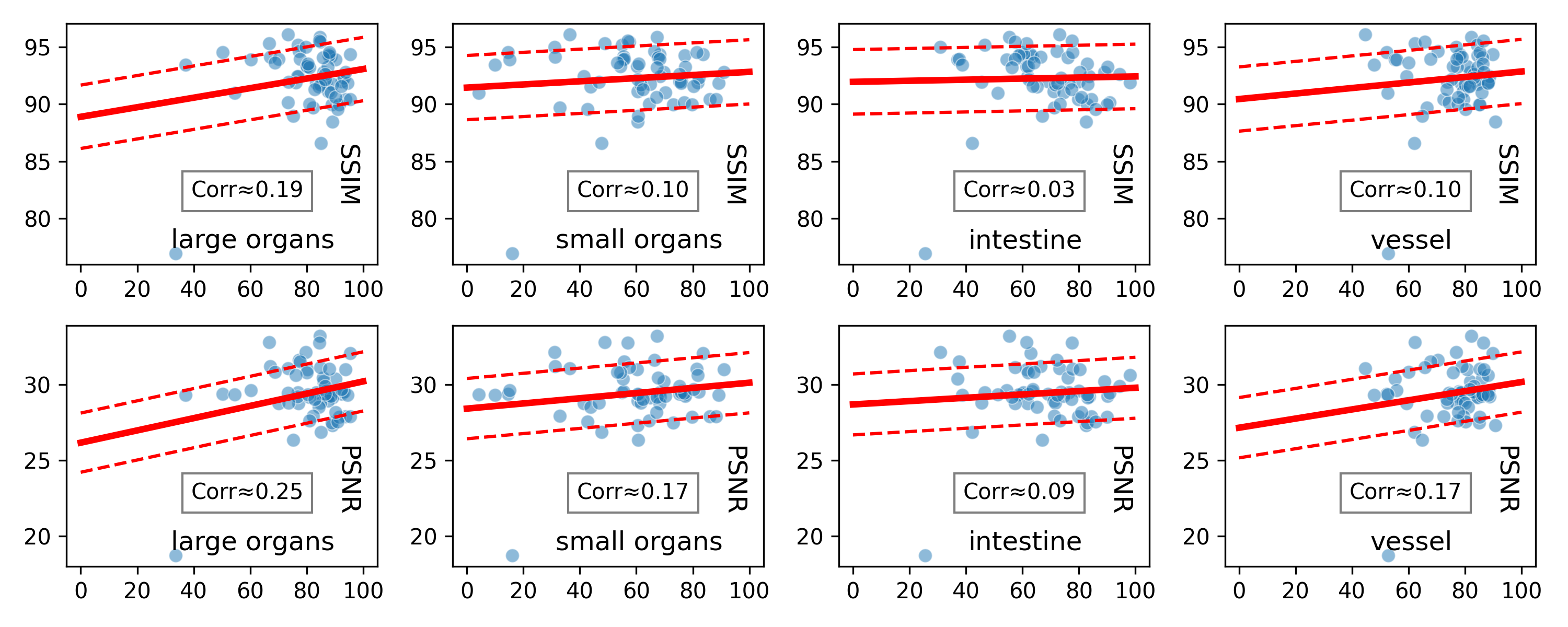}
    \caption{\textbf{Correlation Between Anatomy-Aware Metrics and Pixel-Wise Metrics of SART~\cite{andersen1984simultaneous}.} The solid red line is a linear fit and the dashed lines denote $\pm 1\sigma$ of the residuals. Low correlation indicates that better anatomical preservation does not guarantee higher pixel-wise metrics.}
    \label{fig:supp_pitfall_results-SART}
\end{figure}
\begin{figure}[ht]
    \centering
    \includegraphics[width=\linewidth]{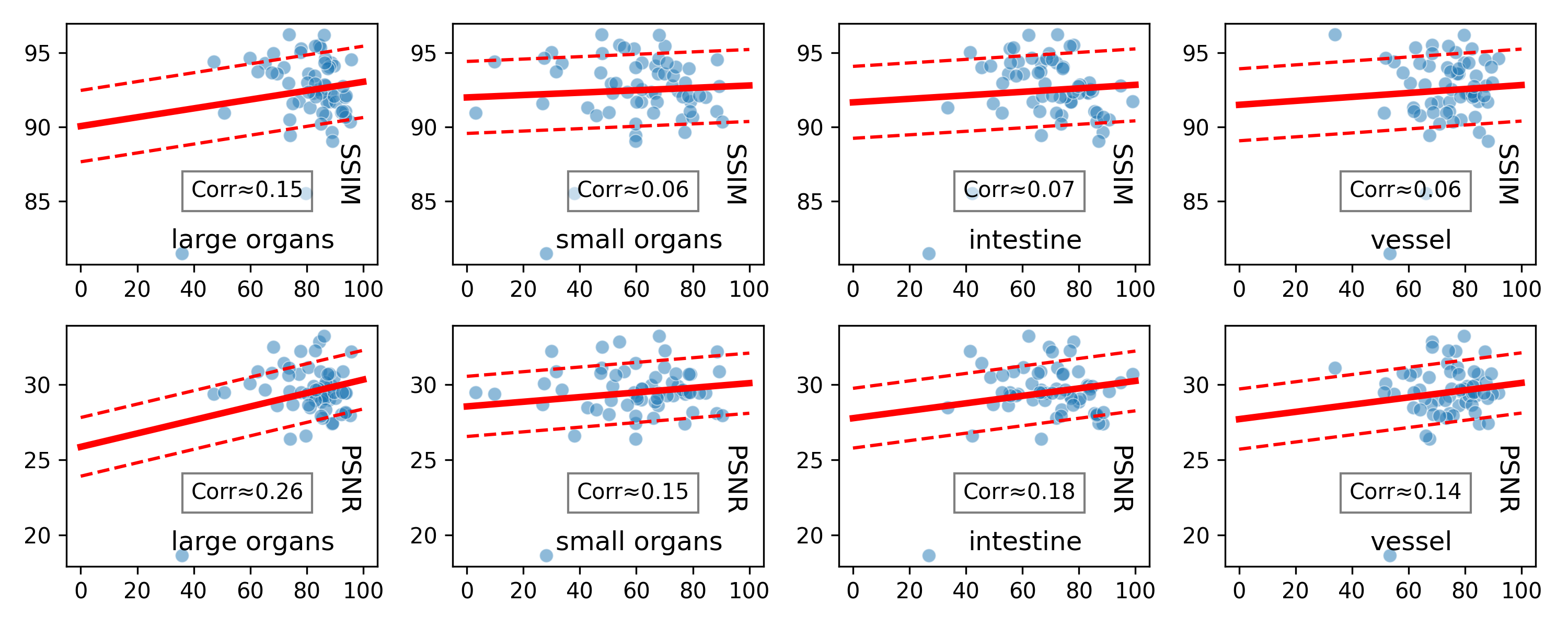}
    \caption{\textbf{Correlation Between Anatomy-Aware Metrics and Pixel-Wise Metrics of ASD-POCS~\cite{sidky2008image}.} The solid red line is a linear fit and the dashed lines denote $\pm 1\sigma$ of the residuals. Low correlation indicates that better anatomical preservation does not guarantee higher pixel-wise metrics.}
    \label{fig:supp_pitfall_results-ASD}
\end{figure}
\begin{figure}[ht]
    \centering
    \includegraphics[width=\linewidth]{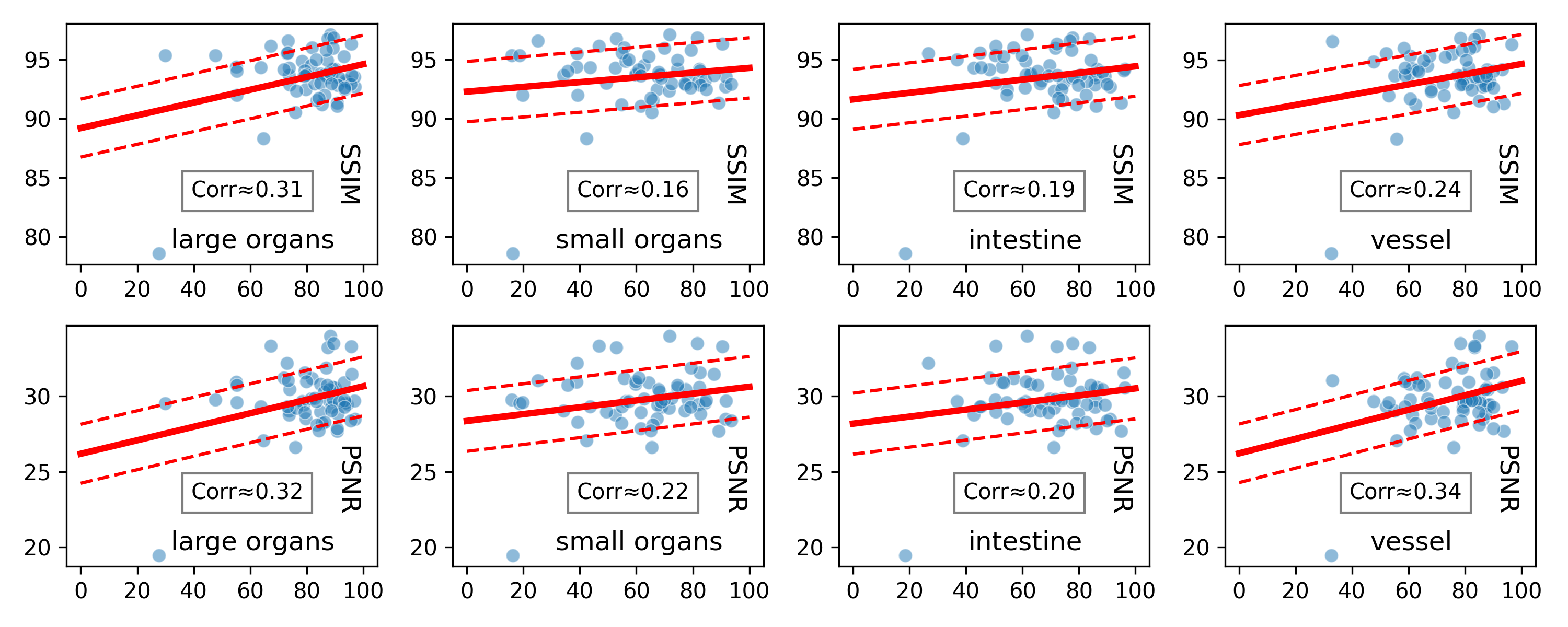}
    \caption{\textbf{Correlation Between Anatomy-Aware Metrics and Pixel-Wise Metrics of SAX-NeRF~\cite{cai2024structure}.} The solid red line is a linear fit and the dashed lines denote $\pm 1\sigma$ of the residuals. Low correlation indicates that better anatomical preservation does not guarantee higher pixel-wise metrics.}
    \label{fig:supp_pitfall_results-sax}
\end{figure}

\clearpage\section{Technical Details of \loss}\label{sec:supp-technical-details}
In \S\ref{sec:method-CARE}, we discuss the \emph{Preliminary} stage of the \loss\ framework: Autoencoder and Latent Diffusion Model. In this section, we provide more technical details of the autoencoder in \S\ref{sec:supp-technical-details-autoencoder}, and provide more technical specifications of the latent diffusion model in \S\ref{sec:supp-technical-details-diffusion}.
\subsection{Autoencoder}\label{sec:supp-technical-details-autoencoder}
In this initial stage, we finetuned a KL-regularized Variational Autoencoder (KL-VAE), using the JHH CT scans. The encoder of the VAE compresses input CT images into latent representations (denoted as $z$), while the decoder reconstructs the original images from these latent variables. The training objective aims to optimize both image reconstruction quality and the regularization imposed by the Kullback-Leibler divergence to ensure meaningful latent representations. These trained autoencoder serve as the foundational components for subsequent diffusion model training.

Specifically, let $X$ denote an input CT image in pixel space. Note that to consider inter-frame consistency and retain the original architecture of the autoencoder network, we stack three adjacent CT slices to be the three channels of the input image, giving $X\in\mathbb R^{H\times W\times 3}$.
The encoder $\mathcal E_{\theta_E}$ maps $X$ to a latent representation $z=\mathcal E_{\theta_E}(X)\in\mathbb R^{h\times w\times c}$.
A paired decoder $\mathcal D_{\theta_D}$ reconstructs the input as $\hat X=\mathcal D_{\theta_D}(z)$.
Training adapts the KL‑VAE pretrained on natural images (checkpoint provided by Stable Diffusion v1.5~\cite{rombach2021highresolution}) to the target CT domain by minimising a weighted sum of three complementary criteria:

\begin{equation}
\begin{aligned}
\mathcal L_{\text{AE}} &=\lambda_{\text{rec}}  \underbrace{\lVert X-\hat X\rVert_1}_{\text{reconstruction}}
+\lambda_{\text{per}} \underbrace{\sum_{l}w_l  \bigl\lVert\phi_l(X)-\phi_l(\hat X)\bigr\rVert_2^2}_{\text{perceptual}}\\
&+\beta \underbrace{{\mathrm{KL}}\bigl(q_{\theta_E}(z\mid X) \| \mathcal N(0,I)\bigr)}_{\text{KL regularisation}}
\end{aligned}
\end{equation}
where:
\begin{itemize}
    \item \textbf{Reconstruction term $\mathcal L_{\text{rec}}$} directly penalises pixel‑wise errors, preserving low‑frequency fidelity.
    \item \textbf{Perceptual term $\mathcal L_{\text{per}}$} computes the $\ell_2$ distance between intermediate feature maps $\phi_l(\cdot)$ of a fixed vision backbone (here VGG‑16 pretrained on ImageNet).  It sharpens textures and maintains semantic consistency that pure pixel metrics miss.
    \item \textbf{KL term $\mathcal L_{\text{KL}}$} pulls the encoder’s approximate posterior $q_{\theta_E}(z\mid X)$ towards the unit Gaussian prior, preventing latent collapse and ensuring the aggregated latent distribution remains close to $\mathcal N(0,I)$.  This normalization is crucial for the subsequent diffusion prior, whose noise schedule assumes unit‑variance latents.
\end{itemize}

The weights $\lambda_{\text{rec}}$, $\lambda_{\text{per}}$, and $\beta$ control the fidelity–regularity trade‑off between generative capacity and roughly standard‑normal latents. Unlike the adversarially‑augmented auto‑encoders used in Stable Diffusion, we fintuned on the three penalizations above directly, simplifying the optimization process and eliminating adversary‑induced instabilities without sacrificing downstream sample quality. 
After fintuning the KL-VAE model on Stable Diffusion's checkpoint, frozen latents $z$ serve as training data for the latent‑space diffusion model described below.

\clearpage\subsection{Latent Diffusion Model}\label{sec:supp-technical-details-diffusion}
This stage pretrains a latent diffusion model with the JHH dataset to enhance low-quality CT scans while preserving anatomical structures.
Each high‑quality CT image $X\in\mathbb{R}^{H\times W\times 3}$ is first compressed by the frozen KL‑VAE encoder $\mathcal{E_{\theta_E}}$ into a latent representation $z=\mathcal{E_{\theta_E}}(X)\in\mathbb{R}^{h\times w\times c}$.  In the forward diffusion process a noisy sequence $\{z_t\}_{t=0}^{T}$ is obtained by
\begin{equation}\label{eq:supp-add-noise}
    z_t=\sqrt{\bar\alpha_t} z_0+\sqrt{1-\bar\alpha_t} \epsilon,
    \qquad
    \epsilon\sim\mathcal{N}(0,I),\;t\in\{1,\dots,T\},
\end{equation}
where $z_0\triangleq z$ and $\{\bar\alpha_t\}_{t=1}^{T}$ is a fixed variance‑preserving schedule.  
To expose structural cues during de-noising we degrade $z$ with a deterministic operator $\mathcal H(\cdot)$ and concatenate it with $z_t$ along the channel dimension,
\begin{equation}\label{eq:concat-degradation}
    \tilde z_t=\operatorname{Concat}[ z_t,\mathcal H(z) ]\;\in\mathbb{R}^{h\times w\times 2c}.
\end{equation}
where $\mathcal H(\cdot)$ is a downsampling process with a factor of 4 and followed by an upsampling process back to the original resolution.

A de-noising UNet $\epsilon_\theta$ receives $\tilde z_t$, the diffusion timestep $t$, and a fixed text embedding $c_{\mathrm{phase}}$ of the contrast phase information, and predicts the noise residual $\hat\epsilon=\epsilon_\theta(\tilde z_t,t,c_{\mathrm{phase}})$.
Then, the network parameters are optimized with the standard noise‑prediction loss
\begin{equation}\label{eq:LDM}
    \mathcal L_{\text{LDM}}
    =\mathbb E_{t,\epsilon,X}
      \left\|
        \epsilon-\epsilon_\theta\bigl(
          \operatorname{Concat}[z_t,\mathcal H(z)], t, c_{\mathrm{phase}}
        \bigr)
      \right\|_2^{2}.
\end{equation}
Because only $\mathcal L_{\text{LDM}}$ is used, learning focuses on accurate latent de-noising; $\mathcal H(z)$ supplies anatomy so the de-noising UNet acquires an intrinsic enhancement capability without extra image‑space supervision.

\clearpage\section{Experimental Results of \loss}\label{sec:supp-exp-results}
\subsection{Benchmarking on Other Segmentation Metrics}
\begin{table}[ht]
    \centering
    \begin{threeparttable}{}
    \scriptsize
    \caption{\textbf{Benchmarking CT Reconstruction Methods on \emph{NSD}.} We evaluate preexisting CT reconstruction methods on high-quality CT scans using both pixel-wise metrics and our anatomy-aware metrics. We report the median and interquartile range (IQR) of these metrics. Cells are marked in \textcolor{blue}{blue}, where a deeper color denotes a greater value.} 
    \label{tab:Benchmarking-results-nsd}
    \begin{tabular}{p{0.13\linewidth}P{0.11\linewidth}P{0.11\linewidth}P{0.11\linewidth}P{0.11\linewidth}P{0.11\linewidth}P{0.11\linewidth}}
    \toprule
      & \multicolumn{2}{c}{pixel-wise metric} & \multicolumn{4}{c}{\textbf{anatomy-aware metric (ours)}}\\
    \cmidrule(lr){2-3}\cmidrule(lr){4-7}
    method  & SSIM & PSNR & \makecell{ NSD$_\text{large}$} & \makecell{NSD$_\text{small}$} & NSD$_\text{intestine}$  & \makecell{ NSD$_\text{vessel}$}\\
    \midrule
    InTomo~\cite{zang2021intratomo}         &\cellcolor{blue!41.0}82.7\tiny{~(80.6,84.7)}  &\cellcolor{blue!36.45322648617175}24.3\tiny{~(23.1,25.2)}  &\cellcolor{blue!1.0}2.7\tiny{~(1.9,4.1)}  &\cellcolor{blue!0.0}0.0\tiny{~(0.0,0.0)}  &\cellcolor{blue!1.0}2.0\tiny{~(0.3,3.6)}  &\cellcolor{blue!0.0}0.0\tiny{~(0.0,0.0)} \\
    NeRF~\cite{mildenhall2021nerf}          &\cellcolor{blue!43.0}86.8\tiny{~(84.4,89.0)}  &\cellcolor{blue!38.581192872975826}25.7\tiny{~(24.7,27.1)}  &\cellcolor{blue!10.0}21.7\tiny{~(9.5,37.6)}  &\cellcolor{blue!0.0}0.1\tiny{~(0.0,3.9)}  &\cellcolor{blue!2.0}5.1\tiny{~(3.4,9.6)}  &\cellcolor{blue!2.0}5.8\tiny{~(1.9,16.6)} \\
    TensoRF~\cite{chen2022tensorf}          &\cellcolor{blue!44.0}88.4\tiny{~(87.0,90.7)}  &\cellcolor{blue!41.64348063564633}27.8\tiny{~(26.9,28.9)}  &\cellcolor{blue!20.0}41.3\tiny{~(20.3,50.5)}  &\cellcolor{blue!5.0}10.9\tiny{~(0.0,28.3)}  &\cellcolor{blue!4.0}9.4\tiny{~(4.8,14.8)}  &\cellcolor{blue!10.0}20.5\tiny{~(10.1,40.2)} \\
    R$^2$-GS~\cite{zha2024r}                &\cellcolor{blue!46.0}93.2\tiny{~(91.9,93.9)}  &\cellcolor{blue!42.84789951357505}28.6\tiny{~(27.0,29.5)}  &\cellcolor{blue!36.0}72.6\tiny{~(57.1,80.7)}  &\cellcolor{blue!27.0}54.0\tiny{~(42.3,67.9)}  &\cellcolor{blue!7.0}15.1\tiny{~(10.5,19.8)}  &\cellcolor{blue!31.0}62.1\tiny{~(47.9,75.6)} \\
    NAF~\cite{zha2022naf}                   &\cellcolor{blue!46.0}92.3\tiny{~(91.1,94.1)}  &\cellcolor{blue!44.1268279757467}29.4\tiny{~(28.8,30.6)}  &\cellcolor{blue!37.0}74.7\tiny{~(69.0,82.2)}  &\cellcolor{blue!30.0}61.1\tiny{~(47.0,68.6)}  &\cellcolor{blue!9.0}18.2\tiny{~(13.2,24.6)}  &\cellcolor{blue!35.0}71.3\tiny{~(64.0,81.9)} \\
    FDK~\cite{feldkamp1984practical}         &\cellcolor{blue!46.0}92.4\tiny{~(91.1,94.2)}  &\cellcolor{blue!44.23771676019473}29.5\tiny{~(28.8,30.8)}  &\cellcolor{blue!37.0}75.3\tiny{~(69.2,82.0)}  &\cellcolor{blue!31.0}62.2\tiny{~(50.4,69.3)}  &\cellcolor{blue!9.0}18.1\tiny{~(14.1,24.6)}  &\cellcolor{blue!36.0}72.3\tiny{~(62.6,81.7)} \\
    SART~\cite{andersen1984simultaneous}    &\cellcolor{blue!46.0}92.5\tiny{~(91.1,94.1)}  &\cellcolor{blue!44.08483814672893}29.4\tiny{~(28.8,30.6)}  &\cellcolor{blue!38.0}76.2\tiny{~(67.2,82.8)}  &\cellcolor{blue!30.0}61.8\tiny{~(49.3,72.5)}  &\cellcolor{blue!8.0}18.0\tiny{~(12.9,24.6)}  &\cellcolor{blue!36.0}73.6\tiny{~(61.3,83.5)} \\
    ASD-POCS~\cite{sidky2008image}          &\cellcolor{blue!46.0}92.5\tiny{~(91.3,94.1)}  &\cellcolor{blue!44.22636306514291}29.5\tiny{~(28.7,30.7)}  &\cellcolor{blue!38.0}76.3\tiny{~(67.4,82.7)}  &\cellcolor{blue!31.0}63.0\tiny{~(47.8,71.8)}  &\cellcolor{blue!9.0}18.6\tiny{~(13.9,24.5)}  &\cellcolor{blue!36.0}72.2\tiny{~(64.6,82.9)} \\
    SAX-NeRF~\cite{cai2024structure}      &\cellcolor{blue!46.0}93.7\tiny{~(92.8,95.0)}  &\cellcolor{blue!44.49139274215548}29.7\tiny{~(29.0,30.8)}  &\cellcolor{blue!37.0}75.9\tiny{~(67.0,82.5)}  &\cellcolor{blue!32.0}64.4\tiny{~(49.8,76.3)}  &\cellcolor{blue!8.0}17.5\tiny{~(13.5,24.2)}  &\cellcolor{blue!35.0}70.5\tiny{~(56.2,85.5)} \\
    \bottomrule
    \end{tabular}
    \end{threeparttable}
\end{table}
\begin{table}[ht]
    \centering
    \begin{threeparttable}{}
    \scriptsize
    \caption{\textbf{Benchmarking CT Reconstruction Methods on \emph{DSC}.} We evaluate preexisting CT reconstruction methods on high-quality CT scans using both pixel-wise metrics and our anatomy-aware metrics. We report the median and interquartile range (IQR) of these metrics. Cells are marked in \textcolor{blue}{blue}, where a deeper color denotes a greater value.} 
    \label{tab:Benchmarking-results-dsc}
    \begin{tabular}{p{0.13\linewidth}P{0.11\linewidth}P{0.11\linewidth}P{0.11\linewidth}P{0.11\linewidth}P{0.11\linewidth}P{0.11\linewidth}}
    \toprule
      & \multicolumn{2}{c}{pixel-wise metric} & \multicolumn{4}{c}{\textbf{anatomy-aware metric (ours)}}\\
    \cmidrule(lr){2-3}\cmidrule(lr){4-7}
    method  & SSIM & PSNR & \makecell{ DSC$_\text{large}$} & \makecell{DSC$_\text{small}$} & DSC$_\text{intestine}$  & \makecell{ DSC$_\text{vessel}$}\\
    \midrule
    InTomo~\cite{zang2021intratomo}      &\cellcolor{blue!41.0}82.7\tiny{~(80.6,84.7)}  &\cellcolor{blue!36.45322648617175}24.3\tiny{~(23.1,25.2)}  &\cellcolor{blue!3.0}7.9\tiny{~(4.6,10.6)}  &\cellcolor{blue!0.0}0.0\tiny{~(0.0,0.0)}  &\cellcolor{blue!1.0}2.4\tiny{~(0.1,4.9)}  &\cellcolor{blue!0.0}0.0\tiny{~(0.0,0.0)} \\
    NeRF~\cite{mildenhall2021nerf}        &\cellcolor{blue!43.0}86.8\tiny{~(84.4,89.0)}  &\cellcolor{blue!38.581192872975826}25.7\tiny{~(24.7,27.1)}  &\cellcolor{blue!16.0}32.1\tiny{~(15.7,49.8)}  &\cellcolor{blue!0.0}0.0\tiny{~(0.0,2.0)}  &\cellcolor{blue!3.0}6.8\tiny{~(3.1,14.4)}  &\cellcolor{blue!4.0}8.3\tiny{~(1.7,20.0)} \\
    TensoRF~\cite{chen2022tensorf}        &\cellcolor{blue!44.0}88.4\tiny{~(87.0,90.7)}  &\cellcolor{blue!41.64348063564633}27.8\tiny{~(26.9,28.9)}  &\cellcolor{blue!27.0}54.5\tiny{~(32.3,68.0)}  &\cellcolor{blue!3.0}7.0\tiny{~(0.0,20.9)}  &\cellcolor{blue!6.0}14.0\tiny{~(4.9,20.8)}  &\cellcolor{blue!11.0}22.9\tiny{~(12.9,37.8)} \\
    R$^2$-GS~\cite{zha2024r}              &\cellcolor{blue!46.0}93.2\tiny{~(91.9,93.9)}  &\cellcolor{blue!42.84789951357505}28.6\tiny{~(27.0,29.5)}  &\cellcolor{blue!39.0}79.3\tiny{~(73.0,86.0)}  &\cellcolor{blue!23.0}46.9\tiny{~(35.0,61.6)}  &\cellcolor{blue!12.0}25.2\tiny{~(16.7,32.5)}  &\cellcolor{blue!28.0}57.1\tiny{~(40.5,65.9)} \\
    NAF~\cite{zha2022naf}                &\cellcolor{blue!46.0}92.3\tiny{~(91.1,94.1)}  &\cellcolor{blue!44.1268279757467}29.4\tiny{~(28.8,30.6)}  &\cellcolor{blue!42.0}84.1\tiny{~(76.5,87.5)}  &\cellcolor{blue!26.0}52.9\tiny{~(39.6,63.9)}  &\cellcolor{blue!14.0}29.7\tiny{~(21.2,40.0)}  &\cellcolor{blue!31.0}63.3\tiny{~(56.6,73.0)} \\
    FDK~\cite{feldkamp1984practical}       &\cellcolor{blue!46.0}92.4\tiny{~(91.1,94.2)}  &\cellcolor{blue!44.23771676019473}29.5\tiny{~(28.8,30.8)}  &\cellcolor{blue!42.0}84.2\tiny{~(77.3,87.8)}  &\cellcolor{blue!26.0}53.5\tiny{~(42.0,64.0)}  &\cellcolor{blue!14.0}29.6\tiny{~(20.6,40.1)}  &\cellcolor{blue!31.0}63.3\tiny{~(56.0,72.9)} \\
    SART~\cite{andersen1984simultaneous}    &\cellcolor{blue!46.0}92.5\tiny{~(91.1,94.1)}  &\cellcolor{blue!44.08483814672893}29.4\tiny{~(28.8,30.6)}  &\cellcolor{blue!41.0}83.8\tiny{~(77.5,87.8)}  &\cellcolor{blue!26.0}52.3\tiny{~(40.4,64.2)}  &\cellcolor{blue!14.0}28.5\tiny{~(20.4,40.5)}  &\cellcolor{blue!31.0}63.6\tiny{~(55.2,74.3)} \\
    ASD-POCS~\cite{sidky2008image}        &\cellcolor{blue!46.0}92.5\tiny{~(91.3,94.1)}  &\cellcolor{blue!44.22636306514291}29.5\tiny{~(28.7,30.7)}  &\cellcolor{blue!42.0}84.2\tiny{~(77.0,88.3)}  &\cellcolor{blue!26.0}52.6\tiny{~(40.3,65.6)}  &\cellcolor{blue!14.0}29.2\tiny{~(21.5,40.4)}  &\cellcolor{blue!31.0}62.4\tiny{~(54.4,73.0)} \\
    SAX-NeRF~\cite{cai2024structure}       &\cellcolor{blue!46.0}93.7\tiny{~(92.8,95.0)}  &\cellcolor{blue!44.49139274215548}29.7\tiny{~(29.0,30.8)}  &\cellcolor{blue!41.0}83.4\tiny{~(75.4,86.7)}  &\cellcolor{blue!27.0}54.2\tiny{~(39.1,70.6)}  &\cellcolor{blue!14.0}29.7\tiny{~(20.5,38.8)}  &\cellcolor{blue!31.0}63.5\tiny{~(50.6,75.3)} \\
    \bottomrule
    \end{tabular}
    \end{threeparttable}
\end{table}

\clearpage\subsection{Qualitative Results of \loss}\label{sec:supp-qualitative}
\begin{figure}[ht]
    \centering
    \includegraphics[width=\linewidth]{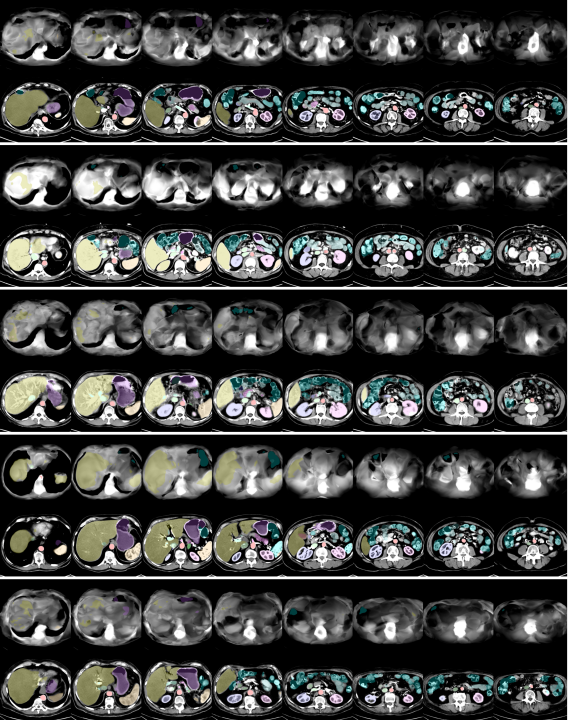}
    \caption{\textbf{Qualitative results of InTomo~\cite{zang2021intratomo} with and without \loss.} Every two rows showcase the results of eight CT slices of a single CT scan. \loss\ performs great anatomical preservation during enhancement. The images are center-cropped to eliminate the background. A soft-tissue window (300/50 HU) was used to enhance abdominal organ visibility.}
    \label{fig:supp_visual_compare-intratomo}
\end{figure}
\begin{figure}[ht]
    \centering
    \includegraphics[width=\linewidth]{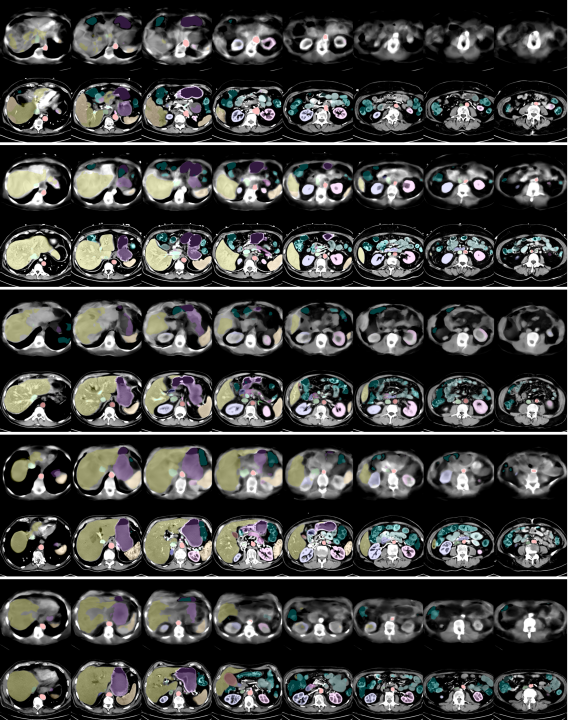}
    \caption{\textbf{Qualitative results of NeRF~\cite{mildenhall2021nerf} with and without \loss.} Every two rows showcase the results of eight CT slices of a single CT scan. \loss\ performs great anatomical preservation during enhancement. The images are center-cropped to eliminate the background. A soft-tissue window (300/50 HU) was used to enhance abdominal organ visibility.}
    \label{fig:supp_visual_compare-nerf}
\end{figure}
\begin{figure}[ht]
    \centering
    \includegraphics[width=\linewidth]{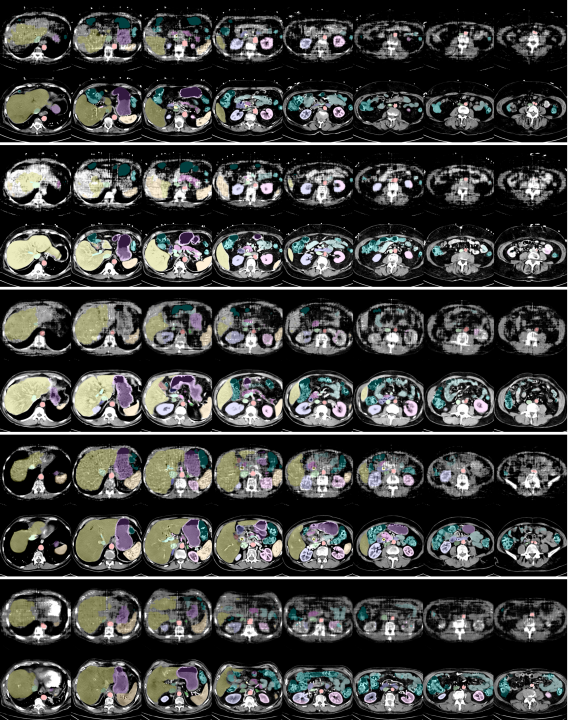}
    \caption{\textbf{Qualitative results of TensoRF~\cite{chen2022tensorf} with and without \loss.} Every two rows showcase the results of eight CT slices of a single CT scan. \loss\ performs great anatomical preservation during enhancement. The images are center-cropped to eliminate the background. A soft-tissue window (300/50 HU) was used to enhance abdominal organ visibility.}
    \label{fig:supp_visual_compare-tensorf}
\end{figure}
\begin{figure}[ht]
    \centering
    \includegraphics[width=\linewidth]{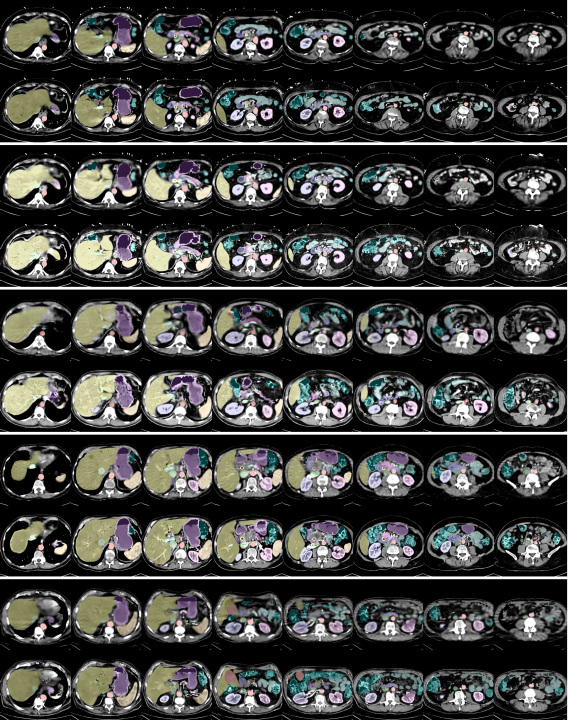}
    \caption{\textbf{Qualitative results of R$^2$-GS~\cite{zha2024r} with and without \loss.} Every two rows showcase the results of eight CT slices of a single CT scan. \loss\ performs great anatomical preservation during enhancement. The images are center-cropped to eliminate the background. A soft-tissue window (300/50 HU) was used to enhance abdominal organ visibility.}
    \label{fig:supp_visual_compare-r2GS}
\end{figure}
\begin{figure}[ht]
    \centering
    \includegraphics[width=\linewidth]{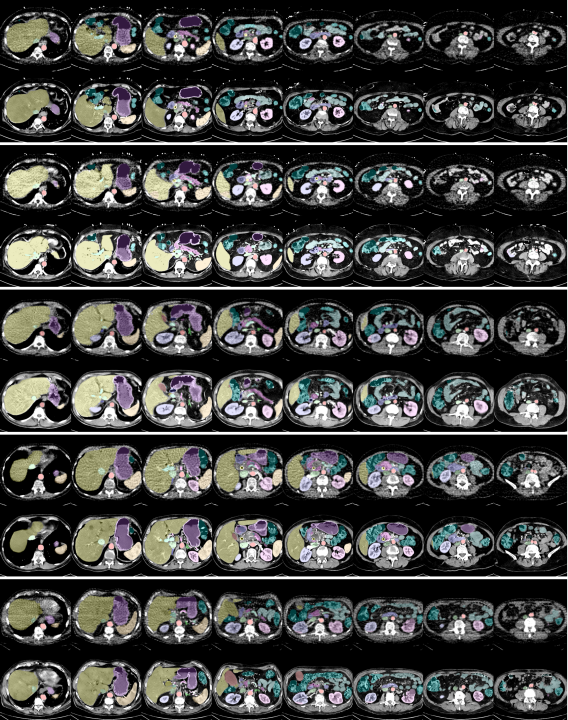}
    \caption{\textbf{Qualitative results of NAF~\cite{zha2022naf} with and without \loss.} Every two rows showcase the results of eight CT slices of a single CT scan. \loss\ performs great anatomical preservation during enhancement. The images are center-cropped to eliminate the background. A soft-tissue window (300/50 HU) was used to enhance abdominal organ visibility.}
    \label{fig:supp_visual_compare-naf}
\end{figure}
\begin{figure}[ht]
    \centering
    \includegraphics[width=\linewidth]{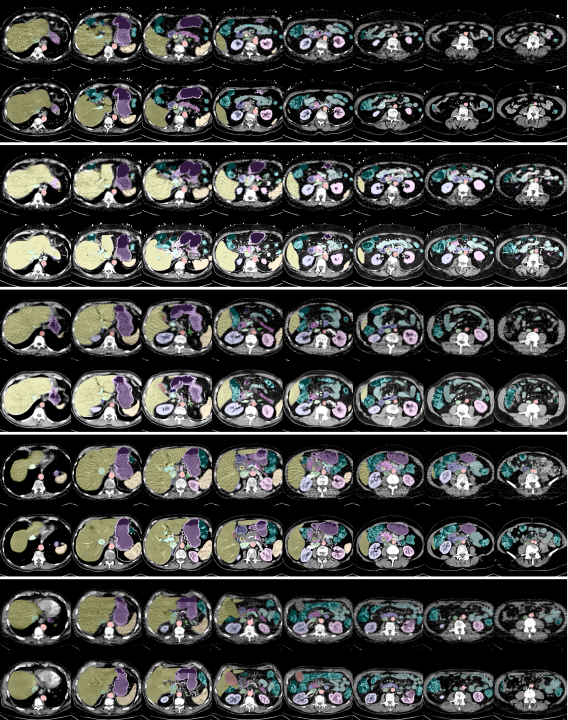}
    \caption{\textbf{Qualitative results of FDK~\cite{feldkamp1984practical} with and without \loss.} Every two rows showcase the results of eight CT slices of a single CT scan. \loss\ performs great anatomical preservation during enhancement. The images are center-cropped to eliminate the background. A soft-tissue window (300/50 HU) was used to enhance abdominal organ visibility.}
    \label{fig:supp_visual_compare-FDK}
\end{figure}
\begin{figure}[ht]
    \centering
    \includegraphics[width=\linewidth]{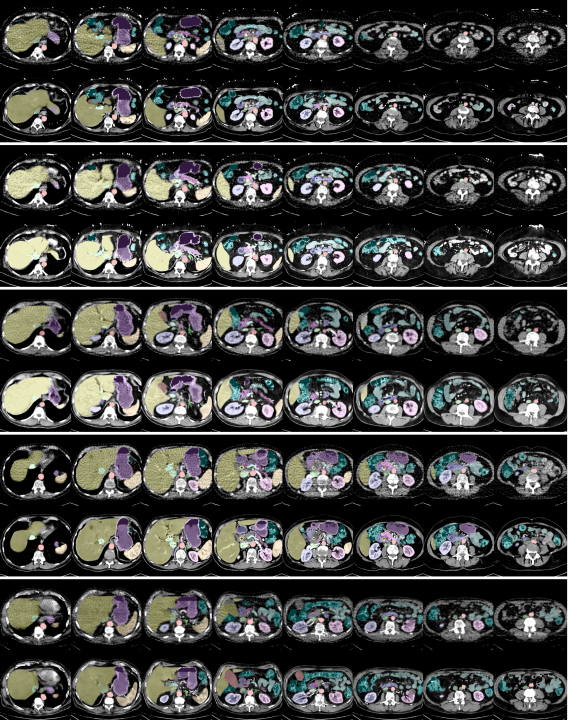}
    \caption{\textbf{Qualitative results of SART~\cite{andersen1984simultaneous} with and without \loss.} Every two rows showcase the results of eight CT slices of a single CT scan. \loss\ performs great anatomical preservation during enhancement. The images are center-cropped to eliminate the background. A soft-tissue window (300/50 HU) was used to enhance abdominal organ visibility.}
    \label{fig:supp_visual_compare-SART}
\end{figure}
\begin{figure}[ht]
    \centering
    \includegraphics[width=\linewidth]{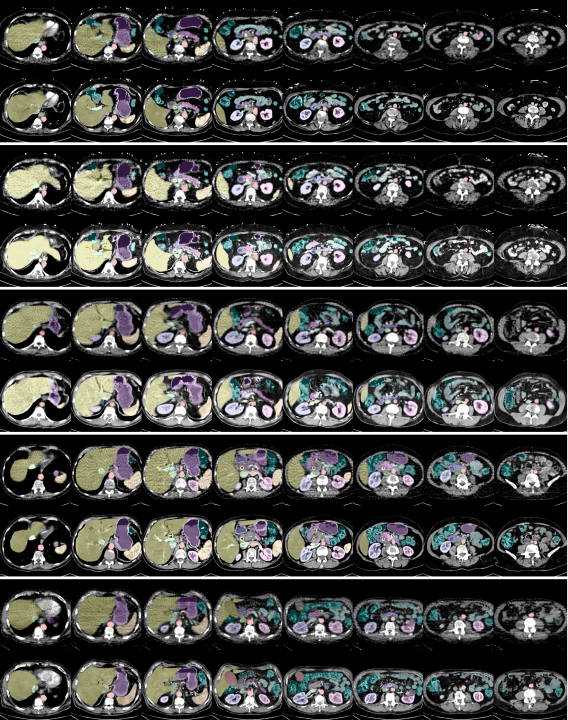}
    \caption{\textbf{Qualitative results of ASD-POCS~\cite{sidky2008image} with and without \loss.} Every two rows showcase the results of eight CT slices of a single CT scan. \loss\ performs great anatomical preservation during enhancement. The images are center-cropped to eliminate the background. A soft-tissue window (300/50 HU) was used to enhance abdominal organ visibility.}
    \label{fig:supp_visual_compare-ASD}
\end{figure}
\begin{figure}[ht]
    \centering
    \includegraphics[width=\linewidth]{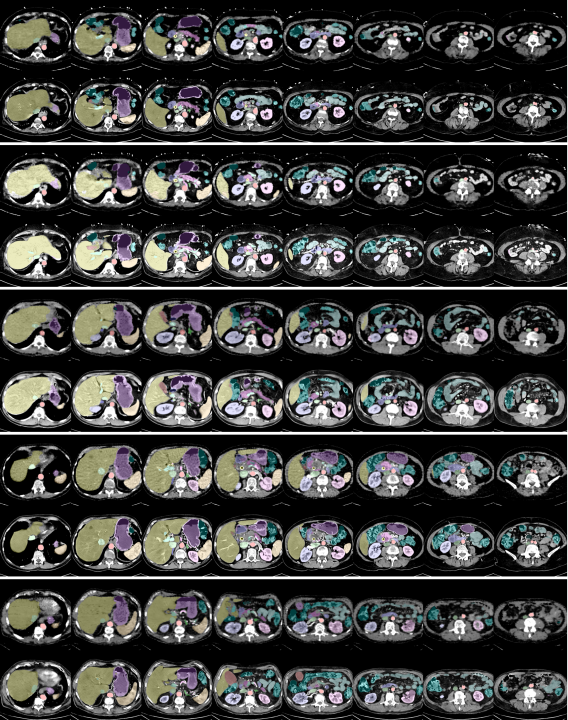}
    \caption{\textbf{Qualitative results of SAX-NeRF~\cite{cai2024structure} with and without \loss.} Every two rows showcase the results of eight CT slices of a single CT scan. \loss\ performs great anatomical preservation during enhancement. The images are center-cropped to eliminate the background. A soft-tissue window (300/50 HU) was used to enhance abdominal organ visibility.}
    \label{fig:supp_visual_compare-sax}
\end{figure}

\clearpage

\subsection{\loss\ on Arterial and Portal-venous Phase CT Scans}\label{sec:supp-AV}
\begin{table*}[ht]
    \centering
    \begin{threeparttable}{}
    \scriptsize
    \caption{\textbf{\loss-enhanced Reconstruction Evaluation on \emph{Arterial} Phase CT Scans.} Evaluate preexisting CT reconstruction methods on high-quality CT scans with \loss\ using both pixel-wise metrics and our anatomy-aware metrics. Note that the results are based on 23 arterial CT scans that \loss\ has never been trained on. We report the median and interquartile range (IQR) of these metrics and perform the Mann-Whitney U test for statistical analysis. Cells are marked in color only if \loss\ brings a significant difference ($p$ < 0.05), while \textcolor{green}{green} if the \loss\ enhancement results have improvement, and \textcolor{red}{red} otherwise. Deeper color represents greater difference.
    } 
    \label{tab:supp-CARE-results-A}
    \begin{tabular}{p{0.08\linewidth}p{0.08\linewidth}P{0.1\linewidth}P{0.1\linewidth}P{0.1\linewidth}P{0.1\linewidth}P{0.1\linewidth}P{0.1\linewidth}}
    \toprule
    &  & \multicolumn{2}{c}{pixel-wise metric} & \multicolumn{4}{c}{\textbf{anatomy-aware metric (ours)}}\\
    \cmidrule(lr){3-4}\cmidrule(lr){5-8}
    method &   & SSIM&PSNR& \makecell{ NSD$_\text{large}$} & \makecell{NSD$_\text{small}$} & clDice$_\text{intestine}$  & \makecell{ clDice$_\text{vessel}$}\\
    
    \midrule
    \multirow{2}{*}{InTomo} 
    & 50 views  &83.4\tiny{~(81.5,84.9)}  &24.7\tiny{~(23.5,25.4)}  &2.4\tiny{~(1.5,3.3)}  &0.0\tiny{~(0.0,0.0)}  &8.1\tiny{~(1.2,27.2)}  &9.3\tiny{~(0.0,18.6)} \\
& \multicolumn{1}{r}{\textbf{+\loss}}  &\cellcolor{red!10}76.3\tiny{~(74.5,80.4)}  &\cellcolor{red!10}22.2\tiny{~(21.1,23.0)}  &\cellcolor{green!30}28.8\tiny{~(20.8,34.8)}  &\cellcolor{green!10}5.5\tiny{~(1.7,10.5)}  &\cellcolor{green!30}30.2\tiny{~(23.7,39.6)}  &\cellcolor{green!40}47.4\tiny{~(34.6,60.8)}\\ 
    
    \midrule
    \multirow{2}{*}{NeRF}   
    & 50 views  &88.1\tiny{~(85.3,90.1)}  &27.0\tiny{~(24.8,27.7)}  &21.9\tiny{~(7.7,31.8)}  &0.1\tiny{~(0.0,2.8)}  &17.1\tiny{~(9.7,25.9)}  &28.9\tiny{~(24.1,44.7)} \\
& \multicolumn{1}{r}{\textbf{+\loss}}  &\cellcolor{red!10}82.1\tiny{~(79.6,83.1)}  &\cellcolor{red!10}24.0\tiny{~(22.5,24.7)}  &\cellcolor{green!40}58.2\tiny{~(46.2,61.0)}  &\cellcolor{green!30}24.4\tiny{~(13.8,28.4)}  &\cellcolor{green!40}50.4\tiny{~(39.3,56.1)}  &\cellcolor{green!30}57.7\tiny{~(45.4,68.8)}\\ 
    
    \midrule
    \multirow{2}{*}{TensoRF}  
    & 50 views  &90.2\tiny{~(87.8,92.3)}  &28.7\tiny{~(27.6,30.2)}  &41.3\tiny{~(21.6,53.3)}  &13.9\tiny{~(2.5,27.8)}  &31.1\tiny{~(20.9,35.8)}  &44.7\tiny{~(33.2,57.4)} \\
& \multicolumn{1}{r}{\textbf{+\loss}}  &\cellcolor{red!0}88.9\tiny{~(87.4,90.2)}  &\cellcolor{red!0}27.9\tiny{~(27.3,28.7)}  &\cellcolor{green!40}73.1\tiny{~(61.3,77.7)}  &\cellcolor{green!30}35.0\tiny{~(21.2,44.6)}  &\cellcolor{green!40}66.3\tiny{~(60.2,73.4)}  &\cellcolor{green!20}58.1\tiny{~(48.7,60.8)}\\
    
    \midrule
    \multirow{2}{*}{{R$^2$-GS}}   
    & 50 views  &93.5\tiny{~(92.3,94.8)}  &29.3\tiny{~(28.0,29.7)}  &75.8\tiny{~(63.5,84.2)}  &60.5\tiny{~(41.3,68.1)}  &53.4\tiny{~(42.9,66.6)}  &64.0\tiny{~(58.8,75.7)} \\
& \multicolumn{1}{r}{\textbf{+\loss}}  &\cellcolor{red!10}89.6\tiny{~(88.4,91.6)}  &\cellcolor{red!10}27.8\tiny{~(26.3,29.0)}  &\cellcolor{green!0}80.8\tiny{~(73.8,86.7)}  &\cellcolor{red!0}56.2\tiny{~(42.0,67.5)}  &\cellcolor{green!20}72.0\tiny{~(66.6,78.8)}  &\cellcolor{green!0}66.8\tiny{~(62.3,71.5)}\\
    
    \midrule
    \multirow{2}{*}{NAF}   
    & 50 views  &93.8\tiny{~(91.4,94.3)}  &29.9\tiny{~(29.0,31.3)}  &83.2\tiny{~(74.5,85.4)}  &60.9\tiny{~(49.2,69.0)}  &71.3\tiny{~(61.5,76.7)}  &75.4\tiny{~(67.0,79.1)} \\
& \multicolumn{1}{r}{\textbf{+\loss}}  &\cellcolor{red!0}92.8\tiny{~(91.3,93.6)}  &\cellcolor{green!0}29.9\tiny{~(28.5,30.4)}  &\cellcolor{green!10}87.0\tiny{~(80.9,90.1)}  &\cellcolor{green!0}63.0\tiny{~(48.7,74.3)}  &\cellcolor{green!10}80.8\tiny{~(73.1,84.2)}  &\cellcolor{red!0}68.2\tiny{~(60.5,74.3)}\\ 

    \midrule
    \multirow{2}{*}{FDK}
    & 50 views  &93.7\tiny{~(91.3,94.5)}  &29.8\tiny{~(29.2,31.3)}  &83.0\tiny{~(75.8,84.6)}  &61.7\tiny{~(48.6,71.2)}  &64.4\tiny{~(58.2,76.5)}  &77.2\tiny{~(64.8,79.7)} \\
& \multicolumn{1}{r}{\textbf{+\loss}}  &\cellcolor{red!0}92.8\tiny{~(91.5,93.7)}  &\cellcolor{green!0}30.3\tiny{~(29.0,31.0)}  &\cellcolor{green!0}85.7\tiny{~(79.1,88.7)}  &\cellcolor{green!0}65.8\tiny{~(54.1,75.5)}  &\cellcolor{green!20}81.4\tiny{~(72.6,84.9)}  &\cellcolor{red!0}67.8\tiny{~(65.8,73.2)}\\ 
    
    \midrule
    \multirow{2}{*}{SART}   
    & 50 views  &93.6\tiny{~(91.6,94.6)}  &30.2\tiny{~(29.4,31.4)}  &83.4\tiny{~(75.3,85.4)}  &60.9\tiny{~(52.1,70.0)}  &66.4\tiny{~(58.4,75.5)}  &78.8\tiny{~(66.8,82.6)} \\
& \multicolumn{1}{r}{\textbf{+\loss}}  &\cellcolor{red!0}93.4\tiny{~(92.7,94.6)}  &\cellcolor{green!0}30.6\tiny{~(29.8,31.2)}  &\cellcolor{green!0}87.1\tiny{~(81.6,90.1)}  &\cellcolor{green!0}68.1\tiny{~(54.0,75.6)}  &\cellcolor{green!20}83.4\tiny{~(75.8,86.1)}  &\cellcolor{red!20}66.5\tiny{~(55.6,75.3)}\\ 
    
    \midrule
    \multirow{2}{*}{ASD-POCS}   
    & 50 views  &93.6\tiny{~(91.8,94.5)}  &29.9\tiny{~(29.2,31.3)}  &83.0\tiny{~(72.9,85.8)}  &66.0\tiny{~(49.7,69.9)}  &70.5\tiny{~(57.6,77.0)}  &74.9\tiny{~(67.8,79.9)} \\
& \multicolumn{1}{r}{\textbf{+\loss}}  &\cellcolor{red!0}92.9\tiny{~(91.9,93.5)}  &\cellcolor{green!0}29.9\tiny{~(28.8,30.6)}  &\cellcolor{green!0}86.1\tiny{~(81.5,89.7)}  &\cellcolor{green!0}67.0\tiny{~(52.3,73.8)}  &\cellcolor{green!10}80.3\tiny{~(72.6,84.6)}  &\cellcolor{red!0}70.0\tiny{~(59.7,76.6)}\\

    \midrule
    \multirow{2}{*}{SAX-NeRF}   
    & 50 views  &93.9\tiny{~(93.0,95.9)}  &30.5\tiny{~(29.5,31.5)}  &82.8\tiny{~(72.9,88.6)}  &65.0\tiny{~(54.3,77.3)}  &66.2\tiny{~(54.4,77.5)}  &79.3\tiny{~(65.3,84.3)} \\
& \multicolumn{1}{r}{\textbf{+\loss}}  &\cellcolor{red!10}92.7\tiny{~(90.9,93.4)}  &\cellcolor{red!10}29.4\tiny{~(28.8,30.3)}  &\cellcolor{green!0}86.2\tiny{~(80.2,90.0)}  &\cellcolor{red!0}61.7\tiny{~(54.2,73.4)}  &\cellcolor{green!20}81.8\tiny{~(72.2,86.3)}  &\cellcolor{red!0}68.1\tiny{~(56.3,75.9)}\\ 
    
    \bottomrule
    \end{tabular}
    \end{threeparttable}
\end{table*}

\begin{table*}[ht]
    \centering
    \begin{threeparttable}{}
    \scriptsize
    \caption{\textbf{\loss-enhanced Reconstruction Evaluation on \emph{Portal Venous} Phase CT Scans.} Evaluate CT reconstruction methods on high-quality CT scans with \loss\ using both pixel-wise metrics and our anatomy-aware metrics. Note that the results are based on 13 portal venous CT scans that \loss\ has never been trained on. We report the median and interquartile range (IQR) of these metrics and perform the Mann-Whitney U test for statistical analysis. Cells are marked in color only if \loss\ brings significant difference ($p$ < 0.05), while \textcolor{green}{green} if the \loss\ enhancement results have improvement, and \textcolor{red}{red} otherwise. Deeper color represents greater difference.
    } 
    \label{tab:supp-CARE-results-V}
    \begin{tabular}{p{0.08\linewidth}p{0.08\linewidth}P{0.1\linewidth}P{0.1\linewidth}P{0.1\linewidth}P{0.1\linewidth}P{0.1\linewidth}P{0.1\linewidth}}
    \toprule
    &  & \multicolumn{2}{c}{pixel-wise metric} & \multicolumn{4}{c}{\textbf{anatomy-aware metric (ours)}}\\
    \cmidrule(lr){3-4}\cmidrule(lr){5-8}
    method &   & SSIM&PSNR& \makecell{ NSD$_\text{large}$} & \makecell{NSD$_\text{small}$} & clDice$_\text{intestine}$  & \makecell{ clDice$_\text{vessel}$}\\
    
    \midrule
    \multirow{2}{*}{InTomo} 
    & 50 views  &81.7\tiny{~(79.9,83.0)}  &23.9\tiny{~(22.8,24.6)}  &2.3\tiny{~(1.9,3.6)}  &0.0\tiny{~(0.0,0.0)}  &5.9\tiny{~(2.7,15.9)}  &0.0\tiny{~(0.0,15.6)} \\
& \multicolumn{1}{r}{\textbf{+\loss}}  &\cellcolor{red!10}77.3\tiny{~(73.3,78.1)}  &\cellcolor{red!10}21.5\tiny{~(20.2,22.3)}  &\cellcolor{green!40}42.0\tiny{~(27.9,47.6)}  &\cellcolor{green!10}9.9\tiny{~(5.9,15.2)}  &\cellcolor{green!40}42.3\tiny{~(33.5,47.1)}  &\cellcolor{green!50}48.6\tiny{~(38.7,63.4)}\\
    
    \midrule
    \multirow{2}{*}{NeRF}   
    & 50 views  &85.6\tiny{~(84.2,88.3)}  &25.3\tiny{~(24.7,26.7)}  &30.7\tiny{~(14.1,43.8)}  &0.2\tiny{~(0.0,7.5)}  &19.1\tiny{~(7.0,32.2)}  &29.2\tiny{~(11.7,40.0)} \\
& \multicolumn{1}{r}{\textbf{+\loss}}  &\cellcolor{red!10}79.0\tiny{~(78.4,81.0)}  &\cellcolor{red!10}22.2\tiny{~(21.1,22.6)}  &\cellcolor{green!30}57.7\tiny{~(56.0,66.5)}  &\cellcolor{green!30}27.3\tiny{~(14.3,31.5)}  &\cellcolor{green!40}56.6\tiny{~(52.6,61.1)}  &\cellcolor{green!30}52.9\tiny{~(45.4,56.7)}\\
    
    \midrule
    \multirow{2}{*}{TensoRF}  
    & 50 views  &87.3\tiny{~(86.1,90.3)}  &27.6\tiny{~(26.8,28.0)}  &47.9\tiny{~(23.7,53.2)}  &8.6\tiny{~(0.0,30.9)}  &22.8\tiny{~(14.7,37.4)}  &51.2\tiny{~(44.5,58.1)} \\
& \multicolumn{1}{r}{\textbf{+\loss}}  &\cellcolor{green!0}88.7\tiny{~(86.5,89.7)}  &\cellcolor{red!0}27.3\tiny{~(26.9,27.5)}  &\cellcolor{green!40}78.1\tiny{~(71.6,82.0)}  &\cellcolor{green!40}40.6\tiny{~(39.2,46.8)}  &\cellcolor{green!50}71.3\tiny{~(65.8,77.4)}  &\cellcolor{green!20}63.0\tiny{~(51.8,75.2)}\\ 
    
    \midrule
    \multirow{2}{*}{{R$^2$-GS}}   
    & 50 views  &92.4\tiny{~(89.7,93.5)}  &27.7\tiny{~(26.3,28.7)}  &80.8\tiny{~(67.4,84.7)}  &53.0\tiny{~(41.8,73.6)}  &54.5\tiny{~(50.9,73.8)}  &71.0\tiny{~(60.1,81.0)} \\
& \multicolumn{1}{r}{\textbf{+\loss}}  &\cellcolor{red!10}88.4\tiny{~(84.8,89.9)}  &\cellcolor{red!0}27.3\tiny{~(24.9,27.8)}  &\cellcolor{green!0}83.7\tiny{~(74.4,89.7)}  &\cellcolor{green!0}55.1\tiny{~(45.6,60.2)}  &\cellcolor{green!30}80.7\tiny{~(76.1,84.9)}  &\cellcolor{green!0}76.3\tiny{~(56.8,86.2)}\\ 
    
    \midrule
    \multirow{2}{*}{NAF}   
    & 50 views  &92.0\tiny{~(90.4,92.8)}  &29.3\tiny{~(28.0,29.6)}  &84.7\tiny{~(81.1,87.6)}  &61.1\tiny{~(52.2,74.9)}  &71.9\tiny{~(63.6,85.6)}  &79.2\tiny{~(73.0,82.5)} \\
& \multicolumn{1}{r}{\textbf{+\loss}}  &\cellcolor{green!0}92.6\tiny{~(92.0,92.8)}  &\cellcolor{green!0}29.5\tiny{~(29.0,30.0)}  &\cellcolor{green!10}91.0\tiny{~(86.4,91.8)}  &\cellcolor{green!0}72.3\tiny{~(59.8,77.6)}  &\cellcolor{green!0}82.9\tiny{~(78.4,90.5)}  &\cellcolor{green!0}83.8\tiny{~(77.9,88.4)}\\ 

    \midrule
    \multirow{2}{*}{FDK}
    & 50 views  &92.0\tiny{~(90.3,93.1)}  &29.1\tiny{~(28.3,29.5)}  &85.7\tiny{~(84.1,86.9)}  &63.3\tiny{~(57.5,69.3)}  &70.1\tiny{~(66.5,83.8)}  &75.9\tiny{~(71.2,82.4)} \\
& \multicolumn{1}{r}{\textbf{+\loss}}  &\cellcolor{green!0}92.2\tiny{~(91.3,92.8)}  &\cellcolor{green!0}29.3\tiny{~(29.0,29.7)}  &\cellcolor{green!0}88.9\tiny{~(86.1,91.4)}  &\cellcolor{green!0}72.9\tiny{~(62.0,79.0)}  &\cellcolor{green!20}81.8\tiny{~(76.3,86.5)}  &\cellcolor{green!0}83.9\tiny{~(74.8,88.0)}\\ 
    
    \midrule
    \multirow{2}{*}{SART}   
    & 50 views  &91.9\tiny{~(90.6,92.6)}  &29.2\tiny{~(28.2,29.6)}  &87.5\tiny{~(77.1,90.2)}  &67.2\tiny{~(55.4,77.2)}  &71.1\tiny{~(63.5,90.0)}  &77.8\tiny{~(64.8,84.7)} \\
& \multicolumn{1}{r}{\textbf{+\loss}}  &\cellcolor{green!10}93.3\tiny{~(92.7,93.9)}  &\cellcolor{green!10}30.0\tiny{~(29.5,30.3)}  &\cellcolor{green!0}89.8\tiny{~(86.5,91.4)}  &\cellcolor{green!0}70.3\tiny{~(61.7,75.2)}  &\cellcolor{green!0}85.7\tiny{~(79.7,88.2)}  &\cellcolor{green!0}83.6\tiny{~(79.5,88.2)}\\ 
    
    \midrule
    \multirow{2}{*}{ASD-POCS}   
    & 50 views  &92.1\tiny{~(90.7,92.9)}  &29.2\tiny{~(28.2,29.7)}  &86.4\tiny{~(79.6,88.8)}  &60.2\tiny{~(59.7,78.4)}  &72.0\tiny{~(67.1,80.3)}  &74.5\tiny{~(67.4,83.5)} \\
& \multicolumn{1}{r}{\textbf{+\loss}}  &\cellcolor{green!0}92.2\tiny{~(91.6,92.8)}  &\cellcolor{red!0}28.8\tiny{~(28.3,29.3)}  &\cellcolor{green!0}88.8\tiny{~(84.2,91.4)}  &\cellcolor{green!0}71.0\tiny{~(60.3,75.7)}  &\cellcolor{green!10}81.5\tiny{~(78.7,86.5)}  &\cellcolor{green!0}83.6\tiny{~(78.7,86.3)}\\

    \midrule
    \multirow{2}{*}{SAX-NeRF}   
    & 50 views  &93.2\tiny{~(92.3,94.1)}  &29.3\tiny{~(28.8,29.6)}  &79.3\tiny{~(75.9,90.0)}  &68.8\tiny{~(56.2,79.1)}  &71.1\tiny{~(61.0,79.8)}  &79.3\tiny{~(58.6,87.2)} \\
& \multicolumn{1}{r}{\textbf{+\loss}}  &\cellcolor{red!0}92.3\tiny{~(91.5,93.3)}  &\cellcolor{green!0}29.7\tiny{~(28.9,29.9)}  &\cellcolor{green!0}90.8\tiny{~(86.3,91.8)}  &\cellcolor{green!0}70.5\tiny{~(63.7,78.9)}  &\cellcolor{green!10}80.5\tiny{~(77.2,90.0)}  &\cellcolor{green!0}83.8\tiny{~(76.5,88.4)}\\ 
    
    \bottomrule
    \end{tabular}
    \end{threeparttable}
\end{table*}

\clearpage

\section*{NeurIPS Paper Checklist}

\begin{enumerate}

\item {\bf Claims}
    \item[] Question: Do the main claims made in the abstract and introduction accurately reflect the paper's contributions and scope?
    \item[] Answer: \answerYes{} 
    \item[] Justification: The abstract and introduction accurately reflect the paper’s contributions by clearly framing the limitations of pixel-wise metrics, introducing our anatomy-aware evaluation framework and CARE enhancement pipeline, and previewing the empirical gains on multiple reconstruction backbones. 

\item {\bf Limitations}
    \item[] Question: Does the paper discuss the limitations of the work performed by the authors?
    \item[] Answer: \answerYes{} 
    \item[] Justification: The paper discusses the limitations of the work performed by the authors in \S~\ref{sec:disscussion-and-conclusion}. 

\item {\bf Theory assumptions and proofs}
    \item[] Question: For each theoretical result, does the paper provide the full set of assumptions and a complete (and correct) proof?
    \item[] Answer: \answerNA{} 
    \item[] Justification: The paper does not include theoretical results. Its contributions lie in proposing a suite of anatomy-aware evaluation metrics, introducing a diffusion-based CT enhancement framework \loss, and providing thorough empirical validation, rather than in formal assumptions or mathematical proofs. 

    \item {\bf Experimental result reproducibility}
    \item[] Question: Does the paper fully disclose all the information needed to reproduce the main experimental results of the paper to the extent that it affects the main claims and/or conclusions of the paper (regardless of whether the code and data are provided or not)?
    \item[] Answer: \answerYes{} 
    \item[] Justification: We provide the experimental details both in \S~\ref{sec:method-CARE} and in the supplementary materials.

\item {\bf Open access to data and code}
    \item[] Question: Does the paper provide open access to the data and code, with sufficient instructions to faithfully reproduce the main experimental results, as described in supplemental material?
    \item[] Answer: \answerYes{} 
    \item[] Justification: Note that we only provide open access to code and model checkpoints; our data would be kept private. The code is attached in the supplementary materials with a link to the model checkpoint attached. 

\item {\bf Experimental setting/details}
    \item[] Question: Does the paper specify all the training and test details (e.g., data splits, hyperparameters, how they were chosen, type of optimizer, etc.) necessary to understand the results?
    \item[] Answer: \answerYes{} 
    \item[] Justification: The experimental setting/details of training and testing are provided in \S~\ref{sec:method-CARE} and the supplementary materials. 

\item {\bf Experiment statistical significance}
    \item[] Question: Does the paper report error bars suitably and correctly defined or other appropriate information about the statistical significance of the experiments?
    \item[] Answer: \answerYes{} 
    \item[] Justification: The paper performs the Mann-Whitney U test at \tableautorefname~\ref{tab:CARE-results} to mark significance difference between our results with baselines. 

\item {\bf Experiments compute resources}
    \item[] Question: For each experiment, does the paper provide sufficient information on the computer resources (type of compute workers, memory, time of execution) needed to reproduce the experiments?
    \item[] Answer: \answerYes{} 
    \item[] Justification: The paper provides sufficient information on the computer resources in the supplementary materials. 
    
\item {\bf Code of ethics}
    \item[] Question: Does the research conducted in the paper conform, in every respect, with the NeurIPS Code of Ethics \url{https://neurips.cc/public/EthicsGuidelines}?
    \item[] Answer: \answerYes{} 
    \item[] Justification: We preserve anonymity in all submitted materials and strictly conform the NeurIPS Code of Ethics in every respect. All patient-identifiable information was anonymized during preprocessing to ensure privacy protection, and the released dataset contains no identifiable information. 

\item {\bf Broader impacts}
    \item[] Question: Does the paper discuss both potential positive societal impacts and negative societal impacts of the work performed?
    \item[] Answer: \answerYes{} 
    \item[] Justification: See \S~\ref{sec:introduction} and \S~\ref{sec:disscussion-and-conclusion}. 
    
\item {\bf Safeguards}
    \item[] Question: Does the paper describe safeguards that have been put in place for responsible release of data or models that have a high risk for misuse (e.g., pretrained language models, image generators, or scraped datasets)?
    \item[] Answer: \answerNA{} 
    \item[] Justification: The paper poses no such risks. 

\item {\bf Licenses for existing assets}
    \item[] Question: Are the creators or original owners of assets (e.g., code, data, models), used in the paper, properly credited and are the license and terms of use explicitly mentioned and properly respected?
    \item[] Answer: \answerYes{} 
    \item[] Justification: The paper has cited, credited and listed the licenses of all the papers of the code bases used in \S~\ref{sec:introduction} and supplementary materials. 

\item {\bf New assets}
    \item[] Question: Are new assets introduced in the paper well documented and is the documentation provided alongside the assets?
    \item[] Answer: \answerYes{} 
    \item[] Justification: The paper introduce new code and models and they are all well documented in the supplementary materials. 

\item {\bf Crowdsourcing and research with human subjects}
    \item[] Question: For crowdsourcing experiments and research with human subjects, does the paper include the full text of instructions given to participants and screenshots, if applicable, as well as details about compensation (if any)? 
    \item[] Answer: \answerNA{} 
    \item[] Justification: The paper only uses pre-acquired CT scans, and does not involve crowdsourcing nor research with human subjects.

\item {\bf Institutional review board (IRB) approvals or equivalent for research with human subjects}
    \item[] Question: Does the paper describe potential risks incurred by study participants, whether such risks were disclosed to the subjects, and whether Institutional Review Board (IRB) approvals (or an equivalent approval/review based on the requirements of your country or institution) were obtained?
    \item[] Answer: \answerNA{} 
    \item[] Justification: The paper only uses pre-acquired CT scans, and does not involve crowdsourcing nor research with human subjects. 

\item {\bf Declaration of LLM usage}
    \item[] Question: Does the paper describe the usage of LLMs if it is an important, original, or non-standard component of the core methods in this research? Note that if the LLM is used only for writing, editing, or formatting purposes and does not impact the core methodology, scientific rigorousness, or originality of the research, declaration is not required.
    \item[] Answer: \answerNA{} 
    \item[] Justification: The core method development in this research does not involve LLMs.

\end{enumerate}

\end{document}